\documentclass[structabstract]{aa}
\usepackage{txfonts}
\usepackage{graphicx}
\usepackage{natbib}
\usepackage{longtable}
\usepackage{subeqnarray}
\usepackage{cases}
\usepackage{supertabular}

\usepackage[colorlinks=true,citecolor=blue]{hyperref}

\begin{document}

\title{ATLASGAL-selected massive clumps in the inner Galaxy:}
\subtitle{VI. Kinetic temperature and spatial density measured with formaldehyde}

\author{X. D. Tang\inst{1,2,3}
\and C. Henkel\inst{1,4}
\and F. Wyrowski\inst{1}
\and A. Giannetti\inst{1,5}
\and K. M. Menten\inst{1}
\and T. Csengeri\inst{1}
\and S. Leurini\inst{1,6}
\and J. S. Urquhart\inst{1,7}
\and C. K\"{o}nig\inst{1}
\and R. G\"{u}sten\inst{1}
\and Y. X. Lin\inst{1}
\and X. W. Zheng\inst{8}
\and J. Esimbek\inst{2,3}
\and J. J. Zhou\inst{2,3}}

\titlerunning{Kinetic temperature and spatial density in ATLASGAL-selected massive clumps}
\authorrunning{X. D. Tang et al.}

\institute{Max-Planck-Institut f\"{u}r Radioastronomie, Auf dem H\"{u}gel 69, 53121 Bonn, Germany\\
\email{xdtang@mpifr-bonn.mpg.de}
\and Xinjiang Astronomical Observatory, Chinese Academy of Sciences, 830011 Urumqi, PR China
\and Key Laboratory of Radio Astronomy, Chinese Academy of Sciences, 830011 Urumqi, PR China
\and Astronomy Department, King Abdulaziz University, PO Box 80203, 21589 Jeddah, Saudi Arabia
\and INAF--Istituto di Radioastronomia \& Italian ALMA Regional Centre, Via P. Gobetti 101, I--40129 Bologna, Italy
\and INAF--Osservatorio Astronomico di Cagliari, Via della Scienza 5, I--09047, Selargius (CA), Italy
\and School of Physical Sciences, University of Kent, Ingram Building, Canterbury, Kent CT2 7NH, UK
\and School of Astronomy and Space Science, Nanjing University, 210093 Nanjing, PR China}


\abstract{Formaldehyde (H$_2$CO) is a reliable tracer to accurately measure
the physical parameters of dense gas in star forming regions.
We aim to directly determine the kinetic temperature and spatial density
with formaldehyde for the $\sim$100 brightest ATLASGAL-selected clumps
(the TOP100 sample) at 870\,$\mu$m representing various evolutionary stages of high-mass star formation.
Ten transitions ($J$\,=\,3--2 and 4--3) of ortho- and para-H$_2$CO
near 211, 218, 225, and 291\,GHz were observed with the Atacama Pathfinder
EXperiment (APEX) 12\,m telescope. Using non-LTE models with RADEX, we derive the
gas kinetic temperature and spatial density using the measured
para-H$_2$CO\,3$_{21}$--2$_{20}$/3$_{03}$--2$_{02}$,
4$_{22}$--3$_{21}$/4$_{04}$--3$_{03}$,
and 4$_{04}$--3$_{03}$/3$_{03}$--2$_{02}$ ratios.
The gas kinetic temperatures derived from the
para-H$_2$CO\,3$_{21}$--2$_{20}$/3$_{03}$--2$_{02}$ and
4$_{22}$--3$_{21}$/4$_{04}$--3$_{03}$ line
ratios are high, ranging from 43 to $>$300\,K
with an unweighted average of 91\,$\pm$\,4\,K. Deduced $T_{\rm kin}$ values
from the $J$\,=\,3--2 and 4--3 transitions are similar.
Spatial densities of the gas derived
from the para-H$_2$CO\,4$_{04}$--3$_{03}$/3$_{03}$--2$_{02}$ line ratios
yield 0.6--8.3\,$\times$\,10$^6$ cm$^{-3}$ with
an unweighted average of 1.5\,($\pm$0.1)\,$\times$\,10$^6$ cm$^{-3}$.
A comparison of kinetic temperatures derived from
para-H$_2$CO, NH$_3$, and the dust emission indicates
that para-H$_2$CO traces a distinctly higher temperature than the
NH$_3$\,(2,2)/(1,1) transitions and the dust, tracing
heated gas more directly associated with the star formation process.
The H$_2$CO linewidths are found to be correlated with bolometric luminosity
and increase with the evolutionary stage of the clumps, which suggests
that higher luminosities tend to be associated with a more turbulent
molecular medium. It seems that the spatial densities measured with H$_2$CO do not
vary significantly with the evolutionary stage of the clumps.
However, averaged gas kinetic temperatures derived from H$_2$CO
increase with time through the evolution of the clumps.
The high temperature of the gas traced by H$_2$CO may be
mainly caused by radiation from embedded young massive stars
and the interaction of outflows with the ambient medium.
For $L_{\rm bol}$/$M_{\rm clump}$\,$\gtrsim$\,10\,L$_{\odot}$/M$_{\odot}$,
we find a rough correlation between gas kinetic temperature
and this ratio, which is indicative of the
evolutionary stage of the individual clumps.
The strong relationship between H$_2$CO line luminosities and
clump masses is apparently linear during the late evolutionary stages of the clumps,
indicating that $L_{\rm H_2CO}$ does reliably trace the mass of warm dense
molecular gas. In our massive clumps H$_2$CO line luminosities
are approximately linearly correlated with
bolometric luminosities over about four orders
of magnitude in $L_{\rm bol}$, which suggests that the mass of dense
molecular gas traced by the H$_2$CO line luminosity is well correlated with star formation.}

\maketitle

\section{Introduction}
\label{sect:Introduction}
In the Galactic disk, star formation appears to occur only in dense regions
(spatial density $n$(H$_2$)\,$\gtrsim$\,10$^4$\,cm$^{-3}$) composed of
molecular gas \citep{Lada2010,Ginsburg2015}.
High mass stars form in massive clumps with typical size of order $\sim$1 pc
(e.g., \citealt{Dunham2010,Dunham2011,Rosolowsky2010,
Urquhart2014,He2015,Wienen2015,Konig2017,Yuan2017}).
High-mass stars influence
the surrounding environment and subsequent star formation through their
feedback such as outflows, winds, as well as UV radiation.
However, the details of the high mass star formation process and
how their feedback may affect the initial conditions of high
mass stars in their formation process are still far from being clear
and require, as a basis, the precise determination of kinetic temperature
and density.

The Atacama Pathfinder EXperiment (APEX) Telescope Large Area
Survey of the GALaxy (ATLASGAL) \citep{Schuller2009},
presenting observations in a Galactic longitude and latitude
range of $\pm$60$^\circ$ and $\pm$1.5$^\circ$, respectively,
introduces a global view on
star formation at 870\,$\mu$m and identifies $\sim$10,000
massive clumps in various stages of evolution undergoing
high mass star formation in the inner Galaxy \citep{Contreras2013,Urquhart2014,Urquhart2017,Csengeri2014}.
The most fundamental physical parameters, kinetic temperature
and spatial density of the clumps, affect chemistry, star formation,
and could also impact the stellar initial mass function.
Accurate measurements of
these physical parameters are indispensable for a general
understanding of the physical processes involved in these massive
star-forming clumps.

Formaldehyde (H$_2$CO) is a ubiquitous molecule in interstellar clouds
\citep{Downes1980,Bieging1982,Henkel1991,Zylka1992,Mangum2008,
Mangum2013a,Ao2013,Tang2013,Ginsburg2015,Ginsburg2016,Ginsburg2017,Guo2016}.
As a slightly asymmetric rotor molecule,
H$_2$CO exhibits a large number of millimeter and submillimeter transitions.
It is a reliable tracer of physical conditions such as temperature
and density \citep{Henkel1980,Henkel1983,Mangum1993a,Muhle2007,Ginsburg2011,
Ginsburg2015,Ginsburg2016,Ao2013}.
Since the relative populations of the $K_{\rm a}$ ladders of H$_2$CO are
predominantly governed by collisions,
ratios of H$_2$CO line fluxes involving different
$K_{\rm a}$ ladders are good tracers of the kinetic
temperature, such as para-H$_2$CO $J_{\rm K_aK_c}$\,=\,3$_{22}$--2$_{21}$/3$_{03}$--2$_{02}$,
4$_{23}$--3$_{22}$/4$_{04}$--3$_{03}$,
and 5$_{23}$--4$_{22}$/5$_{05}$--4$_{04}$ \citep{Mangum1993a}.
Once the kinetic temperature is known, line ratios involving
the same $K_{\rm a}$ ladders yield estimates of the
spatial density of the gas, such as
$J_{\rm K_aK_c}$\,=\,4$_{04}$--3$_{03}$/3$_{03}$--2$_{02}$,
5$_{05}$--4$_{04}$/3$_{03}$--2$_{02}$, and
5$_{24}$--4$_{23}$/3$_{22}$--2$_{21}$
\citep{Mangum1993a,Muhle2007,Immer2016}.
Transitions connecting the same rotational levels (e.g., $J$\,=\,3--2 or 4--3)
and belonging to either the para- or ortho-H$_2$CO subspecies,
but being part of different $K_{\rm a}$ ladders (e.g., $K_{\rm a}$\,=\,0, 2) are
particularly useful. They can be measured simultaneously
with the same receiver system and their relative
strengths (para-H$_2$CO\,3$_{22}$--2$_{21}$/3$_{03}$--2$_{02}$,
3$_{21}$--2$_{20}$/3$_{03}$--2$_{02}$, 4$_{23}$--3$_{22}$/4$_{04}$--3$_{03}$,
and 4$_{22}$--3$_{21}$/4$_{04}$--3$_{03}$
provide sensitive thermometry. Para-H$_2$CO is therefore possibly
the best of the very few molecular tracers that are available
for such an analysis of the dense molecular gas. H$_2$CO line ratios have been
used to measure physical parameters in our Galactic center clouds
\citep{Qin2008,Ao2013,Johnston2014,Ginsburg2016,Immer2016,Lu2017},
star formation regions \citep{Mangum1993a,Hurt1996,Mangum1999,
Mitchell2001,Watanabe2008,Nagy2012,Lindberg2015,Tang2017a,Tang2017c},
as well as in external galaxies \citep{Muhle2007,Tang2017b}.

In this work, we aim to directly measure the kinetic
temperature and spatial density toward massive
star-forming clumps selected from the ATLASGAL survey
making use of the rotational transitions of
H$_2$CO ($J$\,=\,3--2 and 4--3). Our main goals are
(a) comparing kinetic temperatures from the gas to
temperature estimates based on the dust,
(b) searching for a correlation between kinetic temperature
and linewidth, which is expected in the case of conversion of turbulent
energy into heat, (c) seeking for links between kinetic
temperature and star formation rate as well as evolutionary stage
of the massive star forming regions, and (d) testing the
star formation law by correlating the luminosity of
the H$_2$CO lines to infrared luminosity.

In Sections \ref{sect:samp-obs} and \ref{sect:Results},
we describe the measured samples, our H$_2$CO observations and the data
reduction, and introduce the main results.
The discussion is presented in Section \ref{sect:Discussion}.
Our main conclusions are summarized in Section \ref{sect:Summary}.

\section{Sample, observations and data reduction}
\label{sect:samp-obs}
We have selected the 110 brightest clumps from the ATLASGAL
survey (the TOP100 sample) obeying simple IR criteria to cover a range in
evolutionary stages as described in \cite{Giannetti2014} and \cite{Konig2017}.
They consist almost entirely
of clumps that have the potential to form, or are forming,
massive stars. Depending on their IR and radio continuum
properties, the sample of potentially high-mass star forming clumps
at various evolutionary stages can be separated into four
categories: 70\,$\mu$m weak sources (70w), infrared weak
clumps (IRw), infrared bright objects (IRb),
and sources containing compact H\,{\scriptsize II}
regions (H\,{\scriptsize II}) \citep{Giannetti2014,Konig2017}.
Previous work on this sample addressed
SiO emission (for parts of the sample, \citealt{Csengeri2016}),
dust continuum characterization \citep{Konig2017},
millimeter hydrogen recombination lines
(for more evolved (i.e. H\,{\scriptsize II} regions) parts of the sample, \citealt{Kim2017}),
and temperature structure \citep{Giannetti2017}.
The TOP100 is an ideal sample to study the
physical and chemical parameters of the
potentially massive star-forming regions at
various evolutionary stages.

\begin{table}[t]
\small
\caption{Observed H$_2$CO transition parameters.}
\label{table:APEX-receiver}
\centering
\begin{tabular}
{ccccc}
\hline\hline 
Transition &Frequency &$E_{\rm u}$ &Receiver &Beam size \\
&GHz &K & &arcsec \\
\hline 
o-H$_2$CO 3$_{13}$--2$_{12}$ &211.212 &32.06  &PI230 &29.5 \\
p-H$_2$CO 3$_{03}$--2$_{02}$ &218.222 &20.96  &PI230 &28.6 \\
p-H$_2$CO 3$_{22}$--2$_{21}$ &218.476 &68.09  &PI230 &28.6 \\
p-H$_2$CO 3$_{21}$--2$_{20}$ &218.760 &68.11  &PI230 &28.5 \\
o-H$_2$CO 3$_{12}$--2$_{11}$ &225.699 &33.45  &PI230 &27.6 \\
p-H$_2$CO 4$_{04}$--3$_{03}$ &290.623 &34.90  &FLASH &21.5 \\
p-H$_2$CO 4$_{23}$--3$_{22}$ &291.238 &82.07  &FLASH &21.4 \\
o-H$_2$CO 4$_{32}$--3$_{31}$ &291.381 &140.94 &FLASH &21.4 \\
o-H$_2$CO 4$_{31}$--3$_{30}$ &291.384 &140.94 &FLASH &21.4 \\
p-H$_2$CO 4$_{22}$--3$_{21}$ &291.948 &82.12  &FLASH &21.4 \\
\hline 
\end{tabular}
\end{table}

\begin{table}[t]
\small
\caption{Observed H$_2$CO transitions and detection rates.}
\label{table:detection rate}
\centering
\begin{tabular}
{ccccc}
\hline\hline 
Transition & Observed & Detection & Detection rate \\
\hline 
o-H$_2$CO 3$_{13}$--2$_{12}$ &94 &91 &97\% \\
p-H$_2$CO 3$_{03}$--2$_{02}$ &94 &92 &98\% \\
p-H$_2$CO 3$_{22}$--2$_{21}$ &94 &65 &69\% \\
p-H$_2$CO 3$_{21}$--2$_{20}$ &94 &66 &70\% \\
o-H$_2$CO 3$_{12}$--2$_{11}$ &94 &93 &99\% \\
p-H$_2$CO 4$_{04}$--3$_{03}$ &98 &97 &99\% \\
p-H$_2$CO 4$_{23}$--3$_{22}$ &98 &80 &82\% \\
o-H$_2$CO 4$_{32}$--3$_{31}$ &98 &83 &85\% \\
o-H$_2$CO 4$_{31}$--3$_{30}$ &98 &83 &85\% \\
p-H$_2$CO 4$_{22}$--3$_{21}$ &98 &83 &85\% \\
\hline 
\end{tabular}
\end{table}

\begin{figure*}[t]
\vspace*{0.2mm}
\begin{center}
\includegraphics[width=1.00\textwidth]{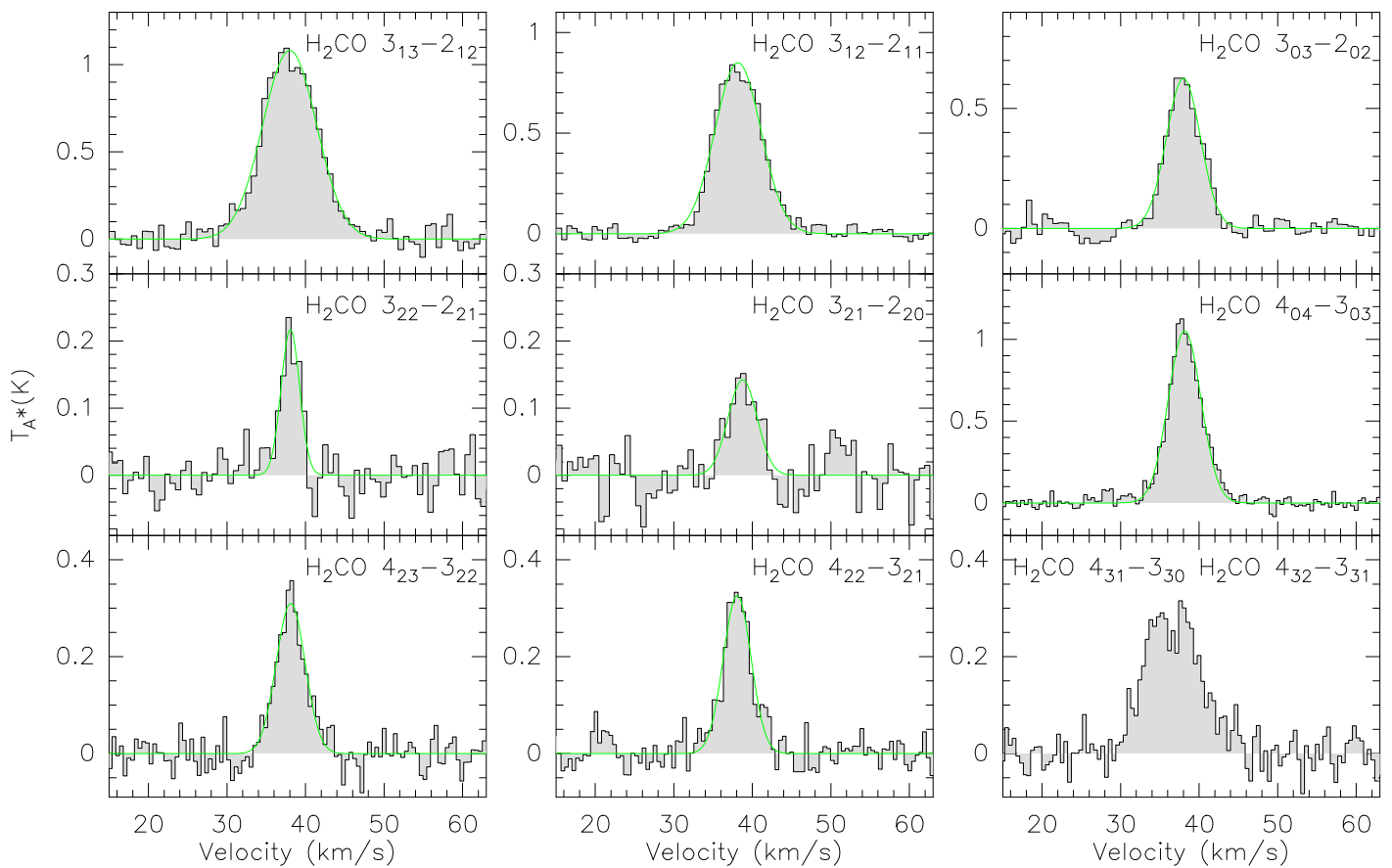}
\end{center}
\caption{Observed H$_2$CO spectra (in grey colour) toward AGAL008.684$-$00.367.
Green lines indicate the Gaussian fit results.}
\label{fig:H2CO-spectra}
\end{figure*}

Sources observed are listed in Table \ref{table:source}.
Our observations were carried out on 2013 July and December,
2014 September and November, and 2015 April, June, July, and
October with the Atacama Pathfinder
EXperiment (APEX\footnote{\tiny This publication is based
on data acquired with the Atacama Pathfinder EXperiment (APEX).
APEX is a collaboration between the Max-Planck-Institut
f\"{u}r Radioastronomie, the European Southern Observatory,
and the Onsala Space Observatory.}) 12\,m telescope located on Chajnantor (Chile).
Specific observational details of the ten measured transitions of
H$_2$CO are listed in Table \ref{table:APEX-receiver}.
Five transitions of H$_2$CO ($J$\,=\,3--2) were observed with the new MPIfR 1-mm
receiver (PI230) with a beam size from 27.6$''$ to 29.5$''$ and
integration times of 1 to 3 minutes.
Five H$_2$CO ($J$\,=\,4--3) transitions were observed with the FLASH
receiver with a beam size $\sim$21.4$''$ and integration
times of 2 to 4 minutes.
For the PI230 receiver, we used a Fast Fourier
Transform Spectrometer (FFTS4G) backend
with two sidebands (Lower and Upper). Each sideband has two spectral
windows of 4\,GHz bandwidth, providing both orthogonal polarizations,
and leading to a total bandwidth of 8\,GHz. An eXtended bandwidth
Fast Fourier Transform Spectrometer (XFFTS) backend with
two spectral windows of 2.5\,GHz bandwidth leading to a total
bandwidth of 4\,GHz was used for the FLASH receiver. These provide velocity
resolutions of $\sim$0.08\,km\,s$^{-1}$ for H$_2$CO ($J$\,=\,3--2)
and $\sim$0.04\,km\,s$^{-1}$ for H$_2$CO ($J$\,=\,4--3).
The observations were performed in position-switching
mode with off-positions offset from the on-position of
the sources by (600$''$,$\pm$600$''$).
We converted the antenna temperatures of the spectra
into main beam brightness temperatures for both
H$_2$CO $J$\,=\,3--2 and 4--3 lines using a factor of 1/0.69.
Observed continuum of Mars, Jupiter, and Saturn were used to calibrate the
spectral line flux. The calibration uncertainty is about 20\%.

Data reduction of spectral lines was performed using
CLASS from the GILDAS package\footnote{\tiny http://www.iram.fr/IRAMFR/GILDAS}.
To enhance signal to noise ratios (S/N) in individual channels,
we smoothed contiguous channels to a velocity resolution of $\sim$0.6\,km\,s$^{-1}$.
The linewidths tend to be $>$\,few km\,s$^{-1}$, so the smoothing
has no impact on our results. The typical noise level is $\sim$0.06\,K
($T_{\rm mb}$ scale) for both H$_2$CO ($J$\,=\,3--2) and
H$_2$CO ($J$\,=\,4--3) at a velocity resolution of $\sim$0.6\,km\,s$^{-1}$.

\section{Results}
\label{sect:Results}
\subsection{Overview}
94 sources with
H$_2$CO ($J$\,=\,3--2) transitions and 98 sources with
H$_2$CO ($J$\,=\,4--3) transitions were observed.
Toward the targeted massive clumps (see Tab.\,\ref{table:detection rate})
nearly all H$_2$CO lines are detected (detection rate $\gtrsim$ 97\%)
for the upper energy above ground state, $E_{\rm u}$, ($<$35\,K).
For high $E_{\rm u}$ ($>$82\,K),
the H$_2$CO detection rate ranges from 82\% to 85\%.
Non-detections are associated with 70w and IRw sources
(see Sect.\,\ref{sect:samp-obs} for the definitions) which
are typically associated with the
early cold evolutionary stages of massive clumps.
Two para-H$_2$CO\,(3$_{22}$--2$_{21}$ and 3$_{21}$--2$_{20}$) transitions
($E_{\rm u}$\,$\sim$\,68\,K) show a lower detection rate ($\sim$70\%),
which is caused by the fact that para-H$_2$CO is the less-abundant of
the two H$_2$CO symmetry species and the source of the weaker $K$\,=\,2 transitions. High detection rates of
H$_2$CO indicate that this species is commonly formed in
massive star-forming clumps and is present during all their evolutionary stages.

Examples of H$_2$CO line spectra are presented in Figure \ref{fig:H2CO-spectra}.
Line parameters are listed in Tables \ref{table:H2CO313-312}, \ref{table:H2CO303-322},
\ref{table:H2CO321-404}, and \ref{table:H2CO423-422},
where velocity-integrated intensity, $\int$$T$$_{\rm mb}$d$v$, local
standard of rest velocity, $V_{\rm lsr}$, full width to half maximum linewidth,
FWHM, and peak main beam brightness temperature, $T$$_{\rm mb}$, were obtained
from Gaussian fits. The rest frequencies of the ortho-H$_2$CO\,4$_{32}$--3$_{31}$
and 4$_{31}$--3$_{30}$ transitions are nearby (see Tab.\,\ref{table:APEX-receiver}
and Fig.\,\ref{fig:H2CO-spectra}). These two lines are blended in all of our sources,
so that Gaussian fits are of limited value and are not part of our tables.

\subsection{Source size correction}
\label{sect:Source-size}
\begin{figure}[t]
\vspace*{0.2mm}
\begin{center}
\includegraphics[width=0.48\textwidth]{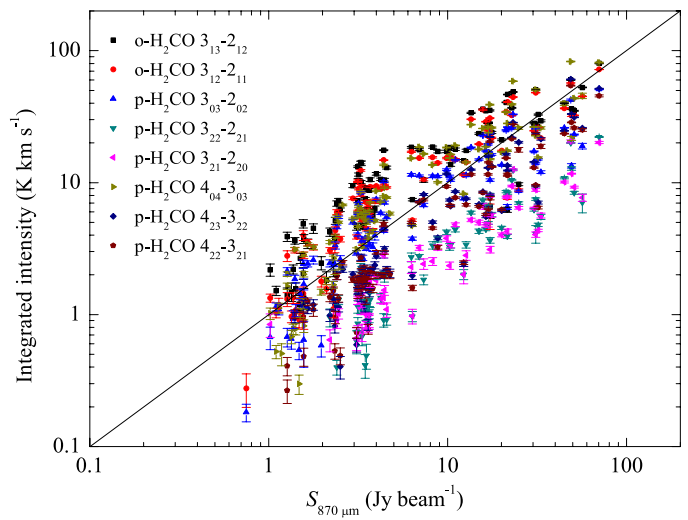}
\end{center}
\caption{Comparison of integrated intensities of H$_2$CO
and 870\,$\mu$m continuum flux densities.
The solid line corresponds to Y\,=\,X in the given units.}
\label{figure:S870um-I(H2CO)}
\end{figure}

The para-H$_2$CO $J$\,=\,3--2 (beam size $\sim$\,28.6$''$) and
4--3 (beam size $\sim$\,21.5$''$) lines we observed were obtained
by single pointing observations with different receivers,
so the area covered by our $J$\,=\,3--2 and 4--3 transitions is slightly different.
We compare the integrated intensities of H$_2$CO, irrespective of the beam size,
with 870\,$\mu$m flux densities in Figure \ref{figure:S870um-I(H2CO)}.
It shows that the H$_2$CO integrated intensities follow the
870\,$\mu$m intensity distribution. Apparently
dense gas traced by H$_2$CO is associated well with the dust
traced by 870\,$\mu$m emission in the massive star-forming clumps.
Mapping observations of para-H$_2$CO\,(3$_{03}$--2$_{02}$, 3$_{22}$--2$_{21}$, and
3$_{21}$--2$_{20}$) towards the Orion molecular cloud 1 (OMC1) with the APEX telescope
also show that para-H$_2$CO integrated intensity distributions agree well
with the dust emission observed at 850\,$\mu$m
(\citealt{Johnstone1999,Tang2017c}). Previous observations of
H$_2$CO\,(4$_{04}$--3$_{03}$, 4$_{23}$--3$_{22}$, 4$_{22}$--3$_{21}$,
4$_{32}$--3$_{31}$, and 4$_{31}$--3$_{30}$) towards massive clumps in
the W33 region with the APEX telescope \citep{Immer2014} also indicate that H$_2$CO
distributions are consistent with the dust emission traced by 870\,$\mu$m.
So here we assume that the source sizes of H$_2$CO are the
same as the full width to half power source sizes of the 870\,$\mu$m dust emission derived from \cite{Csengeri2014}.
We correct for beam dilution by calculating
$T_{\rm mb}^{\prime}$\,=\,$T_{\rm mb}$/$\eta$$_{\rm bf}$ with beam-filling factor
$\eta$$_{\rm bf}$\,=\,$\theta_{\rm s}^2$/($\theta_{\rm s}^2$+$\theta_{\rm beam}^2$).
Here $\theta_{\rm beam}$ and $\theta_{\rm s}$ denote beam and source size,
respectively. The results of $\eta$$_{\rm bf}$ and the
para-H$_2$CO\,4$_{04}$--3$_{03}$/3$_{03}$--2$_{02}$ integrated
intensity ratio ($I'(4_{04}$--3$_{03}$)/$I'(3_{03}$--2$_{02}$))
corrected with $\eta$$_{\rm bf}$ are listed in Table \ref{table:NH2CO-Tkin}.

\subsection{Opacities of H$_2$CO}
\label{sect:column-density-opacities}
To determine the gas kinetic temperatures,
$T_{\rm kin}$, spatial densities, $n$(H$_2$), and para-H$_2$CO column densities,
$N$(H$_2$CO), we use the RADEX non-LTE model \citep{van der Tak2007}
offline code\footnote{\tiny http://var.sron.nl/radex/radex.php}
with collision rates from \cite{Wiesenfeld2013}.
Uncertainties in the collisional excitation rates directly affect the derived
volume densities, while kinetic temperature appears to be less affected by
collisional excitation rate uncertainties (see Sect.\,\ref{sect:temperature}).
The RADEX code needs five input parameters:
background temperature, kinetic temperature, H$_2$ density,
H$_2$CO column density,
and linewidth. For the background temperature, we adopted 2.73\,K.
Model grids for the H$_2$CO lines encompass 40 densities
($n$(H$_2$)\,=\,10$^4$--10$^8$\,cm$^{-3}$), 40 H$_2$CO column
densities ($N$(H$_2$CO)\,=\,10$^{12}$--10$^{16}$\,cm$^{-2}$),
and 40 temperatures ranging from 10 to 400\,K.
For the linewidth, we use the observed linewidth value.

\begin{figure}[t]
\vspace*{0.2mm}
\begin{center}
\includegraphics[width=0.48\textwidth]{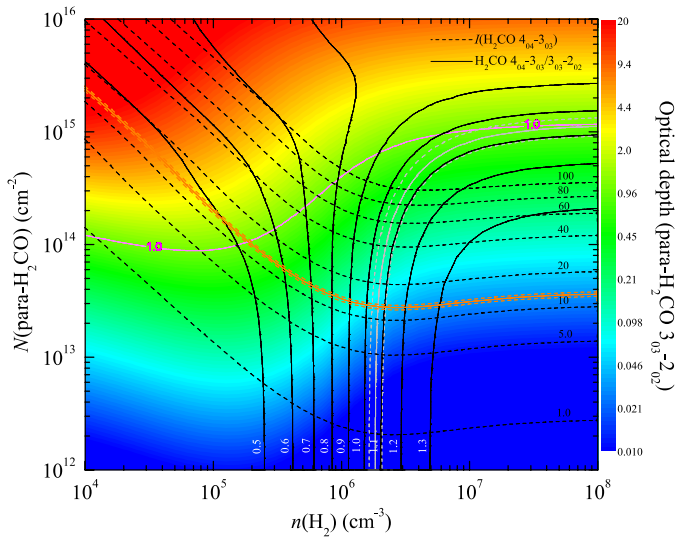}
\end{center}
\caption{Example of RADEX non-LTE modeling of the $N$(H$_2$CO)--$n$(H$_2$)
relation for AGAL008.684$-$00.367 at a kinetic temperature of 55\,K
(see Sect.\,\ref{sect:temperature}).
Black dashed and solid lines are para-H$_2$CO\,4$_{04}$--3$_{03}$ integrated
intensities and para-H$_2$CO\,4$_{04}$--3$_{03}$/3$_{03}$--2$_{02}$
integrated intensity ratios, respectively.
To the measured parameters,
para-H$_2$CO\,4$_{04}$--3$_{03}$ integrated intensity (orange solid
and dashed lines represent observed value and uncertainty)
and para-H$_2$CO\,4$_{04}$--3$_{03}$/3$_{03}$--2$_{02}$ integrated
intensity ratio (white solid and dashed lines) corrected
by the relevant beam-filling factors (see Tab.\,\ref{table:NH2CO-Tkin}).
The colour map shows the optical depth of the para-H$_2$CO\,3$_{03}$--2$_{02}$ line.
The purple line in the upper green area corresponds to optical depth
$\tau$(para-H$_2$CO\,3$_{03}$--2$_{02}$)\,=\,1.0.}
\label{figure:N(H2CO)-n(H2)}
\end{figure}

The value of $N$(para-H$_2$CO) depends on para-H$_2$CO\,3$_{03}$--2$_{02}$
and/or 4$_{04}$--3$_{03}$ integrated intensities and
the para-H$_2$CO\,4$_{04}$--3$_{03}$/3$_{03}$--2$_{02}$ ratio
\citep{Mangum1993a,Tang2017a}.
If the para-H$_2$CO\,3$_{03}$--2$_{02}$ and 4$_{04}$--3$_{03}$
lines are optically thick in our dense massive clumps, this would cause high
para-H$_2$CO\,4$_{04}$--3$_{03}$/3$_{03}$--2$_{02}$,
3$_{21}$--2$_{20}$/3$_{03}$--2$_{02}$,
and 4$_{22}$--3$_{21}$/4$_{04}$--3$_{03}$ ratios.
Higher ratios will imply higher spatial densities and
kinetic temperatures, respectively \citep{Mangum1993a,Ao2013,
Ginsburg2016,Immer2016,Tang2017a,Tang2017b,Tang2017c}.
In order to understand the impact of the line optical depth,
we modelled the optical depth of para-H$_2$CO\,3$_{03}$--2$_{02}$ and
para-H$_2$CO\,4$_{04}$--3$_{03}$ integrated intensities,
and the para-H$_2$CO 4$_{04}$--3$_{03}$/3$_{03}$--2$_{02}$ ratio at a kinetic temperature of 55\,K
(see Sect.\,\ref{sect:temperature}) in Figure \ref{figure:N(H2CO)-n(H2)}
(or see Figure G.2 in \citealt{Immer2016}).
Changing the kinetic temperature, weakly affects the optical depth of
the para-H$_2$CO\,3$_{03}$--2$_{02}$ and para-H$_2$CO\,4$_{04}$--3$_{03}$
lines (less than by a factor of few). The para-H$_2$CO\,4$_{04}$--3$_{03}$/3$_{03}$--2$_{02}$
ratio is then also not greatly changed ($\lesssim$30\%; not shown here).
The figure demonstrates that para-H$_2$CO\,3$_{03}$--2$_{02}$
is optically thin ($\tau$\,<\,1)
at column density $N$(para-H$_2$CO)\,$<$\,1$\times$10$^{14}$\,cm$^{-2}$ and
spatial density 10$^{4-8}$\,cm$^{-3}$. At higher column density
($N$(para-H$_2$CO)\,$>$\,5$\times$10$^{14}$\,cm$^{-2}$),
the para-H$_2$CO\,3$_{03}$--2$_{02}$ becomes optically thick ($\tau$\,>\,5).
The optical depth of para-H$_2$CO\,4$_{04}$--3$_{03}$ shows a similar behavior
(slightly lower values) with respect to that of para-H$_2$CO\,3$_{03}$--2$_{02}$ (not shown here;
or see Figure G.2 in \citealt{Immer2016}).
Considering the observed ranges of integrated intensities of
para-H$_2$CO\,4$_{04}$--3$_{03}$ (typical value $\sim$ 20\,K\,km s$^{-1}$)
and para-H$_2$CO\,4$_{04}$--3$_{03}$/3$_{03}$--2$_{02}$ ratios
(typical value $\sim$\,1.0)
accounting for relevant beam-filling factors from Section \ref{sect:Source-size}
(see Tabs.\,\ref{table:H2CO321-404} and \ref{table:NH2CO-Tkin}),
the optical depths of para-H$_2$CO\,3$_{03}$--2$_{02}$ and 4$_{04}$--3$_{03}$
range from $\sim$0.012 to $\sim$1 in our sample.
Compared to the para-H$_2$CO\,3$_{03}$--2$_{02}$ and 4$_{04}$--3$_{03}$ lines,
para-H$_2$CO\,3$_{22}$--2$_{21}$, 3$_{21}$--2$_{20}$, 4$_{23}$--3$_{22}$,
and 4$_{22}$--3$_{21}$ lines have higher upper energies above the ground state
($E_{\rm u}$\,$>$\,68\,K, see Tab.\,\ref{table:APEX-receiver}), so they
have lower optical depths ($\tau$\,$\ll$\,1).
Therefore, the influence of the para-H$_2$CO\,3$_{03}$--2$_{02}$ and
4$_{04}$--3$_{03}$ optical depths is weak for our determination of
spatial density and kinetic temperature.

In our sample the observed $T_{\rm mb}$(3$_{12}$--2$_{11}$/3$_{03}$--2$_{02}$)
ratios range from 0.74 to 1.83 with an unweighted average of 1.29\,$\pm$\,0.02
(see Tabs.\,\ref{table:H2CO313-312} and \ref{table:H2CO303-322}; errors
given here and elsewhere are standard deviations of the mean).
For the $T_{\rm mb}$(3$_{13}$--2$_{12}$/3$_{03}$--2$_{02}$)
ratio, it ranges from 1.03 to 2.35 with an unweighted average of 1.56\,$\pm$\,0.03
(see Tabs.\,\ref{table:H2CO313-312} and \ref{table:H2CO303-322}).
The relation between $T_{\rm mb}$(3$_{12}$--2$_{11}$/3$_{03}$--2$_{02}$)
and H$_2$CO optical depth, indicated by \cite{Sasselov1990} in their
Figure 2, suggests that for at least 30\% of our sample
($T_{\rm mb}$(3$_{12}$--2$_{11}$/3$_{03}$--2$_{02}$)\,$\lesssim$\,1.19) the
ortho-H$_2$CO\,3$_{12}$--2$_{11}$ and 3$_{13}$--2$_{12}$ lines are
optically thick ($\tau$\,$\gtrsim$\,5).

\subsection{Kinetic temperature}
\label{sect:temperature}
As discussed in Section \ref{sect:Introduction},
the intensity ratios of H$_2$CO
lines involving different $K_{\rm a}$ ladders yield estimates
of the kinetic temperature of the gas \citep{Mangum1993a}.
For our observed transitions of H$_2$CO,
para-H$_2$CO\,3$_{21}$--2$_{20}$/3$_{03}$--2$_{02}$,
3$_{22}$--2$_{21}$/3$_{03}$--2$_{02}$, 4$_{22}$--3$_{21}$/4$_{04}$--3$_{03}$,
and 4$_{23}$--3$_{22}$/4$_{04}$--3$_{03}$ ratios can be useful
thermometers to derive the kinetic temperature.
Para-H$_2$CO\,3$_{22}$--2$_{21}$/3$_{03}$--2$_{02}$
and 3$_{21}$--2$_{20}$/3$_{03}$--2$_{02}$ ratios trace the
kinetic temperature with an uncertainty of
$\lesssim$25\% below 50\,K \citep{Mangum1993a}.
Para-H$_2$CO\,4$_{22}$--3$_{21}$/4$_{04}$--3$_{03}$ and
4$_{23}$--3$_{22}$/4$_{04}$--3$_{03}$ ratios trace the
kinetic temperature with an uncertainty of
$\lesssim$25\% below 75\,K \citep{Mangum1993a}.
The para-H$_2$CO\,3$_{22}$--2$_{21}$/3$_{03}$--2$_{02}$
and 4$_{23}$--3$_{22}$/4$_{04}$--3$_{03}$ line ratios
are slightly affected by the spatial density
(not shown here; for para-H$_2$CO\,3$_{22}$--2$_{21}$/3$_{03}$--2$_{02}$ see
\citealt{Lindberg2015} and \citealt{Tang2017a}). So in this work
we use the para-H$_2$CO\,3$_{21}$--2$_{20}$/3$_{03}$--2$_{02}$ and
4$_{22}$--3$_{21}$/4$_{04}$--3$_{03}$ integrated intensity
ratios to derive the kinetic temperature, which also have been used
for the Galactic central molecular zone (CMZ) clouds \citep{Ginsburg2016,Immer2016}.

\begin{figure}[t]
\vspace*{0.2mm}
\begin{center}
\includegraphics[width=0.48\textwidth]{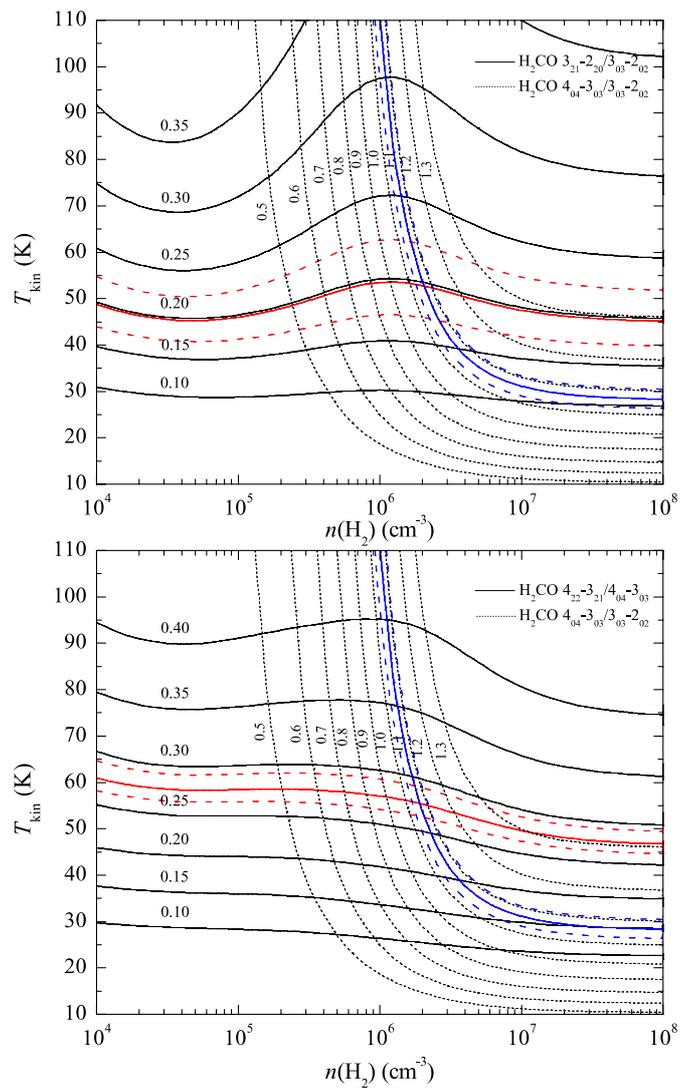}
\end{center}
\caption{Example of RADEX non-LTE modeling of the para-H$_2$CO kinetic
temperature for AGAL008.684$-$00.367.
Black solid and dashed lines are para-H$_2$CO integrated intensity ratios.
Para-H$_2$CO\,4$_{04}$--3$_{03}$/3$_{03}$--2$_{02}$
(blue solid and dashed lines represent observed value and uncertainty,
accounting for different beam-filling factors),
3$_{21}$--2$_{20}$/3$_{03}$--2$_{02}$ and 4$_{22}$--3$_{21}$/4$_{04}$--3$_{03}$
integrated intensity ratios (top and bottom, red solid and dashed lines)
for a para-H$_2$CO column density 2.8\,$\times$\,10$^{13}$\,cm$^{-2}$
derived from the para-H$_2$CO\,4$_{04}$--3$_{03}$ integrated intensity
and para-H$_2$CO\,4$_{04}$--3$_{03}$/3$_{03}$--2$_{02}$ ratio
(see Sect.\,\ref{sect:column-density-opacities}).}
\label{figure:Tkin}
\end{figure}

We ran RADEX to calculate the observed para-H$_2$CO\,3$_{21}$--2$_{20}$/3$_{03}$--2$_{02}$,
4$_{22}$--3$_{21}$/4$_{04}$--3$_{03}$, and 4$_{04}$--3$_{03}$/3$_{03}$--2$_{02}$
integrated intensity ratios corrected by the relevant beam-filling factors
assuming these transitions of para-H$_2$CO are optically thin
(see Sect.\,\ref{sect:column-density-opacities}).
In Figure \ref{figure:Tkin}, an example is presented to show how the
parameters are constrained by the line ratio distribution of para-H$_2$CO,
accounting for different
beam-filling factors in the $T_{\rm kin}$--$n$(H$_2$) parameter space.
We used the column density derived from
the para-H$_2$CO\,4$_{04}$--3$_{03}$ integrated intensity and
para-H$_2$CO\,4$_{04}$--3$_{03}$/3$_{03}$--2$_{02}$ ratio accounting for the
beam-filling factors derived in Section \ref{sect:Source-size}
to constrain the kinetic temperature. It shows that
para-H$_2$CO\,3$_{21}$--2$_{20}$/3$_{03}$--2$_{02}$ and
4$_{22}$--3$_{21}$/4$_{04}$--3$_{03}$ line ratios are sensitive
to the gas kinetic temperature (see the black solid lines in Fig.\,\ref{figure:Tkin}),
while being relatively independent
of spatial density. The integrated intensity
ratio $I'$(4$_{04}$--3$_{03}$)/$I'$(3$_{03}$--2$_{02}$) is sensitive to the
gas spatial density at high temperature ($T_{\rm kin}$\,$>$\,40\,K),
where the $I'$(4$_{04}$--3$_{03}$)/$I'$(3$_{03}$--2$_{02}$) ratio
becomes relatively independent of kinetic temperature.
At low temperature ($T_{\rm kin}$\,$<$\,40\,K), this ratio
is influenced almost entirely by the gas kinetic temperature,
because $T_{\rm kin}$ becomes lower than the excitation difference
of the involved states.
Therefore, para-H$_2$CO\,3$_{21}$--2$_{20}$/3$_{03}$--2$_{02}$ and
4$_{22}$--3$_{21}$/4$_{04}$--3$_{03}$ ratios combined with the
para-H$_2$CO 4$_{04}$--3$_{03}$/3$_{03}$--2$_{02}$ ratio are good
tracers to constrain kinetic temperature and spatial density of
dense gas in warm regions (gas temperature $>$30\,K,
for lower $T_{\rm kin}$ the levels of the $J$\,=\,3--2 and 4--3 $K_{\rm a}$\,>\,0 lines
are too far above the ground state) of massive star-forming clumps.
With the two $J$\,=\,4, $K_{\rm a}$\,=\,2 levels being located 80--90\,K
above the ground state (see Tab.\,\ref{table:APEX-receiver}),
the radiative transfer models start to become
insensitive to temperatures in excess of 150\,K. So temperatures
$>$150\,K have to be considered sceptically and should cautiously
be interpreted as $\gtrsim$150\,K \citep{Mangum1993a,Ginsburg2016,Immer2016}.
The derived kinetic temperatures are listed in Table\,\ref{table:NH2CO-Tkin}.

We note a para-H$_2$CO\,3$_{21}$--2$_{20}$/3$_{03}$--2$_{02}$ ratio
"bump" at kinetic temperature $>$50\,K and spatial density
10$^{5.5-7.0}$\,cm$^{-3}$ in Figure\,\ref{figure:Tkin}
(or see Figure\,13 in \cite{Mangum1993a} and Figure\,F.1 in \cite{Lindberg2015}),
because the excitation temperature of the para-H$_2$CO\,3$_{21}$--2$_{20}$
line rises much faster than that of the para-H$_2$CO\,3$_{03}$--2$_{02}$
line with increasing spatial density and/or kinetic temperature \citep{Mangum1993a}.
Kinetic temperatures obtained for a given
para-H$_2$CO\,3$_{21}$--2$_{20}$/3$_{03}$--2$_{02}$ ratio
vary more than by $\gtrsim$20\% at kinetic temperature $>$60\,K
and spatial density 10$^{5.0-7.0}$\,cm$^{-3}$.
This large "bump" in the para-H$_2$CO\,3$_{21}$--2$_{20}$/3$_{03}$--2$_{02}$
contour (see Fig.\,\ref{figure:Tkin} upper panel) probably leads to
an overestimate of the kinetic temperature from
the para-H$_2$CO\,3$_{21}$--2$_{20}$/3$_{03}$--2$_{02}$ ratio.
Para-H$_2$CO\,4$_{22}$--3$_{21}$/4$_{04}$--3$_{03}$
is also influenced by a "bump", this time at kinetic temperature $>$100\,K
and spatial density 10$^{5.5-7.0}$\,cm$^{-3}$
(see Fig.\,\ref{figure:Tkin} lower panel or Figure 13 in \citealt{Mangum1993a}).
Kinetic temperatures derived from the
para-H$_2$CO\,4$_{22}$--3$_{21}$/4$_{04}$--3$_{03}$
ratio vary less than $\lesssim$20\% for $T_{\rm kin}$\,$<$\,150\,K
and spatial density 10$^{5.0-7.0}$\,cm$^{-3}$.
It appears that the para-H$_2$CO\,4$_{22}$--3$_{21}$/4$_{04}$--3$_{03}$ ratio
is more stable and accurate to trace gas kinetic temperature
than the para-H$_2$CO\,3$_{21}$--2$_{20}$/3$_{03}$--2$_{02}$ ratio
at $T_{\rm kin}$\,$<$\,150\,K and spatial density 10$^{5.0-7.0}$\,cm$^{-3}$.

\begin{figure}[t]
\vspace*{0.2mm}
\begin{center}
\includegraphics[width=0.48\textwidth]{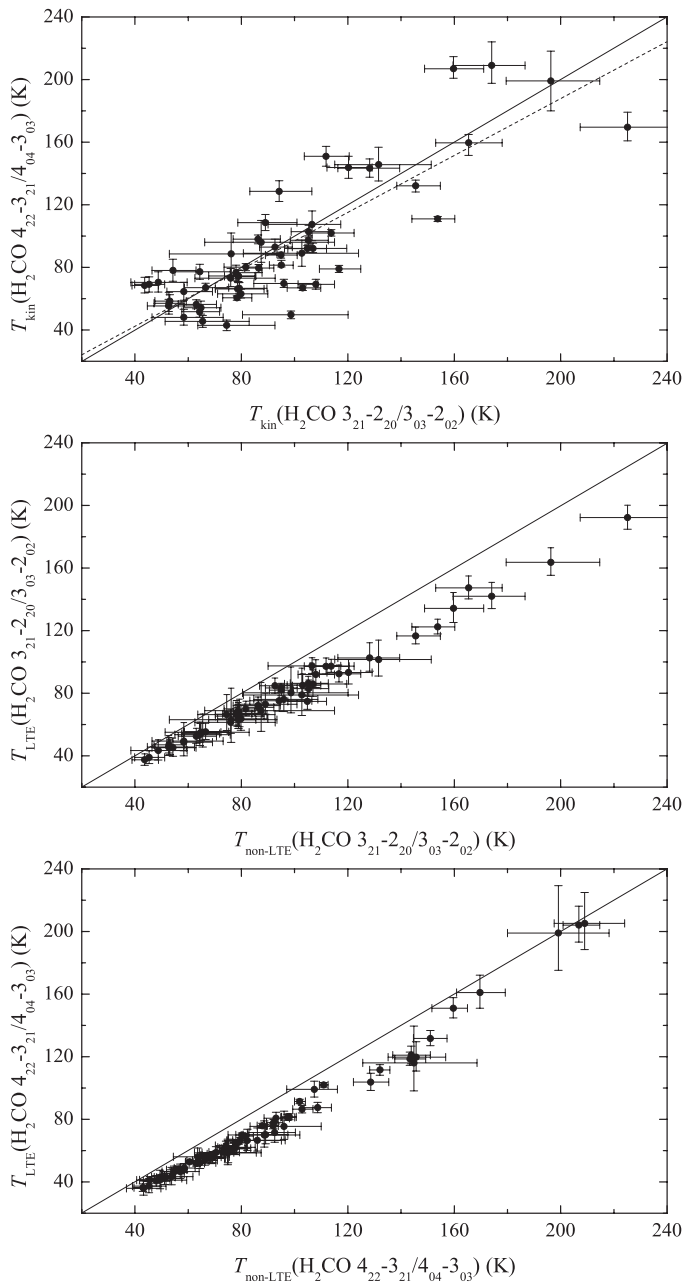}
\end{center}
\caption{Top panel: Comparison of kinetic temperatures
derived from para-H$_2$CO\,3$_{21}$--2$_{20}$/3$_{03}$--2$_{02}$
and 4$_{22}$--3$_{21}$/4$_{04}$--3$_{03}$ ratios. The dashed line is
the result from an unweighed linear fit, $T_{\rm kin}({\rm 4_{22}{-}3_{21}/4_{04}{-}3_{03}})=(0.9\pm0.1) \times T_{\rm kin}({\rm 3_{21}{-}2_{20}/3_{03}{-}2_{02}})+(5.8\pm7.8)$, with a correlation coefficient, $R$, of 0.85. Middle and
bottom panels: Comparisons of kinetic temperatures
derived from LTE and RADEX non-LTE calculations for
para-H$_2$CO\,3$_{21}$--2$_{20}$/3$_{03}$--2$_{02}$ and
4$_{22}$--3$_{21}$/4$_{04}$--3$_{03}$ ratios, respectively.
The temperature uncertainties are obtained from observed
para-H$_2$CO line ratio errors.
Solid lines indicate equal temperatures.}
\label{fig:T303-T404-TLTE}
\end{figure}

A comparison of kinetic temperatures derived from both
para-H$_2$CO\,3$_{21}$--2$_{20}$/3$_{03}$--2$_{02}$ and
4$_{22}$--3$_{21}$/4$_{04}$--3$_{03}$ ratios suggests that the two ratios trace
similar temperatures (see Fig.\,\ref{fig:T303-T404-TLTE}).
It might have been expected, for example by analogy to NH$_3$
(e.g., \citealt{Henkel1987,Mangum2013a,Gong2015a,Gong2015b}),
that higher excited H$_2$CO transitions lead to higher $T_{\rm kin}$ values.
Some of the similar kinetic temperatures derived from the
para-H$_2$CO\,3$_{21}$--2$_{20}$/3$_{03}$--2$_{02}$ and
4$_{22}$--3$_{21}$/4$_{04}$--3$_{03}$ ratios (Fig.\,\ref{fig:T303-T404-TLTE})
might be caused by the para-H$_2$CO\,3$_{21}$--2$_{20}$/3$_{03}$--2$_{02}$
ratio "bump" (Fig.\,\ref{figure:Tkin}, top panel). This excitation effect in the
para-H$_2$CO\,3$_{21}$--2$_{20}$/3$_{03}$--2$_{02}$ ratio may result in an
overestimate of the kinetic temperature derived from this ratio with
large uncertainty ($\gtrsim$20\% at kinetic temperature $>$60\,K) at spatial density 10$^{5.5-7.0}$\,cm$^{-3}$.

The para-H$_2$CO line intensity ratios 3$_{22}$--2$_{21}$/3$_{03}$--2$_{02}$,
3$_{21}$--2$_{20}$/3$_{03}$--2$_{02}$, 4$_{23}$--3$_{22}$/4$_{04}$--3$_{03}$ and
4$_{22}$--3$_{21}$/4$_{04}$--3$_{03}$ can also provide a measurement of the
kinetic temperature of the gas assuming local thermodynamic equilibrium (LTE).
The kinetic temperature can be calculated from these para-H$_2$CO
transition ratios if the lines are optically thin
(see Sect.\,\ref{sect:column-density-opacities}), and originate
from a high density region \citep{Mangum1993a}. Following the method applied by \cite{Mangum1993a} in their Appendix A,
\begin{equation}
T_{\rm LTE}=\frac{47.1}{ln(0.556\frac{I(3_{03}-2_{02})}{I(3_{21}-2_{20})})}\,\,\rm K
\end{equation}
and
\begin{equation}
T_{\rm LTE}=\frac{47.2}{ln(0.750\frac{I(4_{04}-3_{03})}{I(4_{22}-3_{21})})}\,\,\rm K,
\end{equation}
where $I$(3$_{03}$--2$_{02}$)/$I$(3$_{21}$--2$_{20}$) and
$I$(4$_{04}$--3$_{03}$)/$I$(4$_{22}$--3$_{21}$) are the para-H$_2$CO
integrated intensity ratios. The results of the kinetic temperature
calculations from the para-H$_2$CO\,3$_{03}$--2$_{02}$/3$_{21}$--2$_{20}$
and 4$_{04}$--3$_{03}$/4$_{22}$--3$_{21}$
integrated intensity ratios are listed in Table \ref{table:NH2CO-Tkin}.
If the assumption of optically thin emission is correct, the kinetic
temperatures derived from this method have an uncertainty of
$\lesssim$30\% \citep{Mangum1993a}. We also compared the kinetic temperatures
derived from LTE and RADEX non-LTE calculations (see Fig.\,\ref{fig:T303-T404-TLTE}).
It appears that $T_{\rm non-LTE}$ is consistently higher than $T_{\rm LTE}$ by $\lesssim$25\%.
This might be caused by the fact that at densities of 10$^{6.5}$\,cm$^{-3}$ (see Sect.\,\ref{sect:density}) thermalization
is not yet reached \citep{Mangum1993a}.
Therefore higher $T_{\rm kin}$ values are need to compensate for this effect
leading to lower excitation temperatures, and to reproduce data.

\begin{figure*}[t]
\vspace*{0.2mm}
\begin{center}
\includegraphics[width=0.95\textwidth]{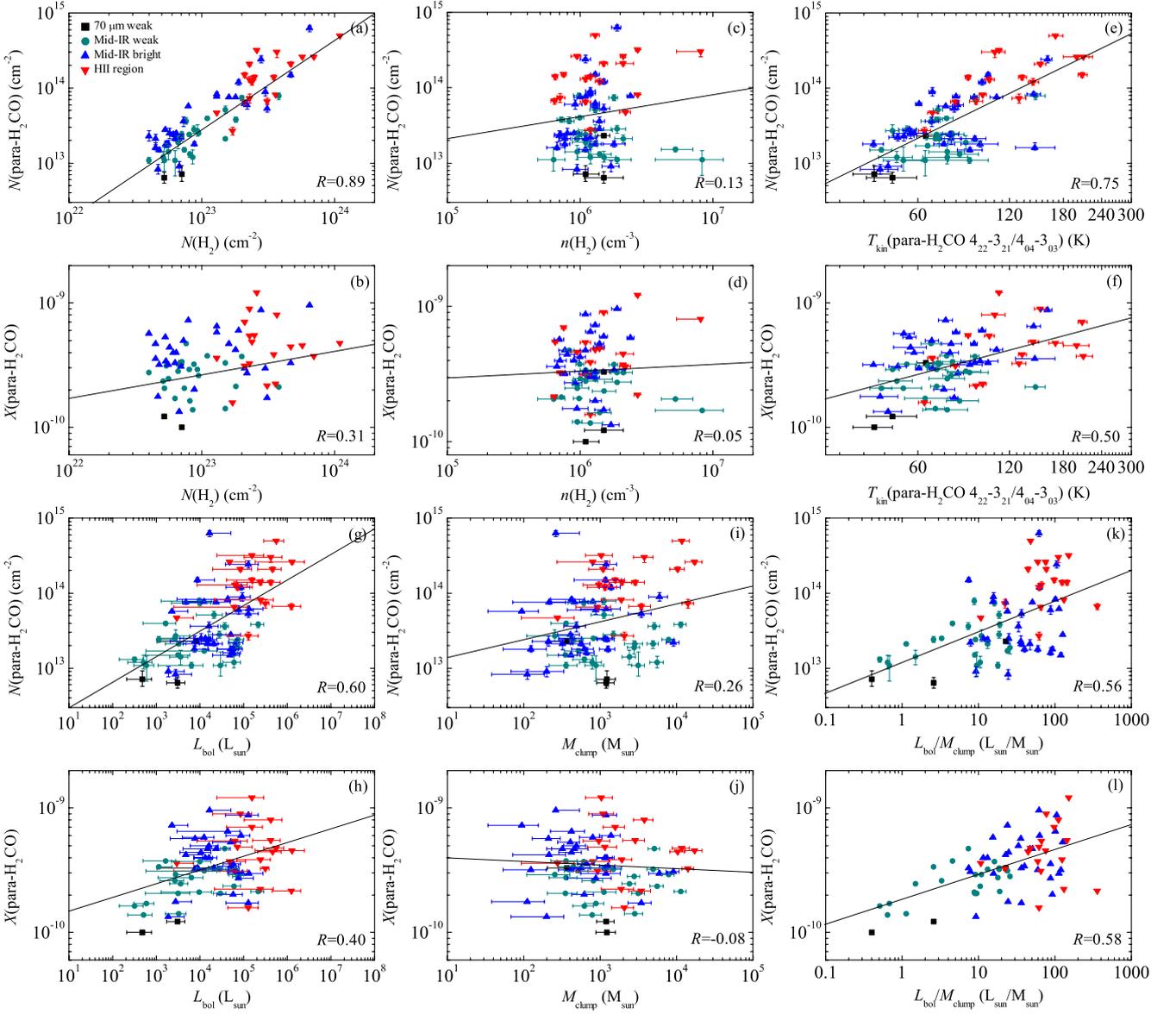}
\end{center}
\caption{Column density $N$(para-H$_2$CO) and fractional abundance
$X$(para-H$_2$CO) vs. column density
$N$(H$_2$) (a, b), spatial density $n$(H$_2$) (c, d), kinetic
temperature $T_{\rm kin}$(para-H$_2$CO\,4$_{22}$--3$_{21}$/4$_{04}$--3$_{03}$) (e, f),
bolometric luminosity (g, h), mass of clump (i, j),
and luminosity-to-mass $L_{\rm bol}$/$M_{\rm clump}$ ratio (k, l).
The column density and spatial density uncertainties are obtained from observed
para-H$_2$CO line brightness temperature and line ratio errors. The straight lines
are the results from unweighed linear fits yielding the given correlation
coefficients, $R$, in the lower right corner of each panel.}
\label{figure:N(H2CO)-N(H2)-Tk}
\end{figure*}

\begin{table*}[t]
\caption{Averaged parameters in different stages of the massive clumps.}
\label{table:average-parameters}
\centering
\begin{tabular}
{cccccc}
\hline\hline 
Stage &$T_{\rm kin}$ &$n$(H$_2$) &$N$(para-H$_2$CO) &$X$(para-H$_2$CO)
&$\Delta v$(para-H$_2$CO\,4$_{04}$--3$_{03}$)\\
&K &$\times$10$^6$ cm$^{-3}$ &$\times$10$^{13}$ cm$^{-2}$ &$\times$10$^{-10}$ &km s$^{-1}$\\
\hline 
70w        &52 $\pm$ 6  &1.2 $\pm$ 0.2 &1.2 $\pm$ 0.6  &1.8 $\pm$ 0.7 &3.9 $\pm$ 0.4\\
IRw        &73 $\pm$ 4  &1.7 $\pm$ 0.4 &2.7 $\pm$ 0.4  &2.7 $\pm$ 0.2 &4.8 $\pm$ 0.2\\
IRb        &81 $\pm$ 6  &1.2 $\pm$ 0.1 &7.2 $\pm$ 2.1  &4.3 $\pm$ 0.4 &4.9 $\pm$ 0.3\\
H\,{\scriptsize II} region &110 $\pm$ 8 &1.8 $\pm$ 0.4 &16.9 $\pm$ 0.3 &4.9 $\pm$ 0.6 &7.4 $\pm$ 0.4\\
average    &91 $\pm$ 4  &1.5 $\pm$ 0.1 &8.0 $\pm$ 1.3  &3.9 $\pm$ 0.2 &5.3 $\pm$ 0.2\\
\hline 
\end{tabular}
\end{table*}

\subsection{Spatial density and column density}
\label{sect:density}
As described in Section \ref{sect:Introduction}, with the kinetic temperature approximately known,
the relative intensity ratio of H$_2$CO
lines involving the same $K_{\rm a}$ ladders yields estimates
of the spatial density of the gas \citep{Henkel1980,Henkel1983,Mangum1993a}.
For our observed transitions of H$_2$CO,
para-H$_2$CO\,4$_{04}$--3$_{03}$/3$_{03}$--2$_{02}$,
4$_{22}$--3$_{21}$/3$_{21}$--2$_{20}$ (or 4$_{22}$--3$_{21}$/3$_{22}$--2$_{21}$),
and 4$_{23}$--3$_{22}$/3$_{22}$--2$_{21}$ (or 4$_{23}$--3$_{22}$/3$_{21}$--2$_{20}$)
ratios are good densitometers to derive the spatial density.
The para-H$_2$CO $3_{03}$-2$_{02}$ and 4$_{04}$--3$_{03}$ lines
are the strongest of the 218 GHz and 291 GHz transitions,
respectively, and they are nearly all detected in
our sample (see Tab.\,\ref{table:detection rate}). So in this work,
we use the para-H$_2$CO\,4$_{04}$--3$_{03}$/3$_{03}$--2$_{02}$ integrated
intensity ratio to derive the spatial density, which has also been used
in molecular clouds of the Galactic CMZ \citep{Immer2016}.

We ran RADEX to obtain para-H$_2$CO column densities and spatial density,
and calculated the observed para-H$_2$CO\,3$_{03}$--2$_{02}$ and
4$_{04}$--3$_{03}$ integrated intensities in K\,km\,s$^{-1}$ units corrected by the
relevant beam-filling factors ($I$$\rightarrow$$I^{\prime}$).
In Figure \ref{figure:N(H2CO)-n(H2)},
an example is presented to show how the parameters are constrained by the
corrected integrated line intensity and integrated line intensity ratio distribution
of para-H$_2$CO in the $N$(para-H$_2$CO)--$n$(H$_2$) parameter
space. This figure shows that at low column density
($N$(para-H$_2$CO)\,$<$\,5\,$\times$\,10$^{14}$\,cm$^{-2}$) the
$I'$(4$_{04}$--3$_{03}$)/$I'$(3$_{03}$--2$_{02}$) ratio
accounting for different beam-filling factors
(see the black solid lines) is sensitive to the gas spatial density and
becomes relatively independent of the para-H$_2$CO column density,
while the kinetic temperature
is kept constant at $\sim$55\,K (which is close to the actual
temperature, see above). At high column density
($N$(para-H$_2$CO)\,$>$\,5\,$\times$\,10$^{14}$\,cm$^{-2}$)
the $I'$(4$_{04}$--3$_{03}$)/$I'$(3$_{03}$--2$_{02}$) ratio
appears to be not sensitive to the gas spatial density and becomes dependent
on the column density, because the para-H$_2$CO 3$_{03}$--2$_{02}$ transition
starts to become optically thick \citep{Mangum1993a}.
The derived results of $N$(para-H$_2$CO) and spatial density are
listed in Table \ref{table:NH2CO-Tkin}. We use the same method to
obtain ortho-H$_2$CO column densities with the observed
ortho-H$_2$CO (3$_{12}$--2$_{11}$ and 3$_{13}$--2$_{12}$)
integrated intensities adopting kinetic temperature and spatial
density derived from para-H$_2$CO line ratios
(see Sect.\,\ref{sect:temperature} and above) and assuming ortho- and
para-H$_2$CO originate from the same region. The obtained results
of $N$(ortho-H$_2$CO) are listed in Table \ref{table:NH2CO-Tkin}.

As mentioned in Section\,\ref{sect:column-density-opacities}, the
para-H$_2$CO\,3$_{22}$--2$_{21}$, 3$_{21}$--2$_{20}$, 4$_{23}$--3$_{22}$,
and 4$_{22}$--3$_{21}$ lines are optically thin, so
the para-H$_2$CO\,4$_{22}$--3$_{21}$/3$_{21}$--2$_{20}$
(or 4$_{23}$--3$_{22}$/3$_{22}$--2$_{21}$)
ratio is weakly affected by optical depths. To further
check how optical depths influence the
para-H$_2$CO\,4$_{04}$--3$_{03}$/3$_{03}$--2$_{02}$ ratio,
we use the above method also with the
para-H$_2$CO\,4$_{22}$--3$_{21}$/3$_{21}$--2$_{20}$ ratio to
constrain spatial density. The spatial densities obtained
both from para-H$_2$CO\,4$_{04}$--3$_{03}$/3$_{03}$--2$_{02}$
(typical value $\sim$\,1.0) and 4$_{22}$--3$_{21}$/3$_{21}$--2$_{20}$
(typical value $\sim$\,1.5) ratios yield similar values
($n$(H$_2$)\,$\sim$\,2\,$\times$\,10$^6$\,cm$^{-3}$),
which confirms that para-H$_2$CO\,3$_{03}$--2$_{02}$ and
4$_{04}$--3$_{03}$ lines are not strongly affected by saturation
effects when trying to constrain spatial density and
kinetic temperature in our sample. However, the ortho-H$_2$CO\,3$_{12}$--2$_{11}$
and 3$_{13}$--2$_{12}$ lines are affected by opacities $\gtrsim$\,1
in parts of our sample (see Sect.\,\ref{sect:column-density-opacities}),
so the $N$(ortho-H$_2$CO) may be underestimated in these sources.

The 870\,$\mu$m continuum source angular sizes range from
22$''$ to 42$''$ with an average of 29$''$ in our sample.
If the sizes of H$_2$CO are much smaller than those of
the 870\,$\mu$m continuum (and/or our beam size;
see Tab.\,\ref{table:APEX-receiver}), the beam-filling factor
is overestimated. If we assume that the H$_2$CO to 870\,$\mu$m
emission size ratio ($\theta_{\rm H_2CO}$/$\theta_{\rm 870\,\mu m}$) is
90\%, 80\%, and 70\%, for AGAL008.684$-$00.367 as an example
(see Fig.\,\ref{figure:N(H2CO)-n(H2)}), $n$(H$_2$) decreases
by 6\%, 17\%, and 23\%. Mapping observations of massive clumps in
CS\,(7-6) \citep{Wu2010} and 350\,$\mu$m continuum emission \citep{Mueller2002}
show that the median ratio of CS\,(7-6) emission size to the 350\,$\mu$m
continuum emission size is $\sim$0.87 \citep{Liu2016}. Also considering
the slightly different beam sizes for para-H$_2$CO\,3$_{03}$--2$_{02}$
and 4$_{04}$--3$_{03}$ lines (see Sect.\,\ref{sect:Source-size}),
we conclude that the beam-filling factor is not strongly influencing our results
for $n$(H$_2$) constrained from
para-H$_2$CO\,4$_{04}$--3$_{03}$/3$_{03}$--2$_{02}$ line ratios.

The statistical weight ratio of
ortho- and para-H$_2$CO and previous H$_2$CO observations
in other star-forming regions suggest that the ortho-to-para H$_2$CO
abundance ratio is $\lesssim$\,3 \citep{Kahane1984,Mangum1993a,Dickens1999,Jorgensen2005,Guzman2011}.
In most of our sample ($\sim$95\%) the obtained ortho-to-para H$_2$CO
abundance ratios ($N$(ortho-H$_2$CO)/$N$(para-H$_2$CO)) range
from 1.0 to 3.0 with an unweighted average of 2.0\,$\pm$\,0.1.
Assuming that H$_2$CO is formed in and expelled from dust grain mantles,
this ratio corresponds to a dust temperature of $\lesssim$\,20\,K \citep{Kahane1984,Dickens1999}.

\section{Discussion}
\label{sect:Discussion}
\subsection{Variations of spatial density and H$_2$CO abundance}
\label{sect:Variation-density-H2CO}
The gas spatial densities, $n$(H$_2$), derived from
para-H$_2$CO 4$_{04}$--3$_{03}$/3$_{03}$--2$_{02}$ ratios range
from 6.3\,$\times$\,10$^5$ to 8.3\,$\times$\,10$^6$\,cm$^{-3}$ with
an unweighted average of 1.5\,($\pm$0.1)\,$\times$\,10$^6$\,cm$^{-3}$ (Tab.\,\ref{table:NH2CO-Tkin}),
which agrees with the results determined with
para-H$_2$CO (5$_{05}$--4$_{04}$/3$_{03}$--2$_{02}$ and
5$_{24}$--4$_{23}$/3$_{22}$--2$_{21}$) and
ortho-H$_2$CO (4$_{13}$--4$_{14}$/3$_{12}$--3$_{13}$) ratios
from other star-forming regions \citep{Mangum1993a,Hurt1996,McCauley2011,Lindberg2015}.
Mapping the same para-H$_2$CO transitions toward the Galactic CMZ clouds
shows that the spatial density of the widespread warm gas is
constrained to 10$^4$--10$^6$\,cm$^{-3}$ \citep{Immer2016}.
The spatial densities derived from para-H$_2$CO line ratios in our massive clumps
overlap with the values found for high density regions
in the Galactic CMZ clouds (see Table \ref{table:NH2CO-Tkin} or
Figure \ref{figure:N(H2CO)-N(H2)-Tk}, and Figure 3 in \citealt{Immer2016}).
The spatial density deduced from the dust indicates 10$^3$--10$^6$\,cm$^{-3}$
in our sample \citep{Giannetti2017}, which is lower than the spatial densities we obtain.
This suggests that H$_2$CO ($J$\,=\,3--2 and 4--3) traces denser gas than the dust emission.

\begin{figure}[t]
\vspace*{0.2mm}
\begin{center}
\includegraphics[width=0.48\textwidth]{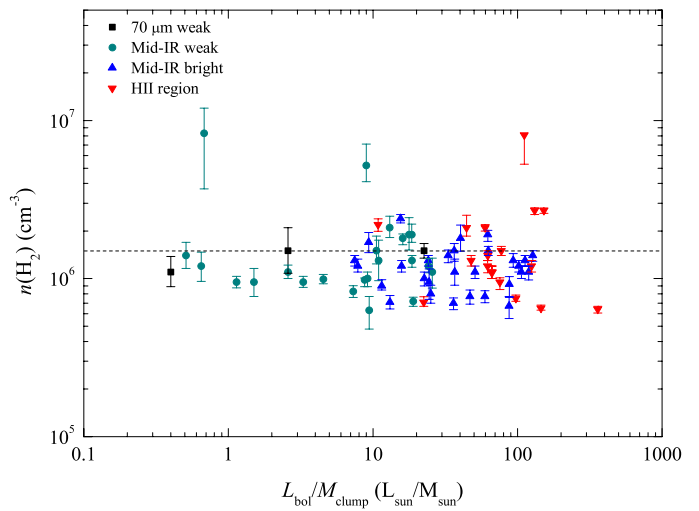}
\end{center}
\caption{The spatial density derived from
para-H$_2$CO\,(4$_{04}$--3$_{03}$/3$_{03}$--2$_{02}$) vs.
luminosity-to-mass ratio $L_{\rm bol}$/$M_{\rm clump}$.
The dashed line indicates the average spatial density.}
\label{figure:n-L-M}
\end{figure}

We derive unweighted averaged spatial densities obtained from
para-H$_2$CO ratios in sources representing four evolutionary stages consisting
of 70 $\mu$m weak (70w), mid-infrared weak (IRw),
and mid-infrared bright (IRb), sources as well as star-forming
clouds with ultracompact H\,{\scriptsize II} regions.
The unweighted averaged spatial densities $n$(H$_2$)
are 1.2\,($\pm$0.2)\,$\times$\,10$^6$, 1.7\,($\pm$0.4)\,$\times$\,10$^6$,
1.2\,($\pm$0.1)\,$\times$\,10$^6$\,cm$^{-3}$, and
1.8\,($\pm$0.4)\,$\times$\,10$^6$ cm$^{-3}$ in 70w, IRw,
IRb, and H\,{\scriptsize II} regions, respectively
(see Tab.\,\ref{table:average-parameters} or Fig.\,\ref{figure:n-L-M}).
It seems that the averaged spatial
densities traced by the para-H$_2$CO 4$_{04}$--3$_{03}$/3$_{03}$--2$_{02}$
ratios do not vary significantly with the evolutionary stage of clumps.
This may indicate that the density structure does not evolve significantly
as the star formation proceeds.
It also suggests that the para-H$_2$CO 4$_{04}$--3$_{03}$/3$_{03}$--2$_{02}$
ratio may be a good densitometer to trace the dense gas at
various stages of massive star formation.

$N$(para-H$_2$CO) derived from the para-H$_2$CO 4$_{04}$--3$_{03}$/3$_{03}$--2$_{02}$
ratio ranges from 6.4\,$\times$\,10$^{12}$ to 6.1\,$\times$\,10$^{14}$\,cm$^{-2}$
with an unweighted average of 8.0\,($\pm$1.3)\,$\times$\,10$^{13}$\,cm$^{-2}$
(Tab.\,\ref{table:average-parameters}), which agrees
with the results from other protostellar cores and star-forming regions \citep{Mangum1993a,Hurt1996,Watanabe2008,Tang2017a}.
We also derive averaged column densities
of para-H$_2$CO for the four evolutionary stages mentioned above.
The unweighted average column densities $N$(para-H$_2$CO)
are 1.2\,($\pm$0.6)\,$\times$\,10$^{13}$, 2.7\,($\pm$0.4)\,$\times$\,10$^{13}$,
7.2\,($\pm$2.1)\,$\times$\,10$^{13}$, and
16.9\,($\pm$0.3)\,$\times$\,10$^{13}$\,cm$^{-2}$ in 70w, IRw,
IRb, and H\,{\scriptsize II} regions, respectively (see Tab.\,\ref{table:average-parameters}).
The fractional abundance $X$(para-H$_2$CO)\,=\,$N$(para-H$_2$CO)/$N$(H$_2$)
becomes 1.0\,$\times$\,10$^{-10}$--1.2\,$\times$\,10$^{-9}$
with an average of 3.9\,($\pm$0.2)\,$\times$\,10$^{-10}$,
where $N$(H$_2$) is derived from the 870\,$\mu$m continuum
emission assuming a dust absorption coefficient
$\kappa_{870}$\,=\,1.85\,cm$^2$\,g$^{-1}$ at 870\,$\mu$m
and adopting the temperature obtained from the dust \citep{Konig2017}.
Therefore the abundance also agrees with the values found
in other star formation regions, Galactic center clouds,
and external galaxies \citep{Gusten1983,Zylka1992,Ao2013,Gerner2014,Tang2017a,Tang2017b}.
The unweighted average fractional abundances
$X$(para-H$_2$CO) are 1.8\,($\pm$0.7)\,$\times$\,10$^{-10}$,
2.7\,($\pm$0.2)\,$\times$\,10$^{-10}$, 4.3\,($\pm$0.4)\,$\times$\,10$^{-10}$,
and 4.9\,($\pm$0.6)\,$\times$\,10$^{-10}$ in 70w, IRw, IRb, and
H\,{\scriptsize II} regions, respectively (see Tab.\,\ref{table:average-parameters}).
Averaged variations of fractional abundances
of $X$(para-H$_2$CO) in different stages of star formation
amount to nearly a factor of 3, which agrees with observed results in
other massive star formation regions
\citep{van der Tak2000a,van der Tak2000b,Gerner2014,Tang2017a}.
Therefore, we confirm that H$_2$CO can be widely used as a probe to
trace the dense gas without drastic changes in abundance during various
stages of star formation.

The column densities of para-H$_2$CO and the fractional abundances of
$X$(para-H$_2$CO) with corresponding H$_2$ column density,
spatial density $n$(H$_2$),
kinetic temperature $T_{\rm kin}$(para-H$_2$CO 4$_{22}$--3$_{21}$/4$_{04}$--3$_{03}$),
bolometric luminosity, clump mass, and luminosity-to-mass
($L_{\rm bol}$/$M_{\rm clump}$) ratio are shown in
Figure \ref{figure:N(H2CO)-N(H2)-Tk}.
It is apparent that the para-H$_2$CO column density increases
proportionally to the H$_2$ column density,
gas kinetic temperature, bolometric luminosity,
and $L_{\rm bol}$/$M_{\rm clump}$ ratio in the massive clumps.
The fractional abundance of $X$(para-H$_2$CO) remains stable with increasing
H$_2$ column density, spatial density, and mass of clump
(Fig.\,\ref{figure:N(H2CO)-N(H2)-Tk}).
Nevertheless, the scatter in $X$(para-H$_2$CO) amounts to 0.1--1.2\,$\times$\,10$^{-9}$,
i.e. to a factor of $\sim$10. The stable (relative to other molecular species; e.g., \citealt{Tang2017b}) para-H$_2$CO
fractional abundances as a function of $N$(H$_2$)
indicate that H$_2$CO is a reliable tracer of the H$_2$ column density.

The luminosity-to-mass ratio is
a good evolutionary tracer for massive and dense cluster-progenitor
clumps \citep{Molinari2008,Molinari2016,Liu2013,Ma2013,Giannetti2017}.
The fractional abundance of $X$(para-H$_2$CO) shows a weak
increasing trend with kinetic temperature, bolometric luminosity, and
$L_{\rm bol}$/$M_{\rm clump}$ ratio (see Fig.\,\ref{figure:N(H2CO)-N(H2)-Tk}).
The H$_2$CO abundances seem to increase with the evolutionary stage of massive clumps.
Similar trends were seen in the massive star formation regions
studied by \cite{Gerner2014} and \cite{Immer2014}.
This indicates that H$_2$CO abundances may be enhanced by high temperature,
infrared radiation, and clump evolution, which would support
a scenario in which H$_2$CO is increasingly released from
dust grains into the gas phase during the evolution of the star-forming region.

\begin{figure}[t]
\vspace*{0.2mm}
\begin{center}
\includegraphics[width=0.48\textwidth]{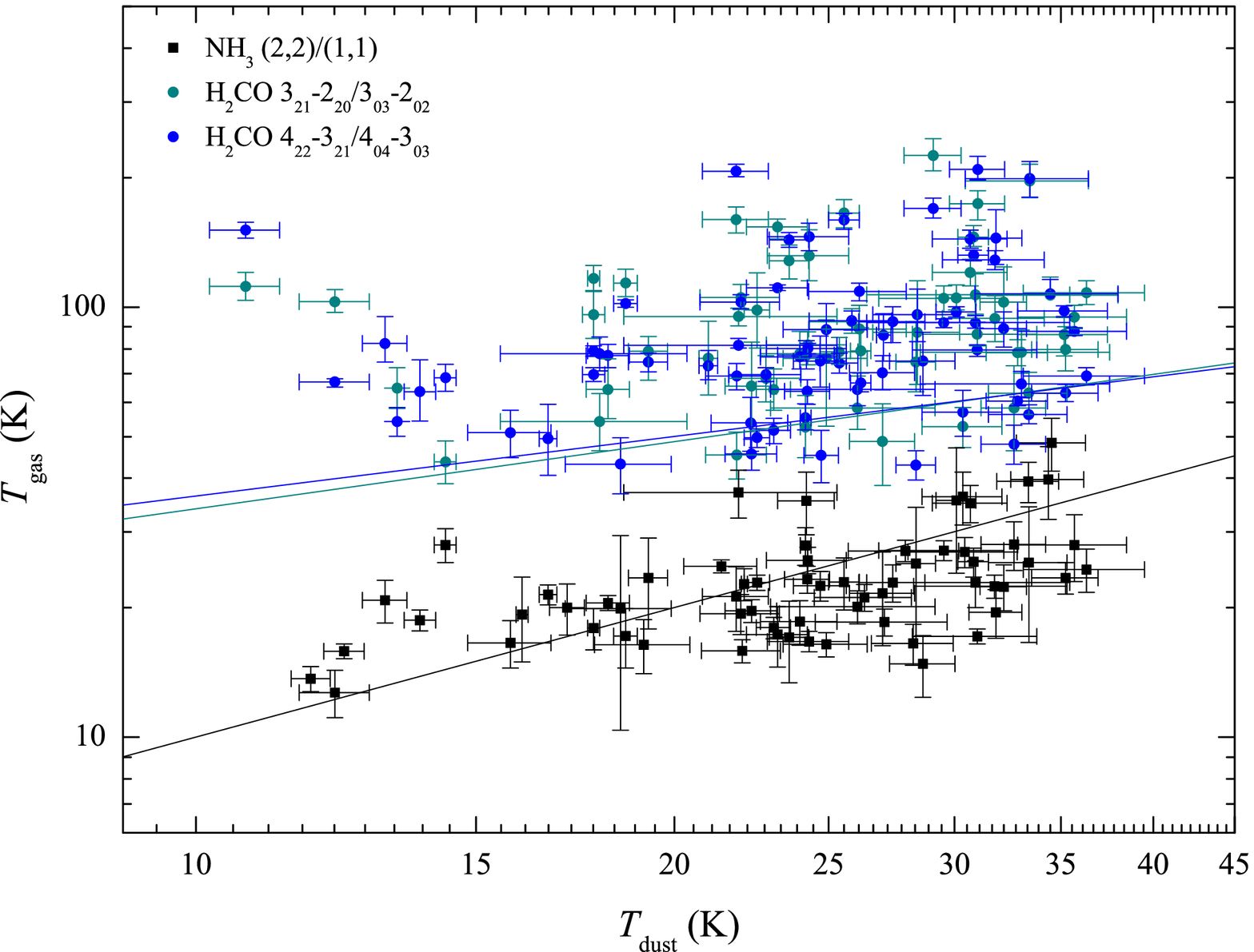}
\includegraphics[width=0.48\textwidth]{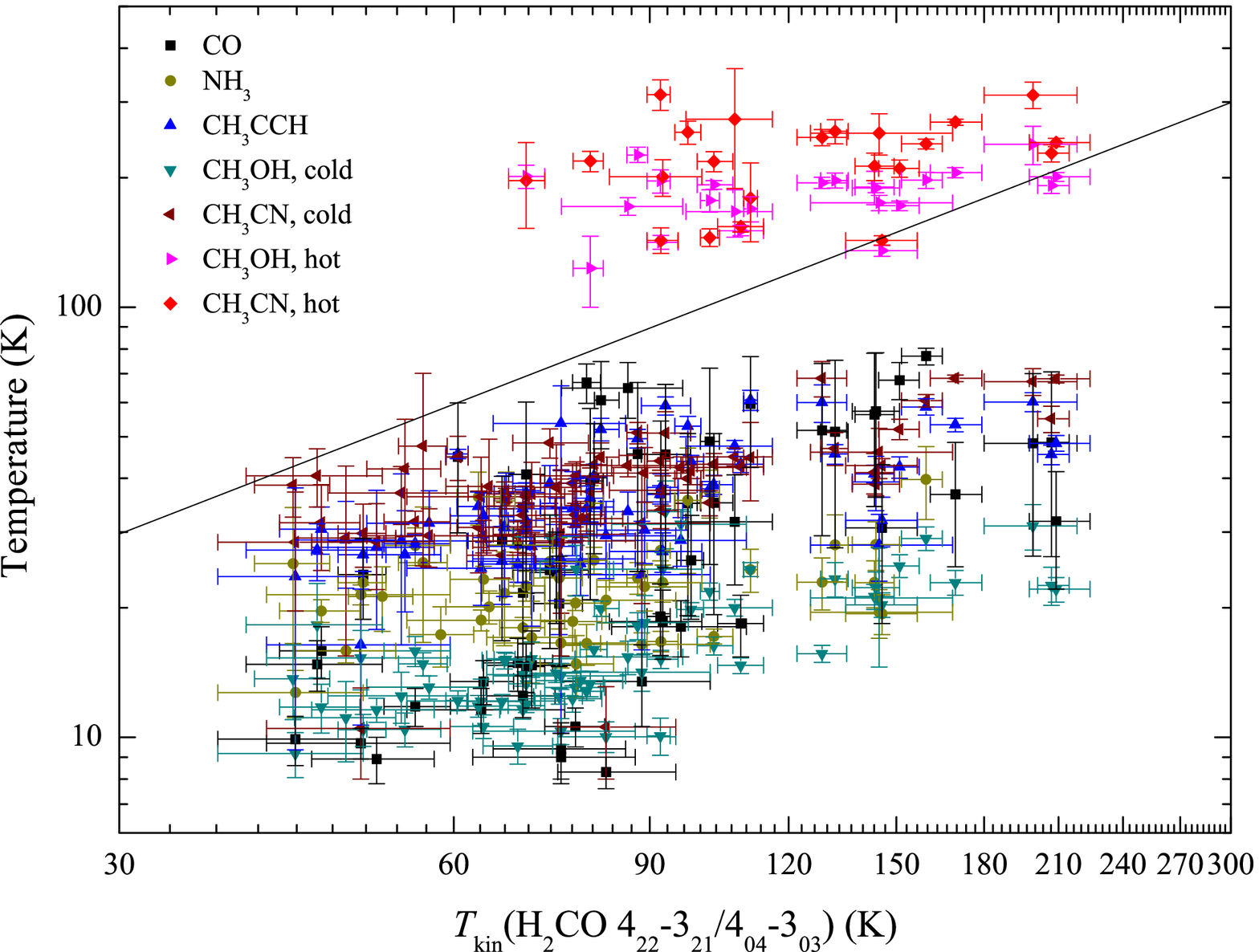}
\end{center}
\caption{Top panel: Comparison of kinetic temperatures derived from
para-H$_2$CO\,(3$_{21}$--2$_{20}$/3$_{03}$--2$_{02}$ and
4$_{22}$--3$_{21}$/4$_{04}$--3$_{03}$, cyan and blue points)
and NH$_3$\,(2,2)/(1,1) (black squares) ratios against the dust temperatures.
NH$_3$ kinetic temperatures are selected from \cite{Wienen2012}.
The cyan and blue straight lines are the results from unweighed linear fits
for gas temperatures derived from para-H$_2$CO\,(3$_{21}$--2$_{20}$/3$_{03}$--2$_{02}$ and
4$_{22}$--3$_{21}$/4$_{04}$--3$_{03}$, repectively.
Bottom panel: Comparisons of gas temperatures derived
from para-H$_2$CO\,(4$_{22}$--3$_{21}$/4$_{04}$--3$_{03}$), CO,
NH$_3$\,(2,2)/(1,1), CH$_3$OH, CH$_3$CN, and CH$_3$CCH.
Temperatures of CO, CH$_3$OH, CH$_3$CN, and CH$_3$CCH are
taken from \cite{Giannetti2017}. The black straight lines in both panels indicate equal temperatures.}
\label{figure:TH2CO-TNH3-Tdust}
\end{figure}

\subsection{Comparison of kinetic temperatures derived from the gas and the dust}
\label{Tgas-Tdust}
The gas kinetic temperatures derived from the
para-H$_2$CO (3$_{21}$--2$_{20}$/3$_{03}$--2$_{02}$ and
4$_{22}$--3$_{21}$/4$_{04}$--3$_{03}$) line
ratios are rather warm, ranging from 43 to $>$300\,K
with an unweighted average of 91\,$\pm$\,4\,K, which agrees with the results
measured with H$_2$CO in other massive star-forming regions and Galactic center clouds
\citep{Mangum1993a,Hurt1996,Mangum1999,Watanabe2008,Nagy2012,Ao2013,Ginsburg2016,Immer2016,Lu2017}.
Most of our clumps, including the detected 70\,$\mu$m weak clumps,
are very warm indicating that there is likely massive star formation
on-going in most of our sample. The average kinetic temperatures $T_{\rm kin}$
are high in early evolutionary stages of the clumps (70w and IRw)
(see Tab.\,\ref{table:average-parameters}), which is consistent with previous
observational results measured with para-H$_2$CO\,(3--2) in star-forming regions
with outflows \citep{Tang2017a}. 16 sources of our sample in early evolutionary
stages have been observed in SiO\,(2--1) and (5--4) \citep{Csengeri2016},
SiO emission is detected in all these sources. This indicates that the dense
gas probed by H$_2$CO may be heated by an outflow/shock. Therefore,
in early evolutionary stages of the clumps, para-H$_2$CO traces higher
temperature gas which may be related to gas excited by star formation
activities (e.g., ourflows, shocks) \citep{Tang2017a}.

Parts of our sample have been measured in NH$_3$\,(2,2)/(1,1) by \cite{Wienen2012}.
We compare gas kinetic temperatures derived from
para-H$_2$CO and NH$_3$\,(2,2)/(1,1)
against dust temperatures in Figure \ref{figure:TH2CO-TNH3-Tdust}.
This comparison shows that the gas temperatures determined from NH$_3$\,(2,2)/(1,1)
agree with the dust temperatures (also see \citealt{Giannetti2017}),
but are lower than those derived from
para-H$_2$CO\,(3$_{21}$--2$_{20}$/3$_{03}$--2$_{02}$ and
4$_{22}$--3$_{21}$/4$_{04}$--3$_{03}$).
Previous observations toward the Galactic CMZ, dense massive clumps,
and star formation regions indicate that
in many cases para-H$_2$CO\,(3$_{21}$--2$_{20}$/3$_{03}$--2$_{02}$ and
4$_{22}$--3$_{21}$/4$_{04}$--3$_{03}$) traces a higher kinetic temperature
than the NH$_3$\,(2,2)/(1,1) transitions and the dust
\citep{Ao2013,Ott2014,Ginsburg2016,Immer2016,Tang2017a,Tang2017c}.
The difference is likely due to the fact that the derived kinetic temperatures
from NH$_3$\,(2,2)/(1,1) may reflect an average temperature
of cooler and more diffuse gas \citep{Henkel1987,Ginsburg2016}, while
para-H$_2$CO\,($J$\,=\,3--2 and 4--3) ratios trace denser and hotter regions
more directly associated with star formation activity \citep{Tang2017a,Tang2017c}.

Temperatures toward our selected massive clumps have been
measured with CO, CH$_3$OH, CH$_3$CN, and CH$_3$CCH \citep{Giannetti2017}.
We compare gas kinetic temperatures derived from
para-H$_2$CO\,(4$_{22}$--3$_{21}$/4$_{04}$--3$_{03}$), CO,
CH$_3$OH, CH$_3$CN, and CH$_3$CCH in Figure \ref{figure:TH2CO-TNH3-Tdust}.
It shows that the gas temperatures determined from para-H$_2$CO
are higher than those derived from CO, CH$_3$OH (cold component),
CH$_3$CN (cold component), and CH$_3$CCH, but are lower than those
obtained from the CH$_3$OH and CH$_3$CN hot components. This indicates
that para-H$_2$CO ($J$\,=\,3--2 and 4--3) ratios may trace dense gas in layers
intermediate between those of CH$_3$CCH and CH$_3$CN (hot component),
with the latter likely most closely related to recently formed massive stars.

The dust temperatures of our sample are obtained from SED fitting to
Herschel HiGal data at 70, 160,
250, 350, and 500\,$\mu$m and ATLASGAL data at 870\,$\mu$m by \cite{Konig2017}.
The results are listed in Table \ref{table:source}. The derived
dust temperature range in our observed sources is 11--41\,K with an unweighted average of 25\,$\pm$\,7\,K.
Previous observations show that the temperatures derived from gas
and dust are often in agreement in the active dense clumps of Galactic disk clouds
\citep{Dunham2010,Giannetti2013,Battersby2014}, but do not agree in the Galactic
CMZ \citep{Gusten1981,Ao2013,Ott2014,Ginsburg2016,Immer2016,Lu2017}.
As in the CMZ, the gas kinetic temperatures derived from para-H$_2$CO
show higher values than the
dust temperature with no apparent correlation (correlation coefficient $R$\,$\sim$\,0.2) between $T_{\rm dust}$
and $T_{\rm gas}$ (see Fig.\,\ref{figure:TH2CO-TNH3-Tdust}).

Commonly it is expected that the gas and dust are thermally coupled in
the densest regions ($n$(H$_2$)\,>\,10$^{4.5}$\,cm$^{-3}$) \citep{Goldsmith2001},
because at such densities
interactions between dust and gas become sufficiently frequent.
The dust emission at mid-infrared (MIR) emission traces primarily
warm dust components \citep{Helou1986}. Dust temperatures derived
from MIR multi-filter data agree with gas temperatures derived from
multi-inversion transitions of NH$_3$ in external galaxies
\citep{Melo2002,Tomono2006,Ao2011,Mauersberger2003}.
Combining the MIR data for our sample, the fit of the warm gas
emission in the SED shows a cold and a warm component (see \citealt{Konig2017}).
Our dust temperatures are taken from the cold component of the SED fitted results.
Dust emission at far-infrared (FIR) emission originates
primarily from colder dust components that may not be directly
associated with star formation activity
\citep{Schnee2009,Bendo2012,Mangum2013a},
so the dust temperatures derived from FIR
measurements rarely exceed 50\,K in star formation regions of
our Galaxy and external galaxies
(e.g., \citealt{Henkel1986,Gao2004a,Bernard2010,Mangum2013a,Guzman2015,Merello2015,
He2016,Lin2016,Konig2017,Yu2016,Tang2017a,Elia2017}).
This suggests that the HiGal dust emission may trace colder
dust components that may not be used as a proxy for dust and
gas kinetic temperatures (at least traced by H$_2$CO) in dense regions
with massive star formation activity.

\subsection{The linewidth-luminosity relation}
\label{sect:linewidth-luminosity}
The observed linewidths of para-H$_2$CO\,(4$_{04}$--3$_{03}$) range
from 1.2 to 12.8\,km\,s$^{-1}$ with an unweighted average of 5.3\,$\pm$\,0.2\,km\,s$^{-1}$.
Using a mean unweighted kinetic temperature $T_{\rm kin}$\,$\sim$\,91\,K and
averaged linewidths of H$_2$CO, the thermal and non-thermal linewidths
($\sigma_{\rm T}$\,=\,$\sqrt{\frac{kT_{\rm kin}}{m_{\rm H_2CO}}}$ and
$\sigma_{\rm NT}$\,=\,$\sqrt{\frac{\Delta v^2}{8{\rm ln}2}-\sigma_{\rm T}^2}$,
where $k$ is the Boltzmann constant, $T_{\rm kin}$ is the kinetic
temperature of the gas, $m_{\rm H_2CO}$ is the mass of the formaldehyde
molecule, and $\Delta v$ is the measured FWHM linewidth of H$_2$CO)
are 0.15 and 2.25\,km\,s$^{-1}$, respectively. The thermal linewidth is significantly
lower than the non-thermal linewidth. The sound speed
($a_{\rm s}$=$\sqrt{\frac{kT_{\rm kin}}{\mu m_{\rm H}}}$,
where $\mu$\,=\,2.37 is the mean molecular weight for molecular
clouds and $m_{\rm H}$ is the mass of the hydrogen atom) is
$\sim$0.54\,km\,s$^{-1}$ at temperature 91\,K, so the Mach number
(given as $M$\,=\,$\sigma_{\rm NT}$/$a_{\rm s}$) is
4.2 which agrees with the results of the high-mass clumps
(mean value $\sim$ 3.5 derived from NH$_3$; \citealt{Wienen2012}) and
the Bolocam Galactic Plane Survey (BGPS) sources (mean value $\sim$ 3.2
derived from NH$_3$; \citealt{Dunham2011}). This indicates that these
massive clumps are turbulent and H$_2$CO linewidths are influenced
strongly by supersonic non-thermal motions in our samples.

\begin{figure}[t]
\vspace*{0.2mm}
\begin{center}
\includegraphics[width=0.48\textwidth]{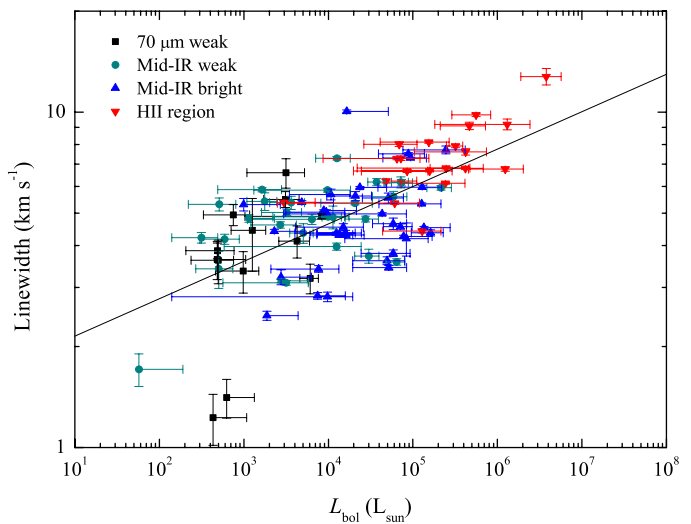}
\end{center}
\caption{The linewidth of the para-H$_2$CO\,(4$_{04}$--3$_{03}$) transition
vs. bolometric luminosity of the measured sources. The straight line is
the result from an unweighed linear fit.}
\label{figure:Width-Lbol}
\end{figure}

Previous observations of NH$_3$
\citep{Wouterloot1988,Myers1991,Harju1993,Ladd1994,Molinari1996,Jijina1999,Wu2006,
Urquhart2011,Urquhart2015}, C$^{18}$O \citep{Saito2001,Ridge2003,Maud2015},
and $^{13}$CO \citep{Wang2009,Lundquist2015} suggest that the linewidth
is correlated with luminosity, which indicates the presence of a link between
the formed stars and the velocity dispersion. We investigate
the linewidth-luminosity relation in the case of the dense gas tracer H$_2$CO.
We plot the linewidth-luminosity relation in Figure \ref{figure:Width-Lbol}.
For the linewidth of para-H$_2$CO\,(4$_{04}$--3$_{03}$)
and bolometric luminosity, the least squares linear fit result is
\begin{eqnarray}
{\rm log}\Delta v({\rm H_2CO~4_{04}{-}3_{03}})=(0.11\pm0.01)\times{\rm log}L_{\rm bol} \nonumber \\+(0.23\pm0.06).
\end{eqnarray}
The correlation coefficient, $R$, is 0.64. Other transitions of
H$_2$CO show similar $\Delta v({\rm H_2CO})$-$L_{\rm bol}$
correlations (not shown here). The slope (0.11\,$\pm$\,0.01)
of the $\Delta v({\rm H_2CO})$-$L_{\rm bol}$ correlation agrees with
previous results found with C$^{18}$O \citep{Saito2001}
and $^{13}$CO \citep{Wang2009}, but is lower than that found with
NH$_3$ \citep{Wouterloot1988,Myers1991,Jijina1999,Wu2006,Urquhart2011,
Urquhart2015}. The correlation appears consistent with the idea
that the internal velocity dispersion of the dense clumps can be used to determine the
mass of the formed stars \citep{Saito2001}.

We also derived averaged linewidths
of para-H$_2$CO\,(4$_{04}$--3$_{03}$) discriminating between
the four evolutionary stages introduced in Section \ref{sect:samp-obs}.
The unweighted averaged linewidths are 3.9\,$\pm$\,0.4, 4.8\,$\pm$\,0.2,
4.9\,$\pm$\,0.3, and 7.4\,$\pm$\,0.4\,km\,s$^{-1}$ in 70w, IRw,
IRb, and H\,{\scriptsize II} regions, respectively
(see Tab.\,\ref{table:average-parameters}).
It seems that the velocity dispersion slightly increases with
the first three evolutionary stages, 70w, IRw, and IRb.
A significant change appears to occur between the first three and the
fourth (H\,{\scriptsize II}) evolutionary stage. This suggests that
the more evolved and more luminous objects tend to be associated
with more turbulent molecular cloud structures \citep{Wang2009}.

\subsection{The non-thermal velocity dispersion-temperature relation}
Previous observations of NH$_3$ and H$_2$CO
(e.g., \citealt{Wouterloot1988,Molinari1996,Jijina1999,Wu2006,Urquhart2011,
Urquhart2015,Wienen2012,Lu2014,Immer2016,Tang2017c}) suggest that the linewidth is
correlated with kinetic temperature. It is suggested that the correlation
between kinetic temperature and linewidth is due to a conversion
of turbulent energy into heat in the Galactic central clouds
(e.g., \citealt{Gusten1985,Ginsburg2016,Immer2016}).

\begin{figure}[t]
\vspace*{0.2mm}
\begin{center}
\includegraphics[width=0.48\textwidth]{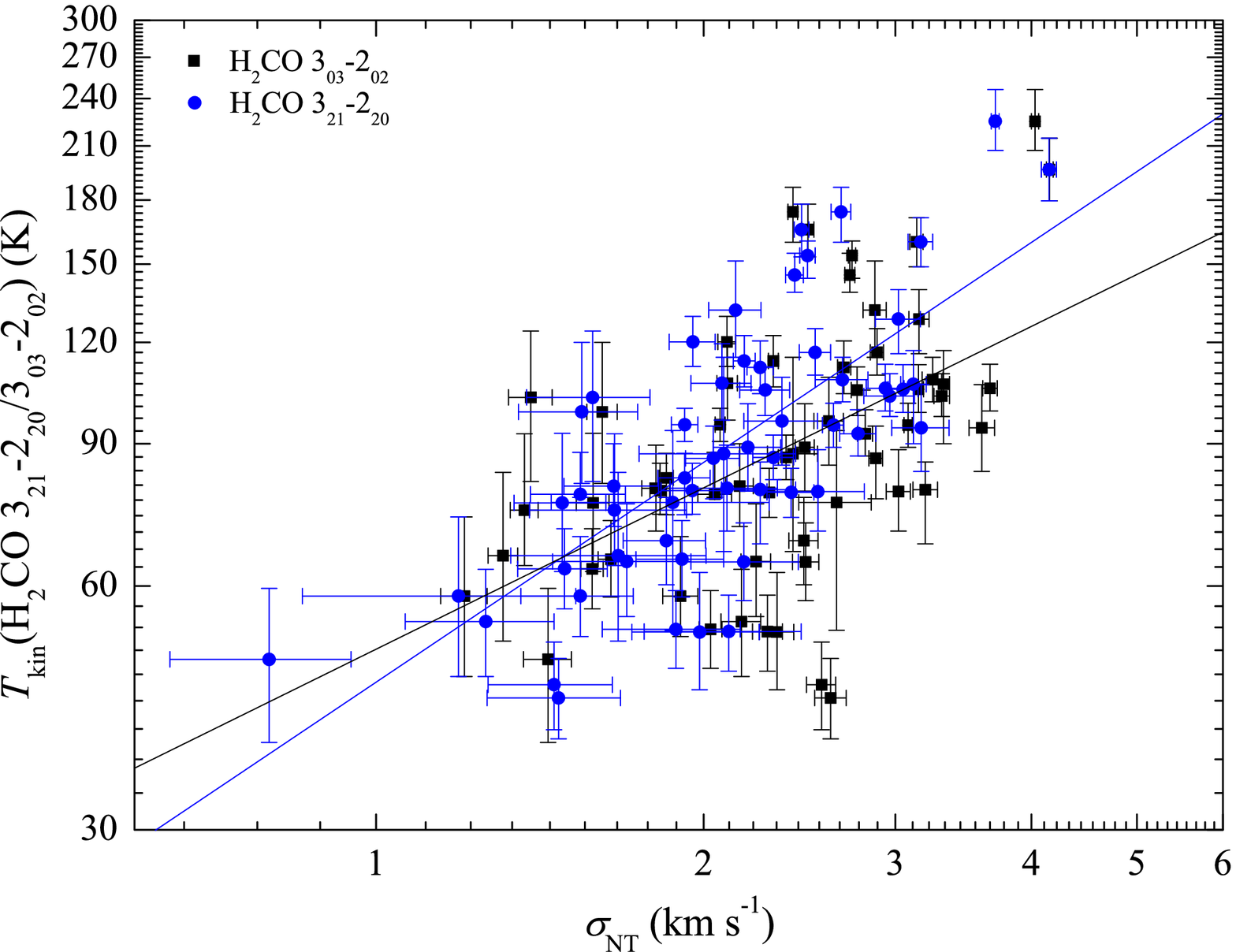}
\includegraphics[width=0.48\textwidth]{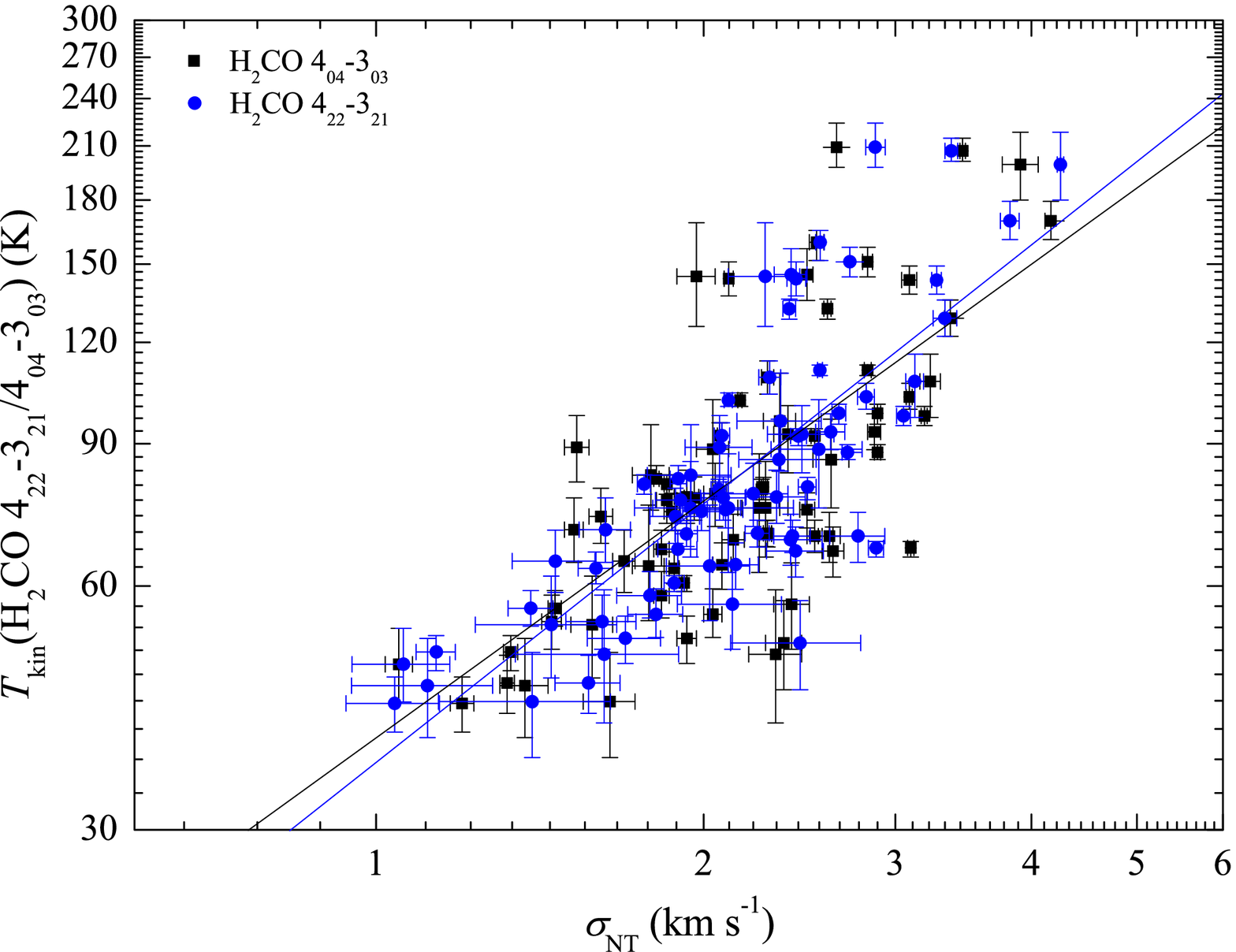}
\end{center}
\caption{Non-thermal velocity dispersion ($\sigma_{\rm NT}$) vs.
gas kinetic temperature for para-H$_2$CO.
For the top panel, gas kinetic temperatures were derived from
para-H$_2$CO\,3$_{21}$--2$_{20}$/3$_{03}$--2$_{02}$ line ratios.
For the bottom panel, the gas kinetic temperatures were derived from
para-H$_2$CO\,4$_{22}$--3$_{21}$/4$_{04}$--3$_{03}$ line ratios.
The straight lines are results from unweighed linear fits.}
\label{figure:Width-Ratio-Tkin}
\end{figure}

Here we examine whether there is a relationship between turbulence
and temperature in our massive clumps. We adopt the non-thermal
velocity dispersion ($\sigma_{\rm NT}$) of para-H$_2$CO in
good approximation as proxy for
the turbulence, and the kinetic temperatures of
para-H$_2$CO\,(3$_{21}$--2$_{20}$/3$_{03}$--2$_{02}$ and
4$_{22}$--3$_{21}$/4$_{04}$--3$_{03}$)
as the gas kinetic temperature (see Fig.\,\ref{figure:Width-Ratio-Tkin}).
For the non-thermal velocity dispersion of para-H$_2$CO
and kinetic temperature, the least squares linear fit results are
listed in Table\,\ref{table:Tk-dv}.
The non-thermal velocity dispersion of
para-H$_2$CO is significantly positively correlated with the
gas kinetic temperature by a power-law of the form
$T_{\rm kin}$\,$\propto$\,$\sigma_{\rm NT}^{0.66-1.06}$,
which is consistent with results
found with NH$_3$ and H$_2$CO in other star formation regions
\citep{Wouterloot1988,Molinari1996,Jijina1999,Wu2006,Urquhart2011,
Urquhart2015,Wienen2012,Lu2014,Tang2017c}.
The gas is heated by turbulent energy according to the
approximate relation $T_{\rm kin}$ $\propto$ $\Delta v^{0.8-1.0}$
(gas kinetic temperature measured with NH$_3$ and H$_2$CO) in molecular clouds
of the Galactic center \citep{Gusten1985,Mauersberger1987,Immer2016},
which is consistent with our result (only in terms
of slope, not of intercept and absolute value).
All this implies that the gas may be heated by
turbulent motions in our massive clumps on scales of $\sim$0.1--1.8\,pc.

\begin{table}[t]
\caption{Kinetic temperature vs. H$_2$CO non-thermal velocity dispersion.}
\centering
\begin{tabular}
{cccccccccccc}
\hline\hline 
Transition & \multicolumn{3}{c}{$T_{\rm kin}$--$\sigma_{\rm NT}$(H$_2$CO)}\\
\cline{2-4}
& Slope & Intercept & $R$ \\
\hline 
p-H$_2$CO 3$_{03}$--2$_{02}$ &0.66 (0.15)  &1.70 (0.06) &0.52\\
p-H$_2$CO 3$_{21}$--2$_{20}$ &0.90 (0.11)  &1.66 (0.04) &0.73\\
p-H$_2$CO 4$_{04}$--3$_{03}$ &0.97 (0.12)  &1.59 (0.04) &0.70\\
p-H$_2$CO 4$_{22}$--3$_{21}$ &1.06 (0.10)  &1.56 (0.04) &0.78\\
\hline 
\end{tabular}
\tablefoot{The format of the regression fits is
${\rm log}T_{\rm kin}\,=\,{\rm Slope}\,\times\,{\rm log}\sigma_{\rm NT}(\rm H_2CO)\,+\,{\rm Intercept}$.
$R$ is the correlation coefficient for the linear fit.}
\label{table:Tk-dv}
\end{table}

Recent para-H$_2$CO mapping observations of molecular clouds
in the Galactic CMZ show that the warm dense gas is heated
most likely by turbulence \citep{Ao2013,Ginsburg2016,Immer2016}.
Following the method applied by \cite{Tang2017c} in their
Equation (2),
\begin{eqnarray}
3.3\times10^{-27}~n~\sigma_{\rm NT}^3~L^{-1} = 4\times10^{-33}~n^2~T_{\rm turb}^{1/2}(T_{\rm turb}-T_{\rm dust}) \nonumber \\
+~6\times10^{-29}~n^{1/2}~T_{\rm turb}^3~{\rm d}v/{\rm d}r,
\end{eqnarray}
where the gas density $n$ is in units of cm$^{-3}$,
the velocity gradient d$v$/d$r$ is in units of km\,s$^{-1}$pc$^{-1}$,
the one-dimensional non-thermal velocity dispersion $\sigma_{\rm NT}$
is in units of km\,s$^{-1}$, and the cloud size $L$ is in units of pc,
we determine the gas kinetic temperature caused by turbulent energy.
We computed the gas kinetic temperature assuming
turbulent heating dominates the heating process. We have
assumed a cloud size of $\sim$1\,pc
(e.g., \citealt{Dunham2010,Dunham2011,Rosolowsky2010,
Urquhart2014,He2015,Wienen2015,Konig2017,Yuan2017}),
a velocity gradient d$v$/d$r$\,=\,1\,km\,s$^{-1}$pc$^{-1}$,
the above mentioned (Sect.\,\ref{sect:linewidth-luminosity})
averaged non-thermal velocity dispersion of 2.25\,km\,s$^{-1}$
measured with H$_2$CO, and an averaged gas spatial
density $\sim$10$^6$\,cm$^{-3}$ derived from H$_2$CO line intensity
ratios (Sect.\,\ref{sect:density}).
If the averaged dust temperature ($T_{\rm dust}$\,$\sim$\,25\,K;
derived from HiGal and ATLASGAL data; see Sect.\,\ref{Tgas-Tdust})
and averaged gas temperature
($T_{\rm kin}$\,$\sim$\,91\,K; derived from the para-H$_2$CO line ratios;
see Sect.\,\ref{Tgas-Tdust}) are adopted as the dust temperatures,
the gas kinetic temperatures due to turbulence motions $T_{\rm turb}$, are 55
and 88\,K, respectively. The obtained $T_{\rm turb}$ values are
slightly lower than the averaged gas kinetic temperature
($T_{\rm kin}$\,$\sim$\,91\,K) derived from the para-H$_2$CO
line ratios. This indicates that turbulent heating significantly
contributes to gas temperature in these massive clumps on scales
of $\sim$0.1--1.8\,pc, which agrees with previous observational
results with H$_2$CO in the Orion molecular cloud 1 (OMC-1; \citealt{Tang2017c}).
Apparently, turbulent heating plays an important role in heating
the dense gas in massive star-forming clumps \citep{Pan2009}.

\subsection{Correlation of gas temperature with luminosity}
Previous observations of our selected massive clump temperatures
determined from CO, NH$_3$, CH$_3$CN, CH$_3$CCH, and CH$_3$OH
\citep{Wienen2012,Giannetti2014,Giannetti2017} suggest
that these clumps are heated by radiation from internal massive stars.
The comparison between the kinetic temperature and luminosity further helps
us to understand the internal heating of embedded infrared
sources upon their surrounding dense gas.

\begin{figure}[t]
\vspace*{0.2mm}
\begin{center}
\includegraphics[width=0.48\textwidth]{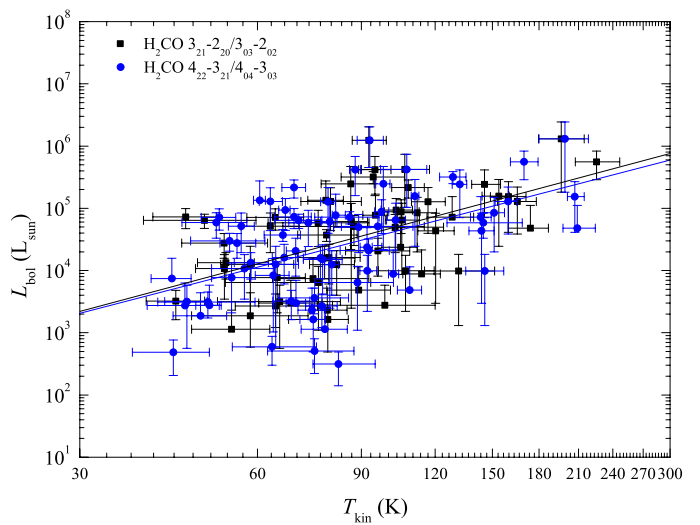}
\end{center}
\caption{The kinetic temperatures derived from
para-H$_2$CO\,(3$_{21}$--2$_{20}$/3$_{03}$--2$_{02}$
and 4$_{22}$--3$_{21}$/4$_{04}$--3$_{03}$, black squares and blue points)
vs. bolometric luminosity.
The straight lines are the results from unweighed linear fits.}
\label{figure:Tkin-Lbol}
\end{figure}

To investigate how the kinetic temperatures traced by para-H$_2$CO
correlate with luminosity in these massive clumps, we compared the gas
kinetic temperature to the bolometric luminosity obtained from
MSX, WISE, Herschel HiGal and ATLASGAL data \citep{Konig2017}.
A comparison between gas kinetic temperatures
derived from para-H$_2$CO\,(3$_{21}$--2$_{20}$/3$_{03}$--2$_{02}$
and 4$_{22}$--3$_{21}$/4$_{04}$--3$_{03}$)
and the bolometric luminosity is shown in Figure \ref{figure:Tkin-Lbol}.
The least squares linear fit results are
\begin{eqnarray}
{\rm log}L_{\rm bol} = (2.53\pm0.54) \times {\rm log}T_{\rm kin}({\rm 3_{21}{-}2_{20}/3_{03}{-}2_{02}}) \nonumber \\-(0.39\pm1.06)
\end{eqnarray}
and
\begin{eqnarray}
{\rm log}L_{\rm bol} = (2.46\pm0.52) \times {\rm log}T_{\rm kin}({\rm 4_{22}{-}3_{21}/4_{04}{-}3_{03}}) \nonumber \\-(0.32\pm0.99),
\end{eqnarray}
with correlation coefficients, $R$, of 0.53 and 0.50, respectively.
It shows that higher temperatures
traced by H$_2$CO are associated with more luminous sources. This
result is expected if dense gas probed by H$_2$CO is illuminated or heated
by massive stars inside or adjacent to the clouds.
The correlations between gas temperature and bolometric luminosity are weak.
The bolometric luminosity and the gas temperature derived from
para-H$_2$CO are related by a power-law of the form
$L_{\rm bol}$\,$\propto$\,$T_{\rm kin}^{2.5\pm0.5}$,
where the power-law index is not very far from that of the Stefan-Boltzmann law
($L$\,$\propto$\,$T_{\rm kin}^4$). This also suggests that the dense gas
is heated most likely by activity from associated massive stars.

\begin{figure}[t]
\vspace*{0.2mm}
\begin{center}
\includegraphics[width=0.48\textwidth]{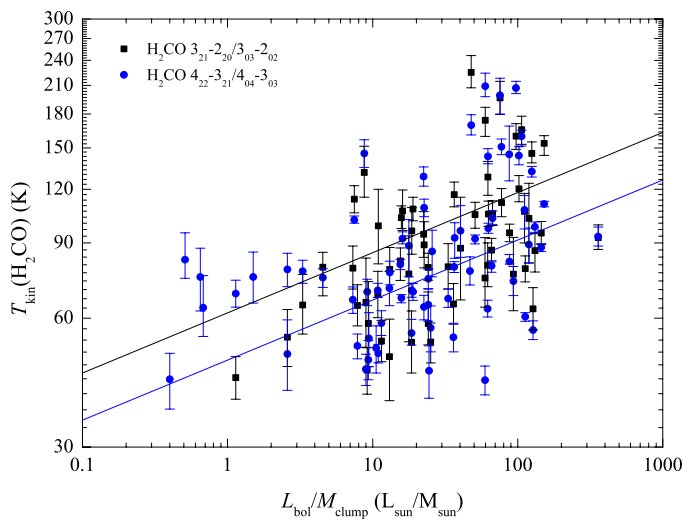}
\end{center}
\caption{The kinetic temperatures derived from
para-H$_2$CO\,(3$_{21}$--2$_{20}$/3$_{03}$--2$_{02}$
and 4$_{22}$--3$_{21}$/4$_{04}$--3$_{03}$, black squares and blue points) vs.
luminosity-to-mass ratio $L_{\rm bol}$/$M_{\rm clump}$.
The straight lines are the results from unweighed linear fits for clumps with
$L_{\rm bol}$/$M_{\rm clump}$ $\gtrsim$ 10 L$_{\odot}$/M$_{\odot}$.}
\label{figure:Tkin-Lbol-M}
\end{figure}

\begin{figure*}[t]
\vspace*{0.2mm}
\begin{center}
\includegraphics[width=0.98\textwidth]{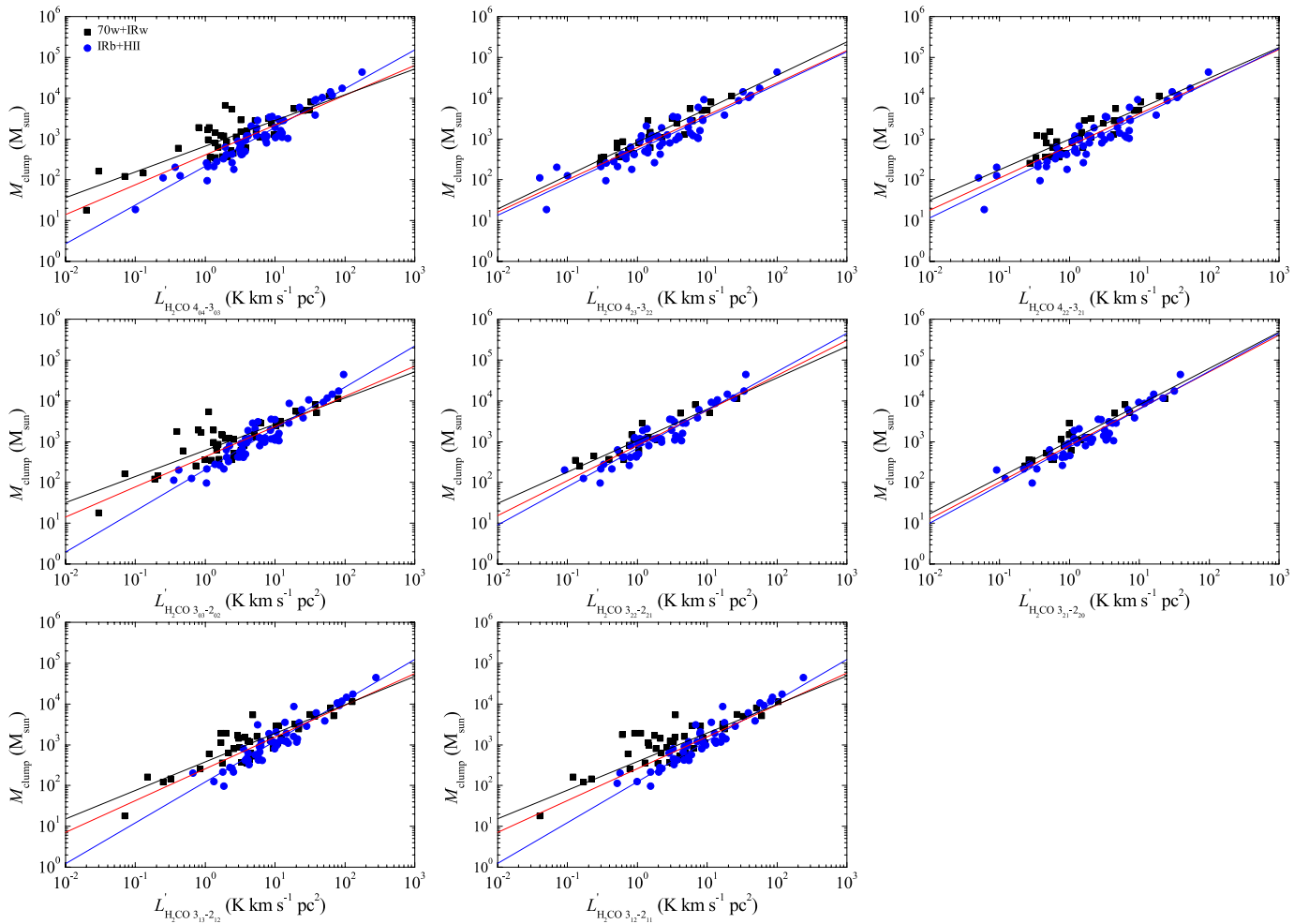}
\end{center}
\caption{$M_{\rm clump}$ vs. $L^{\prime}_{\rm H_2CO}$
for eight transition lines of H$_2$CO. Black squares indicate clumps
classified as early stage (70w and IRw) and blue circles indicate clumps
classified as late stage (IRb and H\,{\scriptsize II} regions) (see Sect.\,\ref{sect:samp-obs}
for an introduction to these sources). The black, blue, and red lines
are the results from linear fits for early stage, late stage,
and all sources, respectively.}
\label{figure:LH2CO-Mclump}
\end{figure*}

Mapping NH$_3$ observations of massive star formation regions
\citep{Lu2014,Urquhart2015} shows that in some cases the gas is
heated by radiation from external sources. Due to a lack of
H$_2$CO source structure information for our sample,
we cannot exclude that external heating is contributing in some sources of our sample.
Therefore, a detailed mapping study of our sample's
temperature structure using formaldehyde is needed.
We intend to map a part of our sample with formaldehyde to reveal
details of the gas heating mechanism in the future.

\subsection{Gas temperature and clump evolution}
To investigate
whether the kinetic temperatures traced by para-H$_2$CO are
influenced by massive young stellar objects (YSOs)
at different evolutionary stages,
we derive averaged kinetic temperatures
obtained from the para-H$_2$CO\,(4$_{22}$--3$_{21}$/4$_{04}$--3$_{03}$)
ratios for the four evolutionary stages outlined in Section \ref{sect:samp-obs}.
The unweighted average kinetic temperatures $T_{\rm kin}$
are 52\,$\pm$\,6, 73\,$\pm$\,4, 81\,$\pm$\,6, and 110\,$\pm$\,8\,K in 70w, IRw,
IRb, and H\,{\scriptsize II} regions, respectively
(see Tab.\,\ref{table:average-parameters}).
From this it is clear that the averaged gas kinetic temperature
increases with the evolutionary stage, which confirms the
trends measured with CO, NH$_3$, CH$_3$CN, CH$_3$CCH, CH$_3$OH,
and dust emission in our and other massive star-forming clumps
\citep{Giannetti2014,Giannetti2017,Guzman2015,Molinari2016,
He2016,Yu2016,Konig2017,Yuan2017,Elia2017}.
It indicates that the gas temperature probed by para-H$_2$CO
is related to the evolution of the clumps.

As mentioned in Section \ref{sect:Variation-density-H2CO},
the luminosity-to-mass ratio, $L_{\rm bol}$/$M_{\rm clump}$,
is a good evolutionary tracer for massive and dense cluster-progenitor
clumps, which defines a continuous evolutionary sequence in time.
We plot the relation between kinetic temperature derived from
para-H$_2$CO\,(3$_{21}$--2$_{20}$/3$_{03}$--2$_{02}$ and
4$_{22}$--3$_{21}$/4$_{04}$--3$_{03}$) ratios and
$L_{\rm bol}$/$M_{\rm clump}$ ratios in Figure \ref{figure:Tkin-Lbol-M}.
The plot shows that the kinetic temperature traced by para-H$_2$CO is indeed a
rising function of the luminosity-to-mass ratio, which is consistent with results
found from CH$_3$CN, CH$_3$CCH, and CH$_3$OH in massive
star-forming clumps \citep{Molinari2016,Giannetti2017}.

\begin{figure*}[t]
\vspace*{0.2mm}
\begin{center}
\includegraphics[width=0.98\textwidth]{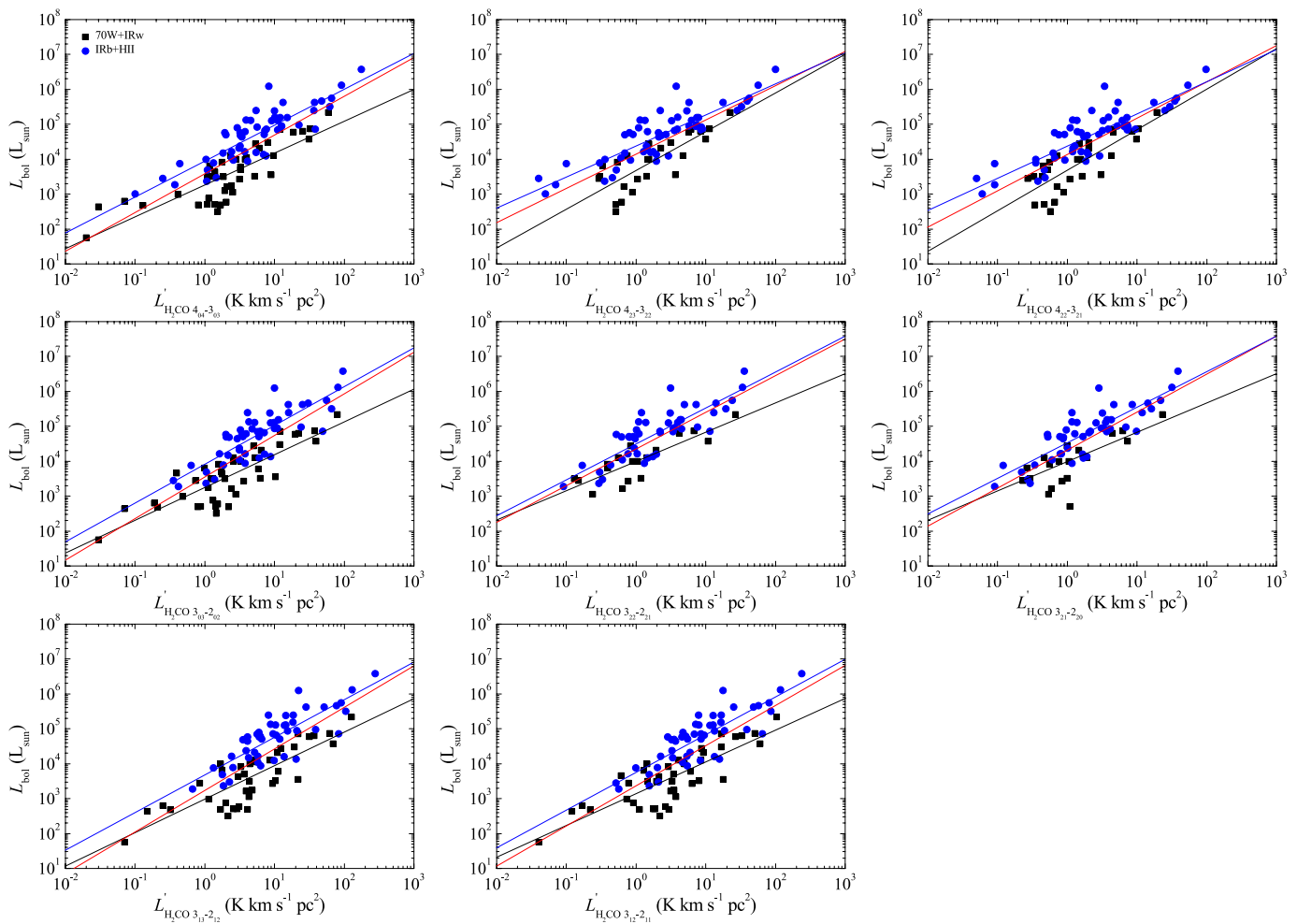}
\end{center}
\caption{$L_{\rm bol}$ vs. $L^{\prime}_{\rm H_2CO}$
for eight transition lines of H$_2$CO. Black squares indicate clumps
classified as early stage (70w and IRw) and blue circles indicate clumps classified
as late stage (IRb and H\,{\scriptsize II} regions). The black, blue, and red lines
are the results from linear fits for early stage, late stage,
and all sources, respectively.}
\label{figure:LH2CO-Lbol}
\end{figure*}

It seems that massive stars reach the main sequence above a threshold
of $L_{\rm bol}$/$M_{\rm clump}$\,$\sim$\,10\,L$_{\odot}$/M$_{\odot}$ \citep{Giannetti2017},
thus strongly increasing their energy output.
$L_{\rm bol}$/$M_{\rm clump}$\,$\gtrsim$\,10\,L$_{\odot}$/M$_{\odot}$ clumps are
associated with IRb and H\,{\scriptsize II} regions in our sample (also see \citealt{Giannetti2017}),
indicating late evolutionary stages (see Tab.\,\ref{table:source}).
For $L_{\rm bol}$/$M_{\rm clump}$\,$\gtrsim$\,10\,L$_{\odot}$/M$_{\odot}$,
the gas temperature and the
luminosity-to-mass ratio are related by power-laws of the form
\begin{eqnarray}
{\rm log}T_{\rm kin}({\rm 3_{21}{-}2_{20}/3_{03}{-}2_{02}})=(0.14\pm0.05)\times{\rm log}(L_{\rm bol}/M_{\rm clump}) \nonumber \\+(1.79\pm0.09)
\end{eqnarray}
and
\begin{eqnarray}
{\rm log}T_{\rm kin}({\rm 4_{22}{-}3_{21}/4_{04}{-}3_{03}})=(0.16\pm0.04)\times{\rm log}(L_{\rm bol}/M_{\rm clump}) \nonumber \\+(1.68\pm0.08),
\end{eqnarray}
with correlation coefficients, $R$, of 0.37 and 0.45, respectively.
The power-law indices are consistent with those derived from
CH$_3$CN, CH$_3$CCH, and CH$_3$OH in our and other massive clumps
($T$\,$\propto$\,($L_{\rm bol}/M_{\rm clump}$)$^{0.12-0.22}$,
\citealt{Molinari2016,Giannetti2017}).
A correlation between gas temperature and $L_{\rm bol}$/$M_{\rm clump}$
ratio indicates that the dense gas appears to be heated by the
newly formed massive stars during the late evolutionary stages of clumps.
$L_{\rm bol}$/$M_{\rm clump}$\,$<$\,10\,L$_{\odot}$/M$_{\odot}$ clumps are well
associated with 70w and IRw indicating earlier evolutionary
stages. For these sources the relation of gas temperature
with the $L_{\rm bol}$/$M_{\rm clump}$ ratio does not follow
the above trend. The temperature in
these sources may not yet be greatly affected by the gas which is heated by
internal power sources \citep{Molinari2016}. Instead it may be related to
gas excited by star formation activities e.g., outflows and shocks \citep{Tang2017a}.

\subsection{Comparisons of H$_2$CO luminosity, bolometric luminosity, and clump mass}
The line luminosities of dense molecular gas tracers (e.g., HCN, CS, and HCO$^+$)
are found to be approximately linearly correlated with far infrared luminosities
($L_{\rm FIR}$\,$\propto$\,$L_{\rm molecule}$) in both
Galactic dense clumps and galaxies
\citep{Gao2004a,Gao2004b,Wu2005,Wu2010,Schenck2011,Ma2013,Zhang2014,Liu2016,Stephens2016},
which indicates a link between star formation rate (SFR) represented
by infrared luminosities and dense molecular gas mass indicated by
molecular line luminosities. Line luminosities of dense molecular gas
(e.g., HCN, CS, HCO$^+$, N$_2$H$^+$, SO) appear to be linearly
related to the mass of dense gas most relevant to star formation \citep{Wu2010,Reiter2011,Liu2016}.

\begin{table*}[t]
\caption{Clump mass and bolometric luminosity vs. H$_2$CO line luminosity.}
\label{table:Lbol-LH2CO}
\centering
\begin{tabular}
{cccccccccccc}
\hline\hline 
Transition & Sample & \multicolumn{3}{c}{$M_{\rm clump}$-$L^{\prime}_{\rm H_2CO}$} && \multicolumn{3}{c}{$L_{\rm bol}$-$L^{\prime}_{\rm H_2CO}$} \\
\cline{3-5} \cline{7-9}
&& Slope & Intercept & $R$ & & Slope & Intercept & $R$ \\
\hline 
                              &70w+IRw &0.70 (0.06) &2.57 (0.06) &0.88 &&0.96 (0.11) &2.98 (0.10) &0.82\\
o-H$_2$CO 3$_{13}$--2$_{12}$  &IRb+H\,{\scriptsize II} &0.99 (0.05) &2.09 (0.06) &0.93 &&1.10 (0.10) &3.66 (0.12) &0.84\\
                              &all     &0.78 (0.05) &2.40 (0.05) &0.88 &&1.19 (0.09) &3.24 (0.10) &0.80\\
\hline
                              &70w+IRw &0.63 (0.07) &2.71 (0.06) &0.83 &&0.91 (0.10) &3.14 (0.09) &0.82\\
o-H$_2$CO 3$_{12}$--2$_{11}$  &IRb+H\,{\scriptsize II} &0.98 (0.05) &2.17 (0.06) &0.94 &&1.08 (0.10) &3.75 (0.11) &0.84\\
                              &all     &0.72 (0.05) &2.53 (0.05) &0.84 &&1.16 (0.09) &3.37 (0.09) &0.81\\
\hline
                              &70w+IRw &0.64 (0.07) &2.79 (0.06) &0.82 &&0.94 (0.10) &3.25 (0.08) &0.83\\
p-H$_2$CO 3$_{03}$--2$_{02}$  &IRb+H\,{\scriptsize II} &1.01 (0.06) &2.32 (0.05) &0.93 &&1.11 (0.11) &3.92 (0.10) &0.83\\
                              &all     &0.74 (0.05) &2.63 (0.04) &0.84 &&1.19 (0.09) &3.55 (0.08) &0.81\\
\hline
                              &70w+IRw &0.77 (0.07) &3.02 (0.05) &0.93 &&0.84 (0.13) &3.99 (0.08) &0.85\\
p-H$_2$CO 3$_{22}$--2$_{21}$  &IRb+H\,{\scriptsize II} &0.94 (0.05) &2.84 (0.03) &0.94 &&1.03 (0.10) &4.49 (0.07) &0.83\\
                              &all     &0.86 (0.04) &2.91 (0.03) &0.93 &&1.05 (0.09) &4.35 (0.06) &0.82\\
\hline
                              &70w+IRw &0.89 (0.08) &3.01 (0.04) &0.93 &&0.95 (0.20) &3.90 (0.11) &0.75\\
p-H$_2$CO 3$_{21}$--2$_{20}$  &IRb+H\,{\scriptsize II} &0.93 (0.05) &2.86 (0.03) &0.94 &&1.02 (0.10) &4.51 (0.06) &0.84\\
                              &all     &0.90 (0.04) &2.91 (0.03) &0.93 &&1.09 (0.11) &4.32 (0.07) &0.79\\
\hline
                              &70w+IRw &0.63 (0.07) &2.82 (0.05) &0.83 &&0.91 (0.10) &3.26 (0.08) &0.83\\
p-H$_2$CO 4$_{04}$--3$_{03}$  &IRb+H\,{\scriptsize II} &0.95 (0.05) &2.33 (0.04) &0.94 &&1.03 (0.08) &3.95 (0.08) &0.86\\
                              &all     &0.73 (0.05) &2.61 (0.04) &0.85 &&1.11 (0.08) &3.59 (0.07) &0.82\\
\hline
                              &70w+IRw &0.82 (0.07) &2.92 (0.04) &0.93 &&1.11 (0.17) &3.67 (0.10) &0.80\\
p-H$_2$CO 4$_{23}$--3$_{22}$  &IRb+H\,{\scriptsize II} &0.80 (0.05) &2.73 (0.04) &0.92 &&0.89 (0.07) &4.38 (0.06) &0.86\\
                              &all     &0.79 (0.04) &2.79 (0.03) &0.91 &&0.98 (0.09) &4.15 (0.06) &0.79\\
\hline
                              &70w+IRw &0.75 (0.09) &2.99 (0.05) &0.86 &&1.16 (0.17) &3.68 (0.09) &0.80\\
p-H$_2$CO 4$_{22}$--3$_{21}$  &IRb+H\,{\scriptsize II} &0.83 (0.05) &2.73 (0.04) &0.92 &&0.92 (0.08) &4.38 (0.06) &0.86\\
                              &all     &0.79 (0.05) &2.83 (0.03) &0.89 &&1.04 (0.09) &4.13 (0.06) &0.80\\
\hline 
\end{tabular}
\tablefoot{The format of the regression fits is
${\rm log}L_{\rm bol} ~({\rm or} ~{\rm log}M_{\rm clump}) = {\rm Slope} \times {\rm log}L^{\prime}_{\rm H_2CO}+{\rm Intercept}$.
$R$ is the correlation coefficient for the linear fit.}
\end{table*}

Observations of H$_2$CO $K$-doublet transitions
($\Delta$$J$\,=\,0, $\Delta$$K_{\rm a}$\,=\,0, $\Delta$$K_{\rm c}$\,=\,$\pm$1) in our
Galaxy and external galaxies show that H$_2$CO traces a denser,
more compact, component of molecular clouds than low-excitation
transitions of CO or HCN \citep{Mangum1993b,Mangum2008,Mangum2013b}.
Our selected H$_2$CO transitions may have similar characteristics.
The critical densities of H$_2$CO 3$_{13}$-2$_{12}$, 3$_{12}$-2$_{11}$,
3$_{03}$--2$_{02}$, and 4$_{04}$--3$_{03}$ transitions are
$\sim$4\,$\times$\,10$^5$, $\sim$6\,$\times$\,10$^5$, $\sim$6\,$\times$\,10$^5$,
and $\sim$1\,$\times$\,10$^6$\,cm$^{-3}$ (at kinetic temperature 50\,K; \citealt{Shirley2015}),
respectively.
Following \cite{Wu2005} and assuming Gaussian brightness
distributions for the sources and a Gaussian beam, H$_2$CO line
luminosities $L^{\prime}_{\rm H_2CO}$ can be derived with
\begin{equation}
L^{\prime}_{\rm H_2CO} = 23.5 \times 10^{-6} \times D^2 \times
\left(\frac{\pi \times \theta^2_{\rm s}}{4{\rm ln}2} \right)
\times \left(\frac{\theta^2_{\rm s}+\theta^2_{\rm beam}}{\theta^2_{\rm s}} \right) \times \int T_{\rm mb}dV.
\end{equation}
Here D is the distance in kpc from \cite{Konig2017}, and
$\theta_{\rm s}$ and $\theta_{\rm beam}$ are the sizes of the
line emission source and of the beam in arcsecond.
As described in Section \ref{sect:density},
we assume that the extent of the H$_2$CO emission is the same as
that of the 870 $\mu$m dust emission derived from \cite{Csengeri2014}.
The resulting $L^{\prime}_{\rm H_2CO}$ values are
listed in Table \ref{table:LH2CO}.

We present the correlations between clump masses and H$_2$CO
luminosities in Figure \ref{figure:LH2CO-Mclump}.
The power-law fitted results are
listed in Table\,\ref{table:Lbol-LH2CO}. These show that
the $M_{\rm clump}$-$L^{\prime}_{\rm H_2CO}$ relations are strongly
correlated, with correlation coefficients ranging
from 0.84 to 0.93 for different H$_2$CO transitions.
The power-law correlations of $M_{\rm clump}$-$L^{\prime}_{\rm H_2CO}$ are
found to be slightly sublinear, and are consistent with results of
e.g., HCN, CS, HCO$^+$, N$_2$H$^+$, and SO found in massive
dense clumps \citep{Wu2010,Reiter2011,Liu2016}. This indicates
that $L^{\prime}_{\rm H_2CO}$ of our observed eight transitions
provides good tracers for the mass of dense gas and confirms
that $L^{\prime}_{\rm molecule}$ of dense molecular tracers does
reliably probe the mass of dense molecular gas.

The $L_{\rm bol}$-$L^{\prime}_{\rm H_2CO}$ relations are
plotted in Figure \ref{figure:LH2CO-Lbol}. We fit power-law
relations of $L_{\rm bol}$-$L^{\prime}_{\rm H_2CO}$ for eight
different H$_2$CO transitions.
The fitted results are listed in
Table \ref{table:Lbol-LH2CO}. The bolometric luminosities and
the H$_2$CO luminosities are related with a slope range of 0.98--1.19 and
correlation coefficients ranging from 0.79 to 0.82 for
different H$_2$CO transitions. Considering the uncertainties, the
correlations are nearly linear. The correlations of
$L_{\rm bol}$-$L^{\prime}_{\rm H_2CO}$
for different H$_2$CO transitions are consistent with previous
observational results of e.g., HCN, CS, HCO$^+$, SiO, HC$_3$N, C$_2$H
in massive dense clumps \citep{Wu2005,Wu2010,Ma2013,Liu2016,Stephens2016}.
This indicates that the mass of dense molecular gas traced by the H$_2$CO
line luminosity is well correlated with star formation.

Observations of dense clumps show that their evolutionary
stage impacts the $L_{\rm IR}$-$L^{\prime}_{\rm molecule}$
relation \citep{Liu2016,Stephens2016}. We distinguish two evolutionary
classes of clumps in their early stage (70w and IRb) and late stage
(IRb and H\,{\scriptsize II} regions), respectively,
in Figure \ref{figure:LH2CO-Lbol}, and the power-law fitted
results are listed in Table \ref{table:Lbol-LH2CO}.
Considering uncertainties of the fitted slopes of
$L_{\rm bol}$-$L^{\prime}_{\rm H_2CO}$ correlations,
we find approximately similar linear correlations of
$L_{\rm bol}$-$L^{\prime}_{\rm H_2CO}$ for different
transitions in both evolutionary stages of the clumps.
This suggests that the $L_{\rm IR}$-$L^{\prime}_{\rm H_2CO}$
relations are only weakly influenced by the evolutionary stage of
the clumps in our sample. We also compared the $M_{\rm clump}$-$L^{\prime}_{\rm H_2CO}$
relations in the two evolutionary stages in Figure \ref{figure:LH2CO-Mclump},
and the power-law fitted results are listed in Table \ref{table:Lbol-LH2CO}.
Apparently, clumps in an early stage are closer to sublinear (slopes
of 0.63--0.89) and clumps in a late stage tend to exhibit more linear slopes
(0.80 to 1.01). For the early stage,
the $M_{\rm clump}$-$L^{\prime}_{\rm H_2CO}$ data show
a larger scatter (see Fig.\,\ref{figure:LH2CO-Mclump}).
This may be due to some clumps with lower luminosity ($<$10$^3$\,L$_\odot$)
which are likely in an early evolutionary stage
with large derived uncertainties of the mass of the clump.
The $M_{\rm clump}$-$L^{\prime}_{\rm H_2CO}$ is found to
be strongly correlated, with correlation coefficients
ranging from 0.92 to 0.94 in the late stages of clumps.
This indicates that $L^{\prime}_{\rm H_2CO}$ ($J$\,=\,3--2 and 4--3)
traces well the mass of warm dense molecular gas associated
with bright infrared emission and H\,{\scriptsize II} regions in massive star-forming clumps.

\section{Summary}
\label{sect:Summary}
We have measured the kinetic temperature and spatial density with
H$_2$CO\,($J$\,=\,4--3) and (3--2) rotational transitions and compare
the derived temperatures with values obtained from NH$_3$, with
dust emission, and with linewidth and bolometric luminosity for the ATLASGAL TOP100
massive star-forming clumps at various evolutionary stages using
the 12-m APEX telescope. The main results are the following:
\begin{enumerate}
\item
Using the RADEX non-LTE model, we derive the gas kinetic
temperature and spatial density, modeling the measured
para-H$_2$CO\,3$_{21}$--2$_{20}$/3$_{03}$--2$_{02}$,
4$_{22}$--3$_{21}$/4$_{04}$--3$_{03}$,
and 4$_{04}$--3$_{03}$/3$_{03}$--2$_{02}$ ratios.
The gas kinetic temperatures derived from the
para-H$_2$CO\,4$_{22}$--3$_{21}$/4$_{04}$--3$_{03}$ and
3$_{21}$--2$_{20}$/3$_{03}$--2$_{02}$ line
ratios are very warm, ranging from 43 to $>$300\,K
with an unweighted average of 91\,$\pm$\,4\,K.
Spatial densities of molecular gas derived
from the para-H$_2$CO\,4$_{04}$--3$_{03}$/3$_{03}$--2$_{02}$ line ratios
yield 0.6--8.3\,$\times$\,10$^6$\,cm$^{-3}$ with
an unweighted average of 1.5\,($\pm$0.1)\,$\times$\,10$^6$\,cm$^{-3}$.

\item
The fractional abundance $X$(para-H$_2$CO) does not vary considerably
during the various stages of massive star formation, ranging from
1.0\,$\times$\,10$^{-10}$ to 1.2\,$\times$\,10$^{-9}$ with an average of
3.9\,($\pm$0.2)\,$\times$\,10$^{-10}$, confirming that H$_2$CO does
reliably trace the H$_2$ column density.

\item
The spatial densities traced by H$_2$CO do not
vary significantly with the evolutionary stage of massive clumps.
This may indicate that the density structure does not evolve significantly
as the star formation proceeds.
\item
A comparison of kinetic temperatures derived from para-H$_2$CO,
NH$_3$\,(2,2)/(1,1), and the dust emission indicates that
para-H$_2$CO traces a distinctly higher temperature than the
NH$_3$\,(2,2)/(1,1) transitions and the dust.

\item
The H$_2$CO linewidths correlate with the bolometric luminosities
and increase with the evolutionary stage of the clumps, which suggests
that high luminosities tend to be associated with more turbulent molecular cloud
structures.

\item
The non-thermal velocity dispersion of H$_2$CO is positively
correlated with the gas kinetic temperature, which indicates that the dense gas
may be heated by dissipation of turbulent energy in those massive clumps.

\item
A weak positive correlation between gas temperature and bolometric luminosity
suggests that the gas might be heated by the activity of the
embedded young massive stars.

\item
The average gas kinetic temperature clearly
increases with the evolutionary stage of the massive clumps.
For $L_{\rm bol}$/$M_{\rm clump}$\,$\gtrsim$\,10\,L$_{\odot}$/M$_{\odot}$,
we find a rough correlation between gas kinetic
temperature and $L_{\rm bol}$/$M_{\rm clump}$ ratio,
which traces the evolutionary stage of the massive clumps \citep{Molinari2016,Giannetti2017}.

\item
The strong correlations between H$_2$CO line luminosities
and clump masses are approximately linear during the late evolutionary
stages of clumps, which indicates that $L_{\rm H_2CO}$\,($J$\,=\,3--2)
and (4--3) reliably trace the mass of warm dense molecular gas associated
with bright infrared emission and H\,{\scriptsize II} regions.
During the earlier stages of evolutionary, the correlation may be
slightly sublinear. The $M_{\rm clump}$-$L^{\prime}_{\rm H_2CO}$ correlation
appears to be influenced by the evolutionary stage of the clumps.

\item
H$_2$CO line luminosities are nearly linearly correlated with
bolometric luminosities over about four orders
of magnitude in $L_{\rm bol}$ of our massive clumps, suggesting that
the mass of dense molecular gas traced by the H$_2$CO line luminosity is
well correlated with star formation. The $L_{\rm bol}$-$L^{\prime}_{\rm H_2CO}$ relation
seems to be weakly affected by the evolutionary stage of the clumps.

\end{enumerate}

\begin{acknowledgements}
The authors are grateful for the valuable comments of the referee Jeff Mangum.
We thank the staff of the APEX telescope for their assistance in observations.
We also thank Nina Brinkmann for her help of data calibration.
This work acknowledges support by The National Natural Science
Foundation of China under grant 11433008, The Program of the Light
in China's Western Region (LCRW) under grant XBBS201424,
and The National Natural Science Foundation of China under grant 11373062.
This work was partially carried out within the Collaborative
Research Council 956, subproject A6, funded by the Deutsche
Forschungsgemeinschaft (DFG). C.H. acknowledges support by a Chinese Academy of Sciences President's
International Fellowship Initiative for visiting scientists (2017VMA0005).
This research has used NASA's Astrophysical Data System (ADS).

\end{acknowledgements}


\Online
\begin{appendix} 

\section{Source and H$_2$CO parameters}

\setcounter{table}{0}
\begin{table*}
\small
\caption{Source parameters.}
\label{table:source}
\centering
\begin{tabular}
{c c c c c c c cccc}
\hline\hline 
Sources &RA(J2000) &DEC(J2000) &Distance &Size &$S$$_{870\mu m}$ &$N$(H$_2$) &$M_{\rm clump}$ &$L$${_{\rm bol}}$ &$T_{\rm dust}$  &Note \\
& $^{h}$ {} $^{m}$ {} $^{s}$ & \degr {} \arcmin {} \arcsec &kpc &
arcsec &Jy beam$^{-1}$ &cm$^{-2}$ &M$_\odot$ &L$_\odot$  &K & \\
\hline 
AGAL008.684$-$00.367 &18:06:23.03 &$-$21:37:10.8 &4.8  &26 &4.56  &8.7$\times10^{22}$ &1.5$\times10^{3}$ &2.8$\times10^{4}$ &24.2 &IRw              \\
AGAL008.706$-$00.414 &18:06:36.65 &$-$21:37:16.3 &4.8  &36 &1.02  &6.1$\times10^{22}$ &1.7$\times10^{3}$ &5.0$\times10^{2}$ &11.8 &IRw              \\
AGAL010.444$-$00.017 &18:08:44.72 &$-$19:54:32.8 &8.6  &26 &2.10  &5.0$\times10^{22}$ &1.6$\times10^{3}$ &1.1$\times10^{4}$ &20.7 &IRw              \\
AGAL010.472$+$00.027 &18:08:37.99 &$-$19:51:47.7 &8.6  &22 &33.22 &4.7$\times10^{23}$ &1.0$\times10^{4}$ &4.7$\times10^{5}$ &30.5 &H{\scriptsize II}\\
AGAL010.624$-$00.384 &18:10:28.62 &$-$19:55:45.5 &5.0  &26 &31.06 &3.7$\times10^{23}$ &3.8$\times10^{3}$ &4.2$\times10^{5}$ &34.5 &H{\scriptsize II}\\
AGAL012.804$-$00.199 &18:14:13.55 &$-$17:55:32.0 &2.4  &33 &31.05 &3.6$\times10^{23}$ &1.9$\times10^{3}$ &2.5$\times10^{5}$ &35.1 &H{\scriptsize II}\\
AGAL013.178$+$00.059 &18:14:00.59 &$-$17:28:38.5 &2.4  &34 &3.69  &7.0$\times10^{22}$ &3.7$\times10^{2}$ &8.3$\times10^{3}$ &24.2 &70w              \\
AGAL013.658$-$00.599 &18:17:24.09 &$-$17:22:10.2 &4.5  &26 &3.92  &6.3$\times10^{22}$ &5.7$\times10^{2}$ &2.1$\times10^{4}$ &27.4 &IRb              \\
AGAL014.114$-$00.574 &18:18:13.03 &$-$16:57:18.6 &2.6  &32 &3.46  &7.4$\times10^{22}$ &3.5$\times10^{2}$ &3.2$\times10^{3}$ &22.4 &IRw              \\
AGAL014.194$-$00.194 &18:16:58.63 &$-$16:42:16.3 &3.9  &28 &3.29  &9.5$\times10^{22}$ &8.2$\times10^{2}$ &2.7$\times10^{3}$ &18.2 &IRw              \\
AGAL014.492$-$00.139 &18:17:22.01 &$-$16:25:01.1 &3.9  &35 &2.30  &1.3$\times10^{23}$ &1.9$\times10^{3}$ &7.5$\times10^{2}$ &12.4 &70w              \\
AGAL014.632$-$00.577 &18:19:14.65 &$-$16:30:02.7 &1.8  &33 &4.40  &9.3$\times10^{22}$ &2.5$\times10^{2}$ &2.8$\times10^{3}$ &22.5 &IRw              \\
AGAL015.029$-$00.669 &18:20:22.45 &$-$16:11:43.8 &2.0  &39 &16.36 &2.1$\times10^{23}$ &1.2$\times10^{3}$ &1.3$\times10^{5}$ &32.9 &IRb              \\
AGAL018.606$-$00.074 &18:25:08.22 &$-$12:45:23.8 &4.3  &29 &1.40  &6.3$\times10^{22}$ &8.8$\times10^{2}$ &5.9$\times10^{2}$ &13.8 &IRw              \\
AGAL018.734$-$00.226 &18:25:56.02 &$-$12:42:49.6 &12.5 &29 &3.64  &8.0$\times10^{22}$ &7.9$\times10^{3}$ &7.3$\times10^{4}$ &21.9 &IRw              \\
AGAL018.888$-$00.474 &18:27:07.41 &$-$12:41:39.8 &4.7  &32 &3.50  &1.5$\times10^{23}$ &2.8$\times10^{3}$ &3.2$\times10^{3}$ &14.4 &IRw              \\
AGAL019.882$-$00.534 &18:29:14.54 &$-$11:50:26.0 &3.7  &26 &6.95  &1.3$\times10^{23}$ &8.0$\times10^{2}$ &1.2$\times10^{4}$ &24.2 &IRb              \\
AGAL022.376$+$00.447 &18:30:24.06 &$-$09:10:39.6 &4.0  &25 &1.65  &8.1$\times10^{22}$ &6.2$\times10^{2}$ &3.2$\times10^{2}$ &13.1 &IRw              \\
AGAL023.206$-$00.377 &18:34:54.90 &$-$08:49:19.1 &4.6  &24 &7.35  &1.6$\times10^{23}$ &1.3$\times10^{3}$ &1.3$\times10^{4}$ &22.1 &IRw              \\
AGAL024.629$+$00.172 &18:35:35.54 &$-$07:18:09.5 &7.7  &28 &1.27  &3.7$\times10^{22}$ &1.5$\times10^{3}$ &5.0$\times10^{3}$ &18.1 &IRw              \\
AGAL028.564$-$00.236 &18:44:17.74 &$-$03:59:42.5 &5.5  &40 &1.97  &1.2$\times10^{23}$ &5.4$\times10^{3}$ &1.8$\times10^{3}$ &11.7 &IRw              \\
AGAL028.861$+$00.066 &18:43:46.04 &$-$03:35:29.9 &7.4  &27 &3.40  &4.1$\times10^{22}$ &1.1$\times10^{3}$ &1.6$\times10^{5}$ &34.5 &IRb              \\
AGAL030.848$-$00.081 &18:47:55.40 &$-$01:53:35.9 &4.9  &31 &1.57  &5.2$\times10^{22}$ &1.2$\times10^{3}$ &3.1$\times10^{3}$ &16.7 &70w              \\
AGAL030.893$+$00.139 &18:47:13.50 &$-$01:45:07.7 &4.9  &33 &1.57  &9.9$\times10^{22}$ &1.9$\times10^{3}$ &5.0$\times10^{2}$ &11.4 &70w              \\
AGAL031.412$+$00.307 &18:47:34.29 &$-$01:12:44.6 &4.9  &23 &21.55 &3.7$\times10^{23}$ &3.1$\times10^{3}$ &6.9$\times10^{4}$ &26.3 &H{\scriptsize II}\\
AGAL034.258$+$00.154 &18:53:18.53 &$+$01:14:57.9 &1.6  &26 &51.03 &7.0$\times10^{23}$ &8.1$\times10^{2}$ &4.8$\times10^{4}$ &31.0 &H{\scriptsize II}\\
AGAL034.401$+$00.226 &18:53:18.62 &$+$01:24:40.4 &1.6  &32 &6.27  &1.3$\times10^{23}$ &2.8$\times10^{2}$ &3.0$\times10^{3}$ &22.8 &H{\scriptsize II}\\
AGAL034.411$+$00.234 &18:53:18.14 &$+$01:25:23.8 &1.6  &25 &10.26 &1.8$\times10^{23}$ &2.1$\times10^{2}$ &4.8$\times10^{3}$ &26.1 &IRb              \\
AGAL034.821$+$00.351 &18:53:38.11 &$+$01:50:27.9 &1.6  &35 &2.51  &4.7$\times10^{22}$ &1.1$\times10^{2}$ &2.7$\times10^{3}$ &24.7 &IRb              \\
AGAL035.197$-$00.742 &18:58:12.93 &$+$01:40:40.6 &2.2  &29 &11.03 &1.6$\times10^{23}$ &4.6$\times10^{2}$ &2.4$\times10^{4}$ &29.5 &IRb              \\
AGAL037.554$+$00.201 &18:59:09.90 &$+$04:12:17.6 &6.7  &27 &3.52  &5.4$\times10^{22}$ &1.3$\times10^{3}$ &5.1$\times10^{4}$ &28.4 &IRb              \\
AGAL043.166$+$00.011 &19:10:13.44 &$+$09:06:15.8 &11.1 &27 &56.59 &...                &4.3$\times10^{4}$ &3.8$\times10^{6}$ &34.0 &H{\scriptsize II}\\
AGAL049.489$-$00.389 &19:23:43.69 &$+$14:30:31.9 &5.4  &24 &70.24 &1.1$\times10^{24}$ &1.2$\times10^{4}$ &5.6$\times10^{5}$ &29.1 &H{\scriptsize II}\\
AGAL053.141$+$00.069 &19:29:17.35 &$+$17:56:21.4 &1.6  &26 &4.39  &7.9$\times10^{22}$ &9.5$\times10^{1}$ &2.3$\times10^{3}$ &25.4 &IRb              \\
AGAL059.782$+$00.066 &19:43:10.90 &$+$23:44:04.4 &2.2  &31 &4.76  &7.4$\times10^{22}$ &2.5$\times10^{2}$ &9.8$\times10^{3}$ &28.2 &IRb              \\
AGAL305.192$-$00.006 &13:11:14.54 &$-$62:47:27.0 &3.8  &29 &3.15  &5.4$\times10^{22}$ &5.2$\times10^{2}$ &1.3$\times10^{4}$ &26.1 &IRw              \\
AGAL305.209$+$00.206 &13:11:13.34 &$-$62:34:38.6 &3.8  &27 &13.51 &1.9$\times10^{23}$ &1.4$\times10^{3}$ &8.8$\times10^{4}$ &30.1 &IRb              \\
AGAL305.562$+$00.014 &13:14:26.41 &$-$62:44:25.0 &3.8  &27 &4.23  &5.3$\times10^{22}$ &4.1$\times10^{2}$ &5.2$\times10^{4}$ &33.4 &IRb              \\
AGAL305.794$-$00.096 &13:16:34.36 &$-$62:49:43.4 &3.8  &33 &1.34  &4.7$\times10^{22}$ &5.9$\times10^{2}$ &9.8$\times10^{2}$ &16.0 &70w              \\
AGAL309.384$-$00.134 &13:47:22.66 &$-$62:18:07.5 &5.3  &29 &3.16  &6.1$\times10^{22}$ &1.2$\times10^{3}$ &1.6$\times10^{4}$ &24.0 &IRb              \\
AGAL310.014$+$00.387 &13:51:38.06 &$-$61:39:15.1 &3.6  &30 &3.65  &4.8$\times10^{22}$ &4.2$\times10^{2}$ &5.0$\times10^{4}$ &32.2 &IRb              \\
AGAL313.576$+$00.324 &14:20:08.33 &$-$60:42:04.9 &3.8  &23 &2.59  &3.9$\times10^{22}$ &1.8$\times10^{2}$ &9.4$\times10^{3}$ &29.2 &IRb              \\
AGAL316.641$-$00.087 &14:44:18.34 &$-$59:55:15.5 &1.2  &24 &2.30  &3.2$\times10^{22}$ &1.8$\times10^{1}$ &9.9$\times10^{2}$ &30.6 &IRb              \\
AGAL317.867$-$00.151 &14:53:16.66 &$-$59:26:34.7 &3.0  &24 &4.00  &1.1$\times10^{23}$ &3.6$\times10^{2}$ &1.6$\times10^{3}$ &19.3 &IRw              \\
AGAL318.779$-$00.137 &14:59:33.19 &$-$59:00:36.8 &2.8  &34 &2.20  &4.0$\times10^{22}$ &3.6$\times10^{2}$ &6.4$\times10^{3}$ &24.9 &IRw              \\
AGAL320.881$-$00.397 &15:14:33.14 &$-$58:11:32.6 &10.0 &27 &1.50  &4.9$\times10^{22}$ &2.9$\times10^{3}$ &6.1$\times10^{3}$ &16.8 &70w              \\
AGAL326.661$+$00.519 &15:45:02.80 &$-$54:09:11.5 &1.8  &29 &3.60  &5.6$\times10^{22}$ &1.3$\times10^{2}$ &7.4$\times10^{3}$ &28.4 &IRb              \\
AGAL326.987$-$00.032 &15:49:07.96 &$-$54:23:05.1 &4.0  &25 &2.40  &7.1$\times10^{22}$ &4.4$\times10^{2}$ &1.1$\times10^{3}$ &17.9 &IRw              \\
AGAL327.119$+$00.509 &15:47:33.16 &$-$53:52:39.7 &5.5  &27 &3.36  &4.5$\times10^{22}$ &6.7$\times10^{2}$ &5.9$\times10^{4}$ &31.8 &IRb              \\
AGAL327.293$-$00.579 &15:53:07.80 &$-$54:37:06.4 &3.1  &23 &49.21 &7.8$\times10^{23}$ &2.8$\times10^{3}$ &8.3$\times10^{4}$ &27.9 &IRb              \\
AGAL327.393$+$00.199 &15:50:19.15 &$-$53:57:04.9 &5.9  &28 &3.15  &6.4$\times10^{22}$ &1.2$\times10^{3}$ &1.3$\times10^{4}$ &23.2 &IRb              \\
AGAL328.809$+$00.632 &15:55:48.56 &$-$52:43:07.8 &3.0  &25 &23.19 &2.6$\times10^{23}$ &1.0$\times10^{3}$ &1.6$\times10^{5}$ &36.3 &H{\scriptsize II}\\
AGAL329.029$-$00.206 &16:00:31.18 &$-$53:12:39.1 &11.5 &35 &8.82  &1.8$\times10^{23}$ &1.1$\times10^{4}$ &2.2$\times10^{5}$ &23.1 &IRw              \\
AGAL329.066$-$00.307 &16:01:09.70 &$-$53:16:06.0 &11.6 &34 &3.18  &7.0$\times10^{22}$ &9.1$\times10^{3}$ &7.1$\times10^{4}$ &21.9 &IRb              \\
AGAL330.879$-$00.367 &16:10:20.31 &$-$52:06:11.2 &4.2  &25 &16.94 &2.1$\times10^{23}$ &1.6$\times10^{3}$ &1.5$\times10^{5}$ &33.4 &H{\scriptsize II}\\
AGAL330.954$-$00.182 &16:09:53.01 &$-$51:54:55.0 &9.3  &23 &45.01 &5.7$\times10^{23}$ &1.7$\times10^{4}$ &1.3$\times10^{6}$ &33.0 &H{\scriptsize II}\\
AGAL331.709$+$00.582 &16:10:05.84 &$-$50:50:29.0 &10.5 &28 &3.27  &7.6$\times10^{22}$ &5.1$\times10^{3}$ &3.7$\times10^{4}$ &21.0 &IRw              \\
AGAL332.094$-$00.421 &16:16:16.56 &$-$51:18:26.2 &3.6  &27 &6.35  &8.8$\times10^{22}$ &6.3$\times10^{2}$ &5.9$\times10^{4}$ &30.8 &IRb              \\
\hline
\end{tabular}
\end{table*}
\setcounter{table}{0}
\begin{table*}
\small
\caption{continued.}
\centering
\begin{tabular}
{c c c c c c c cccc}
\hline\hline 
Sources &RA(J2000) &DEC(J2000) &Distance &Size &$S$$_{870\mu m}$ &$N$(H$_2$) &Mass &$L$${_{\rm bol}}$ &$T_{\rm dust}$  &Note \\
& $^{h}$ {} $^{m}$ {} $^{s}$ & \degr {} \arcmin {} \arcsec &kpc &
arcsec &Jy beam$^{-1}$ &cm$^{-2}$ &M$_\odot$ &L$_\odot$  &K & \\
\hline 
AGAL332.826$-$00.549 &16:20:10.65 &$-$50:53:17.5 &3.6  &25 &30.44 &3.5$\times10^{23}$ &1.9$\times10^{3}$ &2.4$\times10^{5}$ &35.7 &H{\scriptsize II}\\
AGAL333.134$-$00.431 &16:21:01.89 &$-$50:35:12.8 &3.6  &30 &21.68 &2.5$\times10^{23}$ &2.9$\times10^{3}$ &4.2$\times10^{5}$ &35.2 &H{\scriptsize II}\\
AGAL333.284$-$00.387 &16:21:30.34 &$-$50:26:54.5 &3.6  &34 &12.31 &1.7$\times10^{23}$ &2.1$\times10^{3}$ &1.3$\times10^{5}$ &30.4 &H{\scriptsize II}\\
AGAL333.314$+$00.106 &16:19:28.52 &$-$50:04:43.1 &3.6  &28 &3.28  &5.7$\times10^{22}$ &4.3$\times10^{2}$ &1.1$\times10^{4}$ &25.9 &IRb              \\
AGAL333.604$-$00.212 &16:22:09.31 &$-$50:06:02.4 &3.6  &33 &32.08 &3.1$\times10^{23}$ &3.5$\times10^{3}$ &1.2$\times10^{6}$ &41.1 &H{\scriptsize II}\\
AGAL333.656$+$00.059 &16:21:11.56 &$-$49:52:16.7 &5.3  &33 &1.56  &4.7$\times10^{22}$ &1.4$\times10^{3}$ &4.3$\times10^{3}$ &17.8 &70w              \\
AGAL335.789$+$00.174 &16:29:47.27 &$-$48:15:51.7 &3.7  &28 &8.25  &1.5$\times10^{23}$ &1.1$\times10^{3}$ &2.1$\times10^{4}$ &24.7 &IRw              \\
AGAL336.958$-$00.224 &16:36:17.03 &$-$47:40:49.6 &10.9 &23 &1.56  &5.6$\times10^{22}$ &2.4$\times10^{3}$ &3.6$\times10^{3}$ &15.8 &IRw              \\
AGAL337.176$-$00.032 &16:36:18.42 &$-$47:23:24.9 &11.0 &32 &2.36  &5.1$\times10^{22}$ &5.6$\times10^{3}$ &5.9$\times10^{4}$ &22.3 &IRw              \\
AGAL337.258$-$00.101 &16:36:56.41 &$-$47:22:27.2 &11.0 &27 &2.34  &5.2$\times10^{22}$ &3.2$\times10^{3}$ &3.0$\times10^{4}$ &21.7 &IRw              \\
AGAL337.286$+$00.007 &16:36:34.33 &$-$47:16:48.5 &9.4  &34 &1.18  &8.4$\times10^{22}$ &6.6$\times10^{3}$ &1.3$\times10^{3}$ &10.7 &70w              \\
AGAL337.406$-$00.402 &16:38:50.72 &$-$47:27:59.3 &3.3  &26 &17.21 &2.3$\times10^{23}$ &1.1$\times10^{3}$ &8.5$\times10^{4}$ &31.8 &H{\scriptsize II}\\
AGAL337.704$-$00.054 &16:38:29.41 &$-$47:00:38.6 &12.3 &25 &12.71 &2.3$\times10^{23}$ &1.4$\times10^{4}$ &3.2$\times10^{5}$ &25.6 &H{\scriptsize II}\\
AGAL337.916$-$00.477 &16:41:10.42 &$-$47:08:04.4 &3.2  &24 &22.86 &2.8$\times10^{23}$ &1.2$\times10^{3}$ &1.3$\times10^{5}$ &34.4 &IRb              \\
AGAL338.066$+$00.044 &16:39:28.54 &$-$46:40:30.9 &4.7  &35 &1.14  &3.2$\times10^{22}$ &9.6$\times10^{2}$ &3.1$\times10^{3}$ &18.5 &70w              \\
AGAL338.786$+$00.476 &16:40:21.98 &$-$45:51:05.8 &4.5  &33 &1.27  &7.1$\times10^{22}$ &1.2$\times10^{3}$ &4.9$\times10^{2}$ &12.2 &70w              \\
AGAL338.926$+$00.554 &16:40:34.17 &$-$45:41:47.0 &4.4  &35 &15.76 &3.0$\times10^{23}$ &6.0$\times10^{3}$ &9.4$\times10^{4}$ &24.2 &IRb              \\
AGAL339.623$-$00.122 &16:46:06.13 &$-$45:36:47.6 &3.0  &31 &2.95  &4.5$\times10^{22}$ &3.2$\times10^{2}$ &1.5$\times10^{4}$ &28.7 &IRb              \\
AGAL340.374$-$00.391 &16:50:02.57 &$-$45:12:45.6 &3.6  &28 &1.78  &8.5$\times10^{22}$ &7.9$\times10^{2}$ &5.1$\times10^{2}$ &13.4 &IRw              \\
AGAL340.746$-$01.001 &16:54:03.74 &$-$45:18:46.9 &2.8  &29 &2.46  &4.0$\times10^{22}$ &2.1$\times10^{2}$ &7.7$\times10^{3}$ &27.1 &IRb              \\
AGAL340.784$-$00.097 &16:50:15.10 &$-$44:42:30.5 &10.0 &24 &3.82  &6.6$\times10^{22}$ &2.8$\times10^{3}$ &7.2$\times10^{4}$ &26.2 &IRw              \\
AGAL341.217$-$00.212 &16:52:17.92 &$-$44:26:53.5 &3.7  &26 &4.44  &7.3$\times10^{22}$ &4.9$\times10^{2}$ &1.6$\times10^{4}$ &27.0 &IRb              \\
AGAL342.484$+$00.182 &16:55:02.06 &$-$43:12:59.7 &12.6 &25 &3.32  &6.6$\times10^{22}$ &4.9$\times10^{3}$ &6.4$\times10^{4}$ &23.6 &IRw              \\
AGAL343.128$-$00.062 &16:58:17.29 &$-$42:52:08.2 &3.0  &27 &17.67 &2.4$\times10^{23}$ &1.2$\times10^{3}$ &7.2$\times10^{4}$ &30.9 &H{\scriptsize II}\\
AGAL343.756$-$00.164 &17:00:49.94 &$-$42:26:12.8 &2.9  &24 &10.39 &2.0$\times10^{23}$ &6.2$\times10^{2}$ &9.9$\times10^{3}$ &24.3 &IRw              \\
AGAL344.227$-$00.569 &17:04:07.46 &$-$42:18:41.7 &2.5  &25 &17.26 &3.8$\times10^{23}$ &1.1$\times10^{3}$ &9.8$\times10^{3}$ &22.0 &IRw              \\
AGAL345.003$-$00.224 &17:05:11.17 &$-$41:29:05.1 &3.0  &26 &16.86 &2.3$\times10^{23}$ &9.7$\times10^{2}$ &6.5$\times10^{4}$ &31.0 &H{\scriptsize II}\\
AGAL345.488$+$00.314 &17:04:28.06 &$-$40:46:24.4 &2.2  &34 &14.97 &2.1$\times10^{23}$ &9.3$\times10^{2}$ &6.1$\times10^{4}$ &30.7 &H{\scriptsize II}\\
AGAL345.504$+$00.347 &17:04:23.00 &$-$40:44:21.6 &2.3  &30 &9.89  &1.3$\times10^{23}$ &4.2$\times10^{2}$ &4.3$\times10^{4}$ &32.7 &IRb              \\
AGAL345.718$+$00.817 &17:03:06.23 &$-$40:17:05.0 &1.6  &37 &3.10  &6.8$\times10^{22}$ &2.0$\times10^{2}$ &1.9$\times10^{3}$ &22.1 &IRb              \\
AGAL351.131$+$00.771 &17:19:34.56 &$-$35:56:46.1 &1.8  &32 &1.10  &3.1$\times10^{22}$ &1.2$\times10^{2}$ &6.3$\times10^{2}$ &18.6 &70w              \\
AGAL351.161$+$00.697 &17:19:56.68 &$-$35:57:52.9 &1.8  &30 &21.23 &4.7$\times10^{23}$ &1.2$\times10^{3}$ &8.8$\times10^{3}$ &21.9 &IRb              \\
AGAL351.244$+$00.669 &17:20:18.86 &$-$35:54:42.4 &1.8  &39 &16.92 &2.2$\times10^{23}$ &8.9$\times10^{2}$ &7.8$\times10^{4}$ &32.5 &IRb              \\
AGAL351.571$+$00.762 &17:20:51.03 &$-$35:35:23.2 &1.3  &42 &1.48  &4.7$\times10^{22}$ &1.6$\times10^{2}$ &4.3$\times10^{2}$ &17.0 &70w              \\
AGAL351.581$-$00.352 &17:25:25.03 &$-$36:12:45.4 &6.8  &26 &24.96 &4.1$\times10^{23}$ &8.7$\times10^{3}$ &2.5$\times10^{5}$ &27.1 &IRb              \\
AGAL351.774$-$00.537 &17:26:42.55 &$-$36:09:20.0 &1.0  &25 &48.81 &6.5$\times10^{23}$ &2.6$\times10^{2}$ &1.6$\times10^{4}$ &31.8 &IRb              \\
AGAL353.066$+$00.452 &17:26:13.57 &$-$34:31:55.7 &0.9  &28 &1.37  &4.1$\times10^{22}$ &1.8$\times10^{1}$ &5.7$\times10^{1}$ &17.8 &IRw              \\
AGAL353.409$-$00.361 &17:30:26.24 &$-$34:41:48.5 &3.4  &35 &20.03 &3.1$\times10^{23}$ &3.5$\times10^{3}$ &1.3$\times10^{5}$ &28.3 &IRb              \\
AGAL353.417$-$00.079 &17:29:19.10 &$-$34:32:13.1 &6.1  &37 &0.75  &2.4$\times10^{22}$ &1.8$\times10^{3}$ &4.5$\times10^{3}$ &17.1 &70w              \\
AGAL354.944$-$00.537 &17:35:12.03 &$-$33:30:28.9 &1.9  &35 &1.42  &3.8$\times10^{22}$ &1.5$\times10^{2}$ &4.8$\times10^{2}$ &19.1 &70w              \\
\hline 
\end{tabular}
\tablefoot{Parameters related to distance, H$_2$ column density, clump mass, bolometric luminosity, and dust temperature are taken from \cite{Konig2017}.
Source size representing full width to half power values of the 870\,$\mu$m continuum and $S$$_{870\mu m}$ are selected from \cite{Csengeri2014}.
Last column notes are taken from \cite{Konig2017}: 70w = 70 $\mu$m weak, IRw = mid-infrared weak, IRb = mid-infrared bright, and H{\scriptsize II} = H\,{\scriptsize II} region.}
\end{table*}

\setcounter{table}{1}
\begin{table*}
\small
\caption{Ortho-H$_2$CO 3$_{13}$-2$_{12}$ and 3$_{12}$-2$_{11}$ spectral parameters.}
\label{table:H2CO313-312}
\centering
\begin{tabular}
{c c c c c c c cccc}
\hline\hline 
&\multicolumn{4}{c}{ortho-H$_2$CO\,3$_{13}$--2$_{12}$}&&\multicolumn{4}{c}{ortho-H$_2$CO\,3$_{12}$--2$_{11}$}\\
\cline{2-5} \cline{7-10}
Sources &$\int$$T$$_{\rm mb}$d$v$ &$V_{\rm lsr}$ &FWHM &$T$$_{\rm mb}$ &&$\int$$T$$_{\rm mb}$d$v$ &$V_{\rm lsr}$ &FWHM &$T$$_{\rm mb}$\\
&K km\,s$^{-1}$ &km\,s$^{-1}$ &km\,s$^{-1}$ &K &&K km\,s$^{-1}$ & km\,s$^{-1}$ & km\,s$^{-1}$ &K\\
\hline 
AGAL008.684$-$00.367  &13.03 (0.34) &37.96   (0.10) &7.92  (0.24) &1.55 &&9.21  (0.12)  &38.17   (0.05) &6.96  (0.10)  &1.24 \\
AGAL008.706$-$00.414  &2.19  (0.23) &38.21   (0.27) &5.53  (0.73) &0.37 &&1.34  (0.10)  &38.68   (0.12) &3.43  (0.29)  &0.37 \\
AGAL010.444$-$00.017  &1.83  (0.28) &74.67   (0.44) &6.22  (1.22) &0.28 &&1.56  (0.13)  &74.73   (0.19) &4.67  (0.46)  &0.31 \\
AGAL010.472$+$00.027  &29.39 (0.51) &66.13   (0.08) &9.07  (0.18) &3.04 &&21.68 (0.24)  &65.80   (0.05) &8.98  (0.11)  &2.28 \\
AGAL010.624$-$00.384  &50.59 (0.39) &-3.49   (0.03) &7.74  (0.08) &6.14 &&47.96 (0.19)  &-3.65   (0.01) &7.98  (0.04)  &5.65 \\
\hline
\end{tabular}
\end{table*}
\setcounter{table}{1}
\begin{table*}
\small
\caption{continued.}
\centering
\begin{tabular}
{c c c c c c c cccc}
\hline\hline 
&\multicolumn{4}{c}{ortho-H$_2$CO\,3$_{13}$--2$_{12}$}&&\multicolumn{4}{c}{ortho-H$_2$CO\,3$_{12}$--2$_{11}$}\\
\cline{2-5} \cline{7-10}
Sources &$\int$$T$$_{\rm mb}$d$v$ &$V_{\rm lsr}$ &FWHM &$T$$_{\rm mb}$ &&$\int$$T$$_{\rm mb}$d$v$ &$V_{\rm lsr}$ &FWHM &$T$$_{\rm mb}$\\
&K km\,s$^{-1}$ &km\,s$^{-1}$ &km\,s$^{-1}$ &K &&K km\,s$^{-1}$ & km\,s$^{-1}$ & km\,s$^{-1}$ &K\\
\hline 
AGAL332.826$-$00.549  &28.19 (0.30) &-56.18  (0.04) &7.01  (0.09) &3.78 &&24.74 (0.17)  &-56.35  (0.02) &6.85  (0.06)  &3.39 \\
AGAL333.134$-$00.431  &46.54 (0.36) &-52.73  (0.03) &8.11  (0.08) &5.39 &&40.91 (0.18)  &-52.87  (0.02) &7.99  (0.04)  &4.81 \\
AGAL333.284$-$00.387  &14.65 (0.30) &-52.06  (0.06) &5.77  (0.13) &2.39 &&11.42 (0.13)  &-52.11  (0.03) &5.60  (0.07)  &1.91 \\
AGAL333.314$+$00.106  &10.09 (0.29) &-46.08  (0.08) &5.78  (0.21) &1.64 &&8.12  (0.22)  &-46.15  (0.07) &5.51  (0.19)  &1.38 \\
AGAL333.604$-$00.212  &32.84 (0.36) &-46.94  (0.04) &7.36  (0.11) &4.19 &&26.38 (0.28)  &-46.94  (0.04) &7.23  (0.10)  &3.42 \\
\end{tabular}
\tablefoot{"N" in the table indicates that the source has not been observed.}
\end{table*}

\setcounter{table}{2}
\begin{table*}
\small
\caption{Para-H$_2$CO 3$_{03}$--2$_{02}$ and 3$_{22}$--2$_{21}$ spectral parameters.}
\label{table:H2CO303-322}
\centering
\begin{tabular}
{c c c c c c c cccc}
\hline\hline 
&\multicolumn{4}{c}{para-H$_2$CO\,3$_{03}$--2$_{02}$}&&\multicolumn{4}{c}{para-H$_2$CO\,3$_{22}$--2$_{21}$}\\
\cline{2-5} \cline{7-10}
Sources &$\int$$T$$_{\rm mb}$d$v$ &$V_{\rm lsr}$ &FWHM &$T$$_{\rm mb}$ &&$\int$$T$$_{\rm mb}$d$v$ &$V_{\rm lsr}$ &FWHM &$T$$_{\rm mb}$\\
&K km\,s$^{-1}$ &km\,s$^{-1}$ &km\,s$^{-1}$ &K &&K km\,s$^{-1}$ & km\,s$^{-1}$ & km\,s$^{-1}$ &K\\
\hline 
AGAL008.684$-$00.367  &5.50  (0.17)  &37.95   (0.08)  &5.50 (0.20)  &0.94  &&0.92  (0.08)  &38.07  (0.13) &2.75 (0.27)  &0.31 \\
AGAL008.706$-$00.414  &0.67  (0.13)  &38.47   (0.24)  &2.70 (0.69)  &0.23  &&...           &...           &...          &...  \\
AGAL010.444$-$00.017  &...           &...             &...          &      &&...           &...           &...          &...  \\
AGAL010.472$+$00.027  &12.01 (0.28)  &65.77   (0.09)  &8.16 (0.23)  &1.38  &&5.51  (0.24)  &65.50  (0.17) &8.08 (0.41)  &0.64 \\
AGAL010.624$-$00.384  &25.67 (0.24)  &-3.40   (0.04)  &7.83 (0.09)  &3.09  &&7.49  (0.17)  &-3.22  (0.07) &6.71 (0.18)  &1.05 \\
\hline
\end{tabular}
\end{table*}
\setcounter{table}{2}
\begin{table*}
\small
\caption{continued.}
\centering
\begin{tabular}
{c c c c c c c cccc}
\hline\hline 
&\multicolumn{4}{c}{para-H$_2$CO\,3$_{03}$--2$_{02}$}&&\multicolumn{4}{c}{para-H$_2$CO\,3$_{22}$--2$_{21}$}\\
\cline{2-5} \cline{7-10}
Sources &$\int$$T$$_{\rm mb}$d$v$ &$V_{\rm lsr}$ &FWHM &$T$$_{\rm mb}$ &&$\int$$T$$_{\rm mb}$d$v$ &$V_{\rm lsr}$ &FWHM &$T$$_{\rm mb}$\\
&K km\,s$^{-1}$ &km\,s$^{-1}$ &km\,s$^{-1}$ &K &&K km\,s$^{-1}$ & km\,s$^{-1}$ & km\,s$^{-1}$ &K\\
\hline 
AGAL332.826$-$00.549  &17.25 (0.16)  &-56.68  (0.03)  &6.42 (0.07)  &2.52  &&6.23  (0.14)  &-57.17 (0.07) &6.13 (0.16)  &0.96 \\
AGAL333.134$-$00.431  &26.32 (0.19)  &-52.99  (0.02)  &7.26 (0.06)  &3.41  &&8.32  (0.15)  &-53.26 (0.05) &6.04 (0.13)  &1.21 \\
AGAL333.284$-$00.387  &7.60  (0.19)  &-52.14  (0.06)  &5.08 (0.15)  &1.40  &&2.01  (0.18)  &-52.12 (0.28) &4.67 (1.09)  &0.40 \\
AGAL333.314$+$00.106  &5.67  (0.14)  &-46.28  (0.06)  &5.39 (0.17)  &0.99  &&1.16  (0.13)  &-45.83 (0.22) &4.56 (0.48)  &0.23 \\
AGAL333.604$-$00.212  &15.04 (0.18)  &-47.09  (0.04)  &6.63 (0.10)  &2.13  &&4.80  (0.14)  &-47.26 (0.07) &6.04 (0.17)  &0.68 \\
\end{tabular}
\tablefoot{"N" in the table indicates that the source has not been observed.}
\end{table*}

\setcounter{table}{3}
\begin{table*}
\small
\caption{Para-H$_2$CO 3$_{21}$--2$_{20}$ and 4$_{04}$--3$_{03}$ spectral parameters.}
\label{table:H2CO321-404}
\centering
\begin{tabular}
{c c c c c c c cccc}
\hline\hline 
&\multicolumn{4}{c}{para-H$_2$CO\,3$_{21}$--2$_{20}$}&&\multicolumn{4}{c}{para-H$_2$CO\,4$_{04}$--3$_{03}$}\\
\cline{2-5} \cline{7-10}
Sources &$\int$$T$$_{\rm mb}$d$v$ &$V_{\rm lsr}$ &FWHM &$T$$_{\rm mb}$ &&$\int$$T$$_{\rm mb}$d$v$ &$V_{\rm lsr}$ &FWHM &$T$$_{\rm mb}$\\
&K km\,s$^{-1}$ &km\,s$^{-1}$ &km\,s$^{-1}$ &K &&K km\,s$^{-1}$ & km\,s$^{-1}$ & km\,s$^{-1}$ &K\\
\hline 
AGAL008.684$-$00.367  &1.09  (0.14)  &38.76  (0.30)  &4.67 (0.63)  &0.22  &&7.71  (0.13)  &38.16   (0.04) &4.80  (0.10)  &1.51 \\
AGAL008.706$-$00.414  &0.84  (0.17)  &36.84  (0.61)  &6.00 (1.42)  &0.13  &&1.02  (0.10)  &38.71   (0.16) &3.41  (0.43)  &0.28 \\
AGAL010.444$-$00.017  &...           &...            &...          &...   &&1.80  (0.11)  &75.66   (0.15) &4.87  (0.38)  &0.35 \\
AGAL010.472$+$00.027  &5.64  (0.24)  &65.80  (0.17)  &8.08 (0.43)  &0.66  &&25.97 (0.55)  &65.91   (0.09) &9.10  (0.23)  &2.68 \\
AGAL010.624$-$00.384  &8.80  (0.20)  &-2.93  (0.08)  &7.35 (0.20)  &1.13  &&50.55 (0.93)  &-3.44   (0.07) &7.61  (0.17)  &6.23 \\
\hline
\end{tabular}
\end{table*}
\setcounter{table}{3}
\begin{table*}
\small
\caption{continued.}
\centering
\begin{tabular}
{c c c c c c c cccc}
\hline\hline 
&\multicolumn{4}{c}{para-H$_2$CO\,3$_{21}$--2$_{20}$}&&\multicolumn{4}{c}{para-H$_2$CO\,4$_{04}$--3$_{03}$}\\
\cline{2-5} \cline{7-10}
Sources &$\int$$T$$_{\rm mb}$d$v$ &$V_{\rm lsr}$ &FWHM &$T$$_{\rm mb}$ &&$\int$$T$$_{\rm mb}$d$v$ &$V_{\rm lsr}$ &FWHM &$T$$_{\rm mb}$\\
&K km\,s$^{-1}$ &km\,s$^{-1}$ &km\,s$^{-1}$ &K &&K km\,s$^{-1}$ & km\,s$^{-1}$ & km\,s$^{-1}$ &K\\
\hline 
AGAL332.826$-$00.549  &6.40  (0.10)  &-57.39 (0.05)  &5.71 (0.11)  &1.05  &&26.42 (0.15)  &-57.16  (0.02) &6.13  (0.04)  &4.04 \\
AGAL333.134$-$00.431  &7.80  (0.13)  &-53.17 (0.05)  &6.20 (0.14)  &1.26  &&28.13 (0.19)  &-53.29  (0.02) &6.80  (0.06)  &3.88 \\
AGAL333.284$-$00.387  &2.01  (0.30)  &-52.18 (0.17)  &3.89 (0.39)  &0.49  &&8.22  (0.09)  &-52.11  (0.02) &4.42  (0.06)  &1.75 \\
AGAL333.314$+$00.106  &1.11  (0.11)  &-46.60 (0.24)  &4.97 (0.82)  &0.22  &&5.62  (0.16)  &-46.38  (0.08) &5.68  (0.21)  &0.93 \\
AGAL333.604$-$00.212  &4.34  (0.10)  &-47.01 (0.09)  &6.53 (0.25)  &0.69  &&15.49 (0.18)  &-47.22  (0.04) &6.77  (0.10)  &2.14 \\
\end{tabular}
\tablefoot{"N" in the table indicates that the source has not been observed.}
\end{table*}

\setcounter{table}{4}
\begin{table*}
\small
\caption{Para-H$_2$CO 4$_{23}$--3$_{22}$ and 4$_{22}$--3$_{21}$ spectral parameters.}
\label{table:H2CO423-422}
\centering
\begin{tabular}
{c c c c c c c cccc}
\hline\hline 
&\multicolumn{4}{c}{para-H$_2$CO\,4$_{23}$--3$_{22}$}&&\multicolumn{4}{c}{para-H$_2$CO\,4$_{22}$--3$_{21}$}\\
\cline{2-5} \cline{7-10}
Sources &$\int$$T$$_{\rm mb}$d$v$ &$V_{\rm lsr}$ &FWHM &$T$$_{\rm mb}$ &&$\int$$T$$_{\rm mb}$d$v$ &$V_{\rm lsr}$ &FWHM &$T$$_{\rm mb}$\\
&K km\,s$^{-1}$ &km\,s$^{-1}$ &km\,s$^{-1}$ &K &&K km\,s$^{-1}$ & km\,s$^{-1}$ & km\,s$^{-1}$ &K\\
\hline 
AGAL008.684$-$00.367  &2.21  (0.11)  &38.13   (0.11)  &4.54  (0.28)  &0.46  &&2.15  (0.10)  &38.08   (0.09) &4.26  (0.25)  &0.48 \\
AGAL008.706$-$00.414  &...           &...             &...           &...   &&...           &...            &...           &...  \\
AGAL010.444$-$00.017  &...           &...             &...           &...   &&...           &...            &...           &...  \\
AGAL010.472$+$00.027  &21.41 (0.45)  &65.55   (0.11)  &11.02 (0.27)  &1.83  &&18.81 (0.70)  &66.42   (0.18) &9.80  (0.42)  &1.80 \\
AGAL010.624$-$00.384  &23.78 (0.32)  &-3.06   (0.05)  &7.36  (0.12)  &3.04  &&23.55 (0.37)  &-3.04   (0.06) &7.36  (0.14)  &3.00 \\
\hline
\end{tabular}
\end{table*}
\setcounter{table}{4}
\begin{table*}
\small
\caption{continued.}
\centering
\begin{tabular}
{c c c c c c c cccc}
\hline\hline 
&\multicolumn{4}{c}{para-H$_2$CO\,4$_{23}$--3$_{22}$}&&\multicolumn{4}{c}{para-H$_2$CO\,4$_{22}$--3$_{21}$}\\
\cline{2-5} \cline{7-10}
Sources &$\int$$T$$_{\rm mb}$d$v$ &$V_{\rm lsr}$ &FWHM &$T$$_{\rm mb}$ &&$\int$$T$$_{\rm mb}$d$v$ &$V_{\rm lsr}$ &FWHM &$T$$_{\rm mb}$\\
&K km\,s$^{-1}$ &km\,s$^{-1}$ &km\,s$^{-1}$ &K &&K km\,s$^{-1}$ & km\,s$^{-1}$ & km\,s$^{-1}$ &K\\
\hline 
AGAL332.826$-$00.549  &14.31 (0.14)  &-57.68  (0.03)  &6.30  (0.07)  &2.13  &&12.98 (0.15)  &-57.50  (0.03) &5.65  (0.08)  &2.16 \\
AGAL333.134$-$00.431  &12.37 (0.12)  &-53.49  (0.03)  &6.72  (0.08)  &1.72  &&11.33 (0.11)  &-53.35  (0.03) &6.39  (0.08)  &1.67 \\
AGAL333.284$-$00.387  &2.41  (0.09)  &-52.23  (0.07)  &3.74  (0.17)  &0.60  &&2.48  (0.08)  &-52.15  (0.06) &3.75  (0.14)  &0.62 \\
AGAL333.314$+$00.106  &1.41  (0.16)  &-46.59  (0.24)  &4.25  (0.67)  &0.31  &&1.53  (0.14)  &-46.68  (0.23) &5.00  (0.55)  &0.29 \\
AGAL333.604$-$00.212  &7.07  (0.18)  &-47.14  (0.08)  &6.96  (0.23)  &0.95  &&6.47  (0.16)  &-47.13  (0.07) &6.16  (0.19)  &0.99 \\
\end{tabular}
\end{table*}

\setcounter{table}{5}
\begin{table*}
\scriptsize
\caption{Beam filling factors, integrated intensity ratios, para-H$_2$CO column densities, spatial densities, and kinetic temperatures.}
\centering
\begin{tabular}
{cccccccccccccccccccc}
\hline\hline 
&\multicolumn{2}{c}{$\eta_{\rm bf}$}&&$I'$(4$_{04}$--3$_{03}$)&&$N$(p-H$_2$CO) &$N$(o-H$_2$CO)&$n$(H$_2$)&& \multicolumn{5}{c}{Kinetic temperature}\\
\cline{2-3} \cline{5-5} \cline{7-9} \cline{11-15}
Sources &3$_{03}$--2$_{02}$ &4$_{04}$--3$_{03}$ &&$I'$(3$_{03}$--2$_{02}$)&& \multicolumn{3}{c}{4$_{04}$--3$_{03}$/3$_{03}$--2$_{02}$} &&3$_{21}$--2$_{20}$/3$_{03}$--2$_{02}$ & $T_{\rm LTE}$ && 4$_{22}$--3$_{21}$/4$_{04}$--3$_{03}$ & $T_{\rm LTE}$\\
\cline{7-9} \cline{11-12} \cline{14-15}
&&&&&&cm$^{-2}$&cm$^{-2}$&cm$^{-3}$&&K&K&&K&K\\
\hline 
AGAL008.684$-$00.367 &0.452 &0.594 &&1.07$\pm$0.04 &&2.8$\times$10$^{13}$  &7.2$\times$10$^{13}$ &1.9$^{+0.3}_{-0.2}$$\times$10$^6$ &&52.7 $^{+9.7 }_{-8.0 }$ &45.7 $^{+6.1 }_{-5.4 }$ &&55.4 $^{+4.2 }_{-3.5 }$ &47.8 $^{+2.4 }_{-2.3 }$\\
AGAL008.706$-$00.414 &0.613 &0.737 &&1.26$\pm$0.27 &&...                   &...                  &...                               &&...                     &...                     &&...                     &...                    \\
AGAL010.444$-$00.017 &0.452 &0.594 &&...           &&...                   &...                  &...                               &&...                     &...                     &&...                     &...                    \\
AGAL010.472$+$00.027 &0.372 &0.511 &&1.57$\pm$0.05 &&2.1$\times$10$^{14}$  &3.9$\times$10$^{14}$ &2.1$^{+0.4}_{-0.2}$$\times$10$^6$ &&>300                    &278.8$^{+110 }_{-64.0}$ &&>300                    &>300                   \\
AGAL010.624$-$00.384 &0.452 &0.594 &&1.50$\pm$0.03 &&3.0$\times$10$^{14}$  &6.1$\times$10$^{14}$ &8.1$^{    }_{-0.3}$$\times$10$^6$ &&106.6$^{+10.8}_{-16.6}$ &97.5 $^{+5.2 }_{-4.8 }$ &&107.4$^{+8.7 }_{     }$ &99.1 $^{+5.2 }_{-4.9 }$\\
\hline
\end{tabular}
\end{table*}
\setcounter{table}{5}
\begin{table*}
\scriptsize
\caption{continued.}
\centering
\begin{tabular}
{ccccccccccccccccccc}
\hline\hline 
&\multicolumn{2}{c}{$\eta_{\rm bf}$}&&$I'$(4$_{04}$--3$_{03}$)&&$N$(p-H$_2$CO) &$N$(o-H$_2$CO)&$n$(H$_2$)&& \multicolumn{5}{c}{Kinetic temperature}\\
\cline{2-3} \cline{5-5} \cline{7-9} \cline{11-15}
Sources &3$_{03}$--2$_{02}$ &4$_{04}$--3$_{03}$ &&$I'$(3$_{03}$--2$_{02}$)&& \multicolumn{3}{c}{4$_{04}$--3$_{03}$/3$_{03}$--2$_{02}$} &&3$_{21}$--2$_{20}$/3$_{03}$--2$_{02}$ & $T_{\rm LTE}$ && 4$_{22}$--3$_{21}$/4$_{04}$--3$_{03}$ & $T_{\rm LTE}$\\
\cline{7-9} \cline{11-12} \cline{14-15}
&&&&&&cm$^{-2}$&cm$^{-2}$&cm$^{-3}$&&K&K&&K&K\\
\hline 
AGAL332.826$-$00.549 &0.433 &0.575 &&1.15$\pm$0.01 &&1.4$\times$10$^{14}$  &2.1$\times$10$^{14}$ &1.2$^{+0.1}_{-0.1}$$\times$10$^6$ &&145.5$^{+9.3 }_{-7.1 }$ &116.7$^{+5.6 }_{-5.2 }$ &&132.1$^{+3.7 }_{-3.8 }$ &111.5$^{+3.4 }_{-3.3 }$\\
AGAL333.134$-$00.431 &0.524 &0.661 &&0.85$\pm$0.01 &&1.4$\times$10$^{14}$  &3.7$\times$10$^{14}$ &6.5$^{+0.3}_{-0.2}$$\times$10$^5$ &&94.9 $^{+6.1 }_{-5.8 }$ &83.4 $^{+2.9 }_{-2.8 }$ &&87.8 $^{+1.8 }_{-1.9 }$ &75.9 $^{+1.5 }_{-1.4 }$\\
AGAL333.284$-$00.387 &0.586 &0.714 &&0.89$\pm$0.02 &&2.7$\times$10$^{13}$  &7.1$\times$10$^{13}$ &1.2$^{+0.1}_{-0.1}$$\times$10$^6$ &&79.8 $^{+10.2}_{-8.7 }$ &63.6 $^{+8.5 }_{-7.3 }$ &&63.1 $^{+3.0 }_{-2.7 }$ &51.8 $^{+1.9 }_{-1.8 }$\\
AGAL333.314$+$00.106 &0.489 &0.629 &&0.77$\pm$0.03 &&2.5$\times$10$^{13}$  &7.1$\times$10$^{13}$ &8.0$^{+0.1}_{-0.1}$$\times$10$^5$ &&52.8 $^{+5.7 }_{-5.6 }$ &47.1 $^{+5.9 }_{-5.2 }$ &&57.0 $^{+7.1 }_{-6.9 }$ &46.6 $^{+4.7 }_{-4.3 }$\\
AGAL333.604$-$00.212 &0.571 &0.702 &&0.84$\pm$0.01 &&6.7$\times$10$^{13}$  &2.0$\times$10$^{14}$ &6.4$^{+0.3}_{-0.4}$$\times$10$^5$ &&92.6 $^{+6.5 }_{-5.7 }$ &84.7 $^{+5.1 }_{-4.7 }$ &&93.0 $^{+5.0 }_{-4.5 }$ &80.7 $^{+3.8 }_{-3.6 }$\\
\end{tabular}
\label{table:NH2CO-Tkin}
\end{table*}

\setcounter{table}{6}
\begin{table*}
\tiny
\caption{H$_2$CO luminosities.}
\label{table:LH2CO}
\centering
\begin{tabular}
{c c c c c c c cccc}
\hline\hline 
Sources &$L^{\prime}_{\rm H_2CO\ 3_{13}-2_{12}}$ &$L^{\prime}_{\rm H_2CO\ 3_{12}-2_{11}}$ &$L^{\prime}_{\rm H_2CO\ 3_{03}-2_{02}}$ &$L^{\prime}_{\rm H_2CO\ 3_{22}-2_{21}}$ &$L^{\prime}_{\rm H_2CO\ 3_{21}-2_{20}}$ &$L^{\prime}_{\rm H_2CO\ 4_{04}-3_{03}}$ &$L^{\prime}_{\rm H_2CO\ 4_{23}-3_{22}}$ &$L^{\prime}_{\rm H_2CO\ 4_{22}-3_{21}}$ \\
& K km s$^{-1}$ pc$^2$ & K km s$^{-1}$ pc$^2$ &K km s$^{-1}$ pc$^2$ &K km s$^{-1}$ pc$^2$ &K km s$^{-1}$ pc$^2$ &K km s$^{-1}$ pc$^2$ &K km s$^{-1}$ pc$^2$ &K km s$^{-1}$ pc$^2$ \\
\hline 
AGAL008.684$-$00.367  &12.25  &8.66   &5.00  &0.83  &0.99  &5.34   &1.53   &1.49 \\
AGAL008.706$-$00.414  &2.88   &1.76   &0.87  &...   &1.08  &1.09   &...    &...  \\
AGAL010.444$-$00.017  &5.50   &4.70   &...   &...   &...   &3.99   &...    &...  \\
AGAL010.472$+$00.027  &77.46  &57.14  &30.42 &13.97 &14.28 &47.82  &39.42  &34.64\\
AGAL010.624$-$00.384  &51.03  &48.37  &25.01 &7.30  &8.58  &37.53  &17.66  &17.48\\
AGAL012.804$-$00.199  &8.17   &7.90   &4.09  &1.20  &1.18  &5.41   &2.24   &2.27 \\
\hline
\end{tabular}
\end{table*}
\setcounter{table}{6}
\begin{table*}
\tiny
\caption{continued.}
\centering
\begin{tabular}
{c c c c c c c cccc}
\hline\hline 
Sources &$L^{\prime}_{\rm H_2CO\ 3_{13}-2_{12}}$ &$L^{\prime}_{\rm H_2CO\ 3_{12}-2_{11}}$ &$L^{\prime}_{\rm H_2CO\ 3_{03}-2_{02}}$ &$L^{\prime}_{\rm H_2CO\ 3_{22}-2_{21}}$ &$L^{\prime}_{\rm H_2CO\ 3_{21}-2_{20}}$ &$L^{\prime}_{\rm H_2CO\ 4_{04}-3_{03}}$ &$L^{\prime}_{\rm H_2CO\ 4_{23}-3_{22}}$ &$L^{\prime}_{\rm H_2CO\ 4_{22}-3_{21}}$ \\
& K km s$^{-1}$ pc$^2$ & K km s$^{-1}$ pc$^2$ &K km s$^{-1}$ pc$^2$ &K km s$^{-1}$ pc$^2$ &K km s$^{-1}$ pc$^2$ &K km s$^{-1}$ pc$^2$ &K km s$^{-1}$ pc$^2$ &K km s$^{-1}$ pc$^2$ \\
\hline 
AGAL332.826$-$00.549  &14.54  &12.76  &8.59  &3.10  &3.19  &9.91   &5.37   &4.87 \\
AGAL333.134$-$00.431  &28.42  &24.99  &15.60 &4.93  &4.62  &13.22  &5.81   &5.32 \\
AGAL333.284$-$00.387  &10.24  &7.99   &5.17  &1.37  &1.37  &4.59   &1.34   &1.38 \\
AGAL333.314$+$00.106  &5.76   &4.63   &3.14  &0.64  &0.61  &2.42   &0.61   &0.66 \\
AGAL333.604$-$00.212  &22.20  &17.83  &9.90  &3.16  &2.86  &8.29   &3.78   &3.46 \\
AGAL333.656$+$00.059  &3.00   &2.11   &1.75  &...   &...   &1.40   &...    &...  \\
\end{tabular}
\end{table*}

\end{appendix}

\end{document}